\documentclass[aps,pra,reprint,twocolumn,floatfix,groupedaddress,longbibliography]{revtex4-2}

\usepackage{amsmath,amssymb,amstext}
\usepackage{graphicx}
\usepackage{bm}
\usepackage{epstopdf}
\usepackage{mathrsfs} 
\usepackage[usenames,dvipsnames,svgnames]{xcolor}
\usepackage[colorlinks,citecolor=blue]{hyperref} 
\usepackage[normalem]{ulem}
\usepackage{url}

\newcommand{\e}{\mathrm{e}}
\newcommand{\ri}{\mathrm{i}}
\newcommand{\p}{\partial}
\newcommand{\FS}{\mathrm{FS}}
\newcommand{\eff}{\mathrm{eff}}
\newcommand{\m}[1]{\mathcal{#1}}
\newcommand{\co}{\hat{c}}
\newcommand{\ca}{\hat{c}^{\dagger}}

\begin{document}
\title{Flow-equation approach to quantum systems\\ driven by an amplitude-modulated time-periodic force}

\author{Viktor Novi\v{c}enko}
\email[]{viktor.novicenko@tfai.vu.lt}
\homepage[]{http://www.itpa.lt/~novicenko/}
\affiliation{Institute of Theoretical Physics and Astronomy, Vilnius University,
Saul\.{e}tekio Ave.~3, LT-10257 Vilnius, Lithuania}

\author{Giedrius \v{Z}labys}
\email[]{giedrius.zlabys@tfai.vu.lt}
\affiliation{Institute of Theoretical Physics and Astronomy, Vilnius University,
Saul\.{e}tekio Ave.~3, LT-10257 Vilnius, Lithuania}

\author{Egidijus Anisimovas}
\email[]{egidijus.anisimovas@ff.vu.lt}
\affiliation{Institute of Theoretical Physics and Astronomy, Vilnius University,
Saul\.{e}tekio Ave.~3, LT-10257 Vilnius, Lithuania}

\date{\today}

\begin{abstract}
We apply the method of flow equations to describe quantum systems subject to a time-periodic drive with a time-dependent 
envelope. The driven Hamiltonian is expressed in terms of its constituent Fourier harmonics with amplitudes that may 
vary as a function of time. The time evolution of the system is described in terms of the phase-independent effective 
Hamiltonian and the complementary micromotion operator that are generated by deriving and solving the flow equations. 
These equations implement the evolution with respect to an auxiliary flow variable and facilitate a gradual transformation 
of the quasienergy matrix (the Kamiltonian) into a block diagonal form in the extended space. We construct a flow 
generator that prevents the appearance of additional Fourier harmonics during the flow, thus enabling implementation 
of the flow in a computer algebra system. Automated generation of otherwise cumbersome high-frequency expansions 
(for both the effective Hamiltonian and the micromotion) to an arbitrary order thus becomes straightforward for driven 
Hamiltonians expressible in terms of a finite algebra of Hermitian operators. We give several specific examples 
and discuss the possibility to extend the treatment to cover rapid modulation of the envelope. 
\end{abstract}

\maketitle

\section{Introduction}

Periodically driven quantum systems~\cite{Hanggi1998,Holthaus2016tutorial} 
constitute an immensely practical~\cite{Eckardt2017review} and, at the same time, quite tractable intermediate case 
between the two limits of stationary and generic time-dependent systems. The former limit of systems governed by 
time-independent Hamiltonians looks much simpler due to the availability of well-developed notions, methods, 
intuition, and a relatively low numerical complexity. Here, the time evolution is readily available in terms 
of the exponentiated Hamiltonian. On the other hand, the evolution of a generic time-dependent system is described 
by the time-ordered exponential of a Hamiltonian operator that does not commute with itself at different instants 
of time. The evolution operator may be simple to write down, but its evaluation for a nontrivial system requires 
a considerable effort~\cite{Howland1974,Peskin1993,Giscard2020}.

Theoretical description of periodically driven quantum systems, the Floquet
theory~\cite{Shirley1965,Sambe1973,Hanggi1998,Casas2001,Holthaus2016tutorial}, is by now well established 
and forms the basis of Floquet engineering~\cite{Oka2009,Lindner2011,Gu2011,Kitagawa2011,Bukov2015,Eckardt2017review,Oka2019Review,Rudner2020,Bajpai2020,Rodriguez2021} 
that is related to numerous experimental
highlights~\cite{Aidelsburger2011,Aidelsburger2013,Struck2013,Aidelsburger2015,Flaeschner2016,Eckardt2017review,Oka2019Review,Schafer2020,McIver2020},
see also Refs.~\cite{Kitagawa2010,Cayssol2013,FoaTorres2014,Grushin2014,Farrell2016,Huaman2019}. 
The time evolution of a periodically
driven system can be separated into the long-term dynamics governed by a time-independent `effective' Hamiltonian, 
and micromotion that, literally, captures the periodic `micro' motion in the course of a single period 
of the drive~\cite{Rahav2003,Goldman2014,Eckardt2015,Bukov2015,Holthaus2016tutorial,Oka2019Review}. 
Let us emphasize from the outset that two distinct approaches to the formulation of the effective
Hamiltonian are possible~\cite{Bukov2015,Eckardt2017review}. In our paper, as well as in many other
works~\cite{Rahav2003,Goldman2014,Eckardt2015}, we will aim to derive a long-term 
Hamiltonian that is not only stationary but also does not depend on the phase of the drive. We shall reserve the
term \emph{Floquet effective} Hamiltonian to refer to this particular case. In an alternative approach, 
one uses a long-term Hamiltonian that is more appropriately called \emph{Floquet stroboscopic} Hamiltonian
(or just the Floquet Hamiltonian~\cite{Eckardt2017review}). 
The Floquet stroboscopic Hamiltonian~\cite{Casas2001,Mananga2011,Bukov2015,Mananga2016,Holthaus2016tutorial}
is straightforwardly defined as the logarithm of the evolution operator over 
a single period of the drive, and is perfectly suited to describe the evolution between two instants of time
separated by an integer multiple of the period. Note, however, that there exists a whole family of such Hamiltonians, 
parametrized by the phase of the drive or, equivalently, by the initial time.

Systematic construction of Floquet Hamiltonians has been achieved from a number of 
approaches \cite{Verdeny2013,Goldman2014,Eckardt2015,Mikami2016PRB,Vogl2019}, 
such as, for example, various formulations of the 
perturbation theory, transition to the extended space, unitary flows towards diagonalization
(or block diagonalization) defined by a sequence of elemental transformations, and so on. In many cases, 
the Floquet effective Hamiltonian is constructed order by order as an expansion in powers of the inverse frequency 
of the drive~\cite{Bukov2015,Eckardt2015,Goldman2014,Mikami2016PRB,Novicenko2017}. 
Importantly, such expansions support a transparent physical interpretation of the generated terms,
which is a crucial asset in devising synthetic quantum systems that mimic various phenomena of condensed-matter 
physics \cite{Eckardt2015,Holthaus2016tutorial,Eckardt2017review,Oka2019Review}.
For open systems, work on extension to Floquet Lindbladians 
has begun~\cite{Haddafarshi2015,Dai2016,Restrepo2016,Schnell2021,Ikeda2021lindblad,Dai2017}.

An important extension is related to the inclusion of an additional temporal modulation of the envelope 
of the driving signal~\cite{Weinberg2017,Novicenko2017,Zeuch2020,Zeuch2020refuting}. This situation, on the one hand, 
is very practical since it describes a whole class of experimentally relevant setups involving transient 
signals \cite{Desbq2017,Zeuch2020}. On the other hand, although technically the drive is not periodic, it fits well 
into the described scheme of systematic construction of effective Hamiltonians: The high-frequency (or alternative) 
expansions are typically formulated in terms of Fourier components of the driven Hamiltonian and it appears conceptually 
straightforward to endow these components with an additional argument of `slow' time, thus describing the modulation 
of the envelope. Proceeding along this line of thought, the authors of Ref.~\cite{Novicenko2017} obtained 
expressions of the effective Hamiltonian and the micromotion to the second order in the inverse frequency. 
They were able to give a compelling example characterized by a non-Abelian geometric phase \cite{Wilczek1984} 
that emerges precisely from the presence of the time derivative of the Fourier component. 
In another development, a situation relevant to fast coherent manipulation of a qubit was treated 
\cite{Zeuch2020,Zeuch2020refuting}. Here, the authors focused on the accurate stroboscopic description 
and, employing the Magnus-Taylor expansion, were able to proceed to high orders featuring combinations 
of the time derivatives of the Fourier components.

The purpose of the present paper is to explore the field of amplitude-modulated periodically driven 
Hamiltonians relying on the idea of 
unitary flows~\citep{Kehrein2006book,Tomaras2011,Monthus2016,Thomson2018,Vogl2019,Thomson2020,Kelly2020}. 
Originally introduced to eliminate
unwanted couplings in many-body problems~\cite{Wegner1994,Kehrein2006book}, 
these flows draw inspiration from the renormalization semigroup~\cite{Glazek1993,Glazek1994} 
and feature a sequence of elemental (infinitesimal in continuous flows or finite in discrete flows) unitary 
transformations that make the Hamiltonian gradually morph from the original form
to the desired final form, in our case, the block diagonal form in the extended space. 
We set out to systematically construct both the effective Hamiltonian and the micromotion operator. 
As the formulation of the flow is notably nonunique, we develop and use a specific variant,
called the Toda flow, that avoids 
proliferation of constituent Fourier components. When the Hamiltonian can furthermore be expressed
in terms of a finite number of Hermitian (Lie) generators, the flow equations are perfectly suited 
for an automated treatment with the aid of computer algebra systems. We have implemented such a general-purpose
procedure to derive the high-frequency expansion to an arbitrary order and describe several paradigmatic examples.

Our paper has the following structure. In Sec.~\ref{sec:FH}, we introduce the essential notions and tools 
used in the description of periodically driven systems. In Sec.~\ref{sec:FA} we review the idea of unitary 
flows towards diagonalization, and generalize this tool to the block diagonalization of the extended-space
quasienergy operator (also known as the Kamiltonian) in Sec.~\ref{sec:block}. Proceeding to applications, 
in Sec.~\ref{sec:lie} we consider the automated high-frequency expansion and describe several examples 
in Sec.~\ref{sec:ex}. Section~\ref{sec:fast} is devoted to attempts to go beyond the slow modulation 
of the envelope. A number of technical issues, such as derivation of equations and proofs, are relegated 
to Appendices.

%
\section{Floquet Hamiltonian: stroboscopic vs effective}
\label{sec:FH}

\subsection{Floquet stroboscopic Hamiltonian}

Let us consider a quantum system described by a time-periodic Hamilton operator $h(\omega t + 2\pi)=h(\omega t)$ 
and obeying the time-dependent Schr\"{o}dinger equation
\begin{equation}
\label{eq:tdse}
  \ri \hbar \frac{\mathrm{d} \left| \psi (t) \right\rangle}{\mathrm{d} t} = h(\omega t) \left| \psi (t) \right\rangle.
\end{equation}
The time-evolution operator $U(t_{\mathrm{fn}}, t_{\mathrm{in}})$, which defines the unitary evolution of the state-vector 
$\left| \psi (t_{\mathrm{fn}}) \right\rangle = U(t_{\mathrm{fn}},t_{\mathrm{in}}) \left| \psi (t_{\mathrm{in}}) \right\rangle$ 
from the initial time instant $t_{\mathrm{in}}$ to the final time $t_{\mathrm{fn}}$, can be formally written in terms 
of $h(\omega t)$ as a time-ordered exponential
\begin{equation}
\label{eq:teexp}
  U(t_{\mathrm{fn}}, t_{\mathrm{in}})
  = \m{T} \exp\left[-\frac{\ri}{\hbar} \int\limits_{t_{\mathrm{in}}}^{t_{\mathrm{fn}}} h(\omega t) \, \mathrm{d}t \right].
\end{equation}
If the duration of time interval is an exact multiple of the period of the drive $T = 2\pi/\omega$, that is
$t_{\mathrm{fn}} - t_{\mathrm{in}} = NT$, the periodicity of the Hamilton operator ensures that the full evolution 
fulfills $U(t_{\mathrm{fn}},t_{\mathrm{in}}) = \left[U(t_{\mathrm{in}}+T,t_{\mathrm{in}})\right]^N$. 
It is thus natural to introduce the Floquet stroboscopic (FS) Hamiltonian defined as the logarithm 
of the stroboscopic evolution operator 
\begin{equation}
\begin{aligned}
  U(t_{\mathrm{in}} + T,t_{\mathrm{in}}) &= \m{T} \exp\left[-\frac{\ri}{\hbar} 
  \int\limits_{t_{\mathrm{in}}}^{t_{\mathrm{in}}+T} h(\omega t) \mathrm{d}t \right] \\
  &=\exp\left[-\frac{\mathrm{i}}{\hbar} h_{\mathrm{FS}}(t_{\mathrm{in}}) T \right].
\label{eq:fs}
\end{aligned}
\end{equation}
We note that the FS Hamiltonian $h_{\FS}( t_{\mathrm{in}}+T) = h_{\FS}(t_{\mathrm{in}})$, 
albeit stationary, periodically depends on the initial time, or equivalently, on the phase of the drive
in a parametric way.
Thus, when the initial phase is not fixed one needs to deal with all possible initial phases, that is, 
with a whole family of FS Hamiltonians. Another downside of the FS Hamiltonian is that it is unambiguously 
defined only for systems driven in a purely periodic way. In specific cases it was extended to incorporate 
a modulation of the envelope by introducing a custom-made definition~\cite{Weinberg2017,Zeuch2020}. 
The general redefinition of the FS Hamiltonian for modulated systems is not available.

Equation~(\ref{eq:fs}) gives a definition, but does not yet provide a recipe how the FS Hamiltonian 
can be found analytically. For an arbitrary Hamiltonian $h(\omega t)$ an exact analytical expression 
of $h_{\FS}$ cannot be found, thus approximate methods are applicable. One of well-established methods 
relies on the series expansion in powers of the inverse frequency of the drive. If matrix elements 
$h_{ij}(t) = \left\langle \psi_i \right| h(\omega t) \left| \psi_j \right\rangle$ are smaller than 
the characteristic energy of the periodic drive $\hbar \omega$, one can apply the Magnus expansion~\cite{Magnus1954,Blanes2009}
to the unitary evolution featured in Eq.~(\ref{eq:fs}) and obtain the high-frequency expansion 
$h_{\mathrm{FS}} = h_{\mathrm{FS}(0)} + h_{\mathrm{FS}(1)} + h_{\mathrm{FS}(2)}+\mathcal{O}(\omega^{-3})$ with
\begin{subequations}
\label{eq:magfs}
\begin{align}
h_{\mathrm{FS}(0)} &= \frac{1}{2\pi} \int\limits_{\theta_{\mathrm{in}}}^{\theta_{\mathrm{in}}+2\pi} h(\theta) \mathrm{d}\theta, \label{eq:magfs_1} \\
h_{\mathrm{FS}(1)} &= \frac{1}{2(\mathrm{i}\hbar \omega)}\frac{1}{2\pi} \int\limits_{\theta_{\mathrm{in}}}^{\theta_{\mathrm{in}}+2\pi} \int\limits_{\theta_{\mathrm{in}}}^{\theta_1} [h(\theta_1),h(\theta_2)] \mathrm{d}\theta_2 \mathrm{d}\theta_1, \label{eq:magfs_2} \\
h_{\mathrm{FS}(2)} &= \frac{1}{6(\mathrm{i}\hbar \omega)^2}\frac{1}{2\pi} \int\limits_{\theta_{\mathrm{in}}}^{\theta_{\mathrm{in}}+2\pi} \int\limits_{\theta_{\mathrm{in}}}^{\theta_1} \int\limits_{\theta_{\mathrm{in}}}^{\theta_2} \left( \left[ h(\theta_1),[h(\theta_2),h(\theta_3)] \right] \right. \nonumber \\
&\left.+ \left[ h(\theta_3),[h(\theta_2),h(\theta_1)] \right] \right) \mathrm{d}\theta_3 \mathrm{d}\theta_2 \mathrm{d}\theta_1, \label{eq:magfs_3}
\end{align}
\end{subequations}
where $\theta_{\mathrm{in}}=\omega t_{\mathrm{in}}$ is the initial phase. Further terms of the Magnus expansion 
can be constructed recursively~\cite{Blanes2009}.

A different method to obtain the FS Hamiltonian employs a continuous flow of unitary transformations that gradually 
transform the time-periodic $h(\omega t)$ into the stationary $h_{\mathrm{FS}}$~\cite{Vogl2019}. This approach
involves solving the flow equation, which typically can not be done analytically as well. Instead, the flow equation
is used as a starting point for further approximations. In this paper, we present a similar flow approach, 
formulated not for the FS Hamiltonian but for a phase-independent Floquet effective Hamiltonian that also includes
the effects of a time-dependent drive envelope.

\subsection{Floquet effective Hamiltonian}

To proceed, let us consider a quantum system described by the Hamiltonian $h(\omega t, t)$ which is a $2\pi$-periodic 
function of the first argument, and also allows for an additional modulation of the envelope through the dependence
on time in the second argument. Following Refs.~\cite{Guerin2003,Novicenko2017} 
we will study the whole family of solutions 
$\left| \psi_{\theta} (t) \right\rangle$ where $\theta \in [0,2\pi)$ represents the initial phase. The Hamiltonian 
$h(\omega t+\theta,t)$ thus also becomes dependent on the initial phase. The evolution is governed by the time-dependent 
Schr\"{o}dinger equation
\begin{equation}
\label{eq:tdse_ip}
  \ri \hbar \frac{\mathrm{d} \left| \psi_{\theta} (t) \right\rangle}{\mathrm{d} t}
  = h(\omega t+\theta,t) \left| \psi_{\theta} (t) \right\rangle,
\end{equation}
and the initial condition at $t_{\mathrm{in}}$ is assumed to be $\theta$-periodic, i.e.\
$\left| \psi_{\theta+2\pi} (t_{\mathrm{in}}) \right\rangle=\left| \psi_{\theta} (t_{\mathrm{in}}) \right\rangle$. 
Thus, it remains periodic throughout the evolution, and can be expanded in a Fourier series as
\begin{equation}
\label{eq:stvec_fourier}
  \left| \psi_{\theta} (t) \right\rangle
  = \sum\limits_{n=-\infty}^{+\infty} \mathrm{e}^{\mathrm{i}n\theta} \left| \psi^{(n)} (t) \right\rangle.
\end{equation}
The corresponding expansion of the Hamiltonian reads
\begin{equation}
\label{eq:ham_fourier}
h(\omega t+\theta,t)=\sum\limits_{n=-\infty}^{+\infty} \mathrm{e}^{\mathrm{i}n(\omega t+\theta)} h^{(n)} (t),
\end{equation}
with $\left[ h^{(n)} (t)\right]^{\dagger}=h^{(-n)} (t)$ imposed by hermiticity. Note that the Fourier components
$h^{(n)} (t)$ are time dependent as a consequence of the second argument in the initial Hamiltonian, $h(\omega t+\theta,t)$.

The state vector $\left| \psi_{\theta} (t) \right\rangle$ parametrically depends on $\theta$ and is assumed to belong 
to a physical Hilbert space $\mathscr{H}$. Our next step is to introduce an extended vector space and reformulate 
Eq.~(\ref{eq:tdse_ip}) in a new formalism. The idea to introduce the extended vector space for purely time-periodic 
quantum systems first appeared in Ref.~\cite{Sambe1973}. Because of the $\theta$-periodicity, it is natural to introduce 
the space $\mathscr{T}$ of square-integrable functions periodic on the interval~$[0,2\pi)$. 
The exponents $\e^{\ri n \theta}$ with $n \in \mathbb{Z}$ form the standard orthonormal basis $|n\rangle$ 
with a dot product defined as 
$\left\langle m \middle| n \right\rangle=(2\pi)^{-1}\int_{-\pi}^{\pi} \e^{-\ri m \theta} \e^{\ri n \theta} \mathrm{d}\theta=\delta_{mn}$. 
Now one can build the extended space $\mathscr{L} =  \mathscr{T} \otimes \mathscr{H}$ as the tensor product of the physical 
space and the space of $\theta$-periodic functions. The state vector in Eq.~(\ref{eq:stvec_fourier}) can be interpreted 
as a time-dependent vector (we will use double bra-ket notations for the vectors in the extended space and calligraphic letters 
for operators acting in the extended space)
\begin{equation}
\label{eq:stvec_ext}
  \left. \left| \psi (t) \right\rangle\!\right\rangle
  =  \sum\limits_{n=-\infty}^{+\infty} \left| n \right\rangle \otimes \left| \psi^{(n)} (t) \right\rangle.
\end{equation}
belonging to the space $\mathscr{L}$ where $\theta$ is no longer a parameter but an intrinsic variable of the space. 
The Hamiltonian~(\ref{eq:ham_fourier}) can also be interpreted as an operator acting in $\mathscr{L}$:
\begin{equation}
\label{eq:ham_ext}
\mathcal{H}(\omega t,t)=\sum\limits_{n,m=-\infty}^{+\infty}  \left| n+m \right\rangle \mathrm{e}^{\mathrm{i}n \omega t} \left\langle m \right| \otimes h^{(n)}(t).
\end{equation}
However, the above procedure did not yet simplify the analysis as the operator~(\ref{eq:ham_ext}) still 
contains the time-periodic argument $\omega t$. To eliminate this dependence one applies the time-dependent 
unitary transformation
\begin{equation}
\label{eq:unit}
\mathcal{U}=\mathrm{e}^{\omega t \frac{\mathrm{d}}{\mathrm{d}\theta}}=\sum\limits_{n=-\infty}^{+\infty} \left| n \right\rangle \mathrm{e}^{\mathrm{i}n\omega t} \left\langle n \right| \otimes \mathbf{1}_{\mathscr{H}}
\end{equation}
to Eq.~(\ref{eq:tdse_ip}), which should be interpreted in the extended space. From Eq.~(\ref{eq:unit}) 
one can see that the unitary transformation shifts the phase variable: 
$\mathcal{U}^{\dagger} \theta \mathcal{U}=\theta-\omega t$. 
Therefore the argument $\omega t +\theta$ simplifies to just $\theta$. The transformed state vector 
$\left. \left| \phi (t) \right\rangle\!\right\rangle=\mathcal{U}^{\dagger}\left. \left| \psi (t) \right\rangle\!\right\rangle$ 
obeys the time-dependent Schr\"{o}dinger equation
\begin{equation}
\label{eq:tdse_kam}
  \ri \hbar \frac{\mathrm{d} \left. \left| \phi (t) \right\rangle\!\right\rangle}{\mathrm{d} t}
  = \mathcal{K}(t) \left. \left| \phi (t) \right\rangle\!\right\rangle,
\end{equation}
with
\begin{equation}
\label{eq:h_to_k}
  \m{K}(t) = \m{U}^{\dagger} \m{H}(\omega t,t) \m{U} - \ri\hbar \m{U}^{\dagger} \frac{\mathrm{d}\m{U}}{\mathrm{d}t},
\end{equation}
which is often referred to as the Kamiltonian. In the differential form, the obtained Kamiltonian reads
\begin{equation}
\label{eq:kam_diff_form}
\mathcal{K}(t)=-\mathrm{i}\hbar \omega \frac{\mathrm{d}}{\mathrm{d}\theta}\otimes \mathbf{1}_{\mathscr{H}}+\sum\limits_{n=-\infty}^{+\infty} \mathrm{e}^{\mathrm{i}n\theta} \otimes h^{(n)} (t),
\end{equation}
while in the bra-ket notation it reads
\begin{equation}
\begin{aligned}
\mathcal{K}(t)=&\sum\limits_{n=-\infty}^{+\infty} \left|n\right\rangle n\hbar \omega \left\langle n \right|\otimes \mathbf{1}_{\mathscr{H}} \\
 &+\sum\limits_{n,m=-\infty}^{+\infty} \left|m\right\rangle \left\langle n \right| \otimes h^{(m-n)}(t).
\end{aligned}
\label{eq:kam_basis_form}
\end{equation}
From Eqs.~(\ref{eq:kam_diff_form}) and (\ref{eq:kam_basis_form}) one can see that the Kamiltonian does not contain 
the time-periodic argument $\omega t$, and depends on time only through the second argument in the original Hamiltonian, 
$h(\omega t,t)$. If the original Hamiltonian is purely periodic, the Kamiltonian becomes time-independent, and one can 
rely on methods developed for time-independent operators, e.g.\ various formulations of the perturbation theory. 
However, this comes at the cost of working in the extended space $\mathscr{L}$, while the original problem was 
formulated in the simpler physical space $\mathscr{H}$.

The Kamiltonian $\mathcal{K}(t)$ can be represented as an infinite matrix where the matrix elements 
$K_{mn}(t)=\left\langle m \right| \mathcal{K}(t) \left|n \right\rangle$ are operators in $\mathscr{H}$. 
Such a matrix possesses some obvious symmetries, for example, the first upper diagonal is filled with 
copies of the same operator, $K_{n,n+1}=h^{(-1)}(t)$. In order to make this symmetry explicit, we introduce 
the shift operator $P_m = \sum_n \left|n+m \right\rangle \left\langle n\right|$ and the `number' operator 
$N = \sum_n \left|n \right\rangle n\left\langle n\right|$ (both acting in $\mathscr{T}$) 
and rewrite Eq.~(\ref{eq:kam_basis_form}) in terms of these operators
\begin{equation}
  \mathcal{K}(t) = \hbar\omega N \otimes \mathbf{1}_{\mathscr{H}} 
  + \sum\limits_{m=-\infty}^{+\infty}P_m \otimes h^{(m)}(t).
\label{eq:kam_sh_form}
\end{equation}
This expression constitutes a concise statement of the problem.

Next step is to block diagonalize the Kamiltonian. More concretely, we assume that there exists a time-dependent unitary 
operator $\mathcal{D}(t)$ such that the transformed state vector 
$\left. \left| \chi (t) \right\rangle\!\right\rangle=\mathcal{D}^{\dagger}\left. \left| \phi (t) \right\rangle\!\right\rangle$ 
obeys the time-dependent Schr\"{o}dinger equation
\begin{equation}
\label{eq:tdse_dkam}
 \ri \hbar \frac{\mathrm{d} \left. \left| \chi (t) \right\rangle\!\right\rangle}{\mathrm{d} t}
 = \mathcal{K}_D(t) \left. \left| \chi (t) \right\rangle\!\right\rangle,
\end{equation}
with the block-diagonal Kamiltonian
\begin{equation}
\begin{aligned}
\mathcal{K}_D(t) &= \mathcal{D}^{\dagger} \mathcal{K}(t) \mathcal{D}- \mathrm{i}\hbar\mathcal{D}^{\dagger} \frac{\mathrm{d}\mathcal{D}}{\mathrm{d}t}  \\
&= \hbar\omega N \otimes \mathbf{1}_{\mathscr{H}} +P_0 \otimes h_{\mathrm{eff}}(t).
\end{aligned}
\label{eq:dkam}
\end{equation}
The operator $h_{\mathrm{eff}}(t)$ is the Floquet effective (FE) Hamiltonian~\cite{Rahav2003,Goldman2014,Eckardt2015}
acting in the physical space $\mathscr{H}$. Crucially, $h_{\mathrm{eff}}(t)$ does not depend on the initial phase. 
Indeed, we started from $h(\omega t+ \theta,t)$ which depends on the initial phase $\theta$ and arrived at
the $\theta$-independent $h_{\mathrm{eff}}(t)$.

The block-diagonalization procedure and symmetries allow us to restrict the analysis to the zeroth Floquet 
subspace $\mathscr{L}_0$ spanned by the vectors $\left| 0 \right\rangle \otimes \left| \psi_i \right\rangle$, 
where $\left| \psi_i \right\rangle$ are the basis vectors of the physical space $\mathscr{H}$ and 
$i\in \left\lbrace 1,2,\ldots,\mathrm{dim}[\mathscr{H}] \right\rbrace$. Thus the subspace $\mathscr{L}_0$ is isomorphic to the physical space $\mathscr{H}$. The solutions for other Floquet subspaces 
$\mathscr{L}_n$ with $n\neq 0$ can be easily recovered from the solution in the subspace $\mathscr{L}_0$. 
Indeed, let us assume that at the initial time the state vector in Eq.~(\ref{eq:tdse_dkam}) reads 
$\left. \left| \chi (t_{\mathrm{in}}) \right\rangle\!\right\rangle 
= \sum_{n=-\infty}^{+\infty} \left| n \right\rangle \otimes \left| \chi_n(t_{\mathrm{in}}) \right\rangle$. 
Then Eq.~(\ref{eq:tdse_dkam}) decouples into copies of the Schr\"{o}dinger equation
defined in the subspaces $\mathscr{L}_n$ as
\begin{equation}
\label{eq:tdse_sub_n}
  \ri \hbar \frac{\mathrm{d} \left| \chi_n (t) \right\rangle}{\mathrm{d} t}
  = \left[n\hbar \omega+ h_{\mathrm{eff}}(t) \right] \left| \chi_n (t) \right\rangle.
\end{equation}
By defining the evolution operator of Eq.~(\ref{eq:tdse_sub_n}) for $n=0$ as a time-ordered exponential
\begin{equation}
\label{eq:evo}
U_{\mathrm{eff}}(t_{\mathrm{fn}},t_{\mathrm{in}})=\mathcal{T} \exp \left[-\frac{\mathrm{i}}{\hbar} \int\limits_{t_{\mathrm{in}}}^{t_{\mathrm{fn}}} h_{\mathrm{eff}}(t) \mathrm{d}t \right],
\end{equation}
one can write down the full evolution of Eq.~(\ref{eq:tdse_dkam}) as
\begin{equation}
\begin{aligned}
  &\left. \left| \chi (t_{\mathrm{fn}}) \right\rangle\!\right\rangle 
  = \mathcal{U}_{\mathrm{eff}}(t_{\mathrm{fn}},t_{\mathrm{in}})\left. \left| \chi (t_{\mathrm{in}}) \right\rangle\!\right\rangle \\
  &= \sum\limits_{n=-\infty}^{+\infty} 
  \,\mathrm{e}^{-\mathrm{i}n\omega (t_{\mathrm{fn}}-t_{\mathrm{in}})}\left| n \right\rangle
  \otimes U_{\mathrm{eff}}(t_{\mathrm{fn}},t_{\mathrm{in}})
  \left| \chi_n(t_{\mathrm{in}}) \right\rangle.
\end{aligned}
\label{eq:evo_ex}
\end{equation}

The Kamiltonians before the block diagonalization $\mathcal{K}(t)$ and after the block diagonalization $\mathcal{K}_D(t)$ 
possess the same symmetry allowing them to be written in terms of the shift operator $P_m$ and the number operator $N$. 
It implies~\cite{Novicenko2017}, that the unitary operator $\mathcal{D}(t)$ can be written as 
$\mathcal{D}(t) = \sum_{m=-\infty}^{+\infty} P_m \otimes D^{(m)}(t)$, and because of unitarity, 
the operators $D^{(m)}(t)$ satisfy
\begin{equation}
\sum\limits_{m=-\infty}^{+\infty}\left( D^{(m)}(t)\right)^{\dagger} D^{(m+l)}(t)=\delta_{0l}\mathbf{1}_{\mathscr{H}}.
\label{eq:unyt_d}
\end{equation}
Let us recall that starting from the state vector $\left| \psi_{\theta} (t) \right\rangle$ in the physical space $\mathscr{H}$ 
we reformulated the task in the extended space $\mathscr{L}$ for the state vector $\left. \left| \psi (t) \right\rangle\!\right\rangle$, 
then performed the unitary transformation~(\ref{eq:unit}) and obtained the state vector $\left. \left| \phi (t) \right\rangle\!\right\rangle$,
subsequently applied the unitary transformation $\mathcal{D}(t)$ and arrived at the state vector 
$\left. \left| \chi (t) \right\rangle\!\right\rangle$ for which we are able to write down the solution~(\ref{eq:evo_ex}). 
The full evolution in the extended space reads
\begin{equation}
\begin{aligned}
  &\left. \left| \psi (t_{\mathrm{fn}}) \right\rangle\!\right\rangle \\
  &= \mathcal{U}(t_{\mathrm{fn}}) \mathcal{D}(t_{\mathrm{fn}})\mathcal{U}_{\mathrm{eff}}(t_{\mathrm{fn}},t_{\mathrm{in}}) \mathcal{D}^{\dagger}(t_{\mathrm{in}}) \mathcal{U}^{\dagger}(t_{\mathrm{in}}) \left. \left| \psi (t_{\mathrm{in}}) \right\rangle\!\right\rangle.
\end{aligned}
\label{eq:evo_full_ex}
\end{equation}
However, one needs to transform back to the physical space. In order to write down the solution in the physical space, 
for simplicity, let us assume that at the initial time moment $\left| \psi_{\theta} (t_{\mathrm{in}}) \right\rangle$ has 
only the zeroth Fourier component 
$\left| \psi_{\theta} (t_{\mathrm{in}}) \right\rangle=\left| \psi^{(0)} (t_{\mathrm{in}}) \right\rangle$. 
It means that analyzing the $\theta$-dependent family of solutions we have the same initial state vector 
$\left| \psi^{(0)} (t_{\mathrm{in}}) \right\rangle$ for all $\theta$. 
Substituting such an initial state vector into Eq.~(\ref{eq:evo_full_ex}) gives
\begin{equation}
\begin{aligned}
&\left. \left| \psi (t_{\mathrm{fn}}) \right\rangle\!\right\rangle 
= \sum\limits_{m,n=-\infty}^{+\infty} \left| n-m \right\rangle 
\otimes \left( D^{(n)}(t_{\mathrm{fn}}) \,\e^{\ri n\omega t_{\mathrm{fn}}} \right)\\
& \times U_{\mathrm{eff}}(t_{\mathrm{fn}},t_{\mathrm{in}}) \left( D^{(m)}(t_{\mathrm{in}}) \,\e^{\ri m\omega t_{\mathrm{in}}} \right)^{\dagger} \left| \psi^{(0)} (t_{\mathrm{in}}) \right\rangle.
\end{aligned}
\label{eq:evo_full_ex1}
\end{equation}
Translating the ket $\left| n-m \right\rangle$ into the corresponding exponent  $\e^{\ri (n-m)\theta}$, this expression 
can be interpreted as a vector in the physical space which parametrically depends on $\theta$
\begin{equation}
\begin{aligned}
\left| \psi_{\theta} (t_{\mathrm{fn}}) \right\rangle =& U_{\mathrm{micro}}(\omega t_{\mathrm{fn}}+\theta,t_{\mathrm{fn}}) U_{\mathrm{eff}}(t_{\mathrm{fn}},t_{\mathrm{in}}) \\
&U_{\mathrm{micro}}^{\dagger}(\omega t_{\mathrm{in}}+\theta,t_{\mathrm{in}}) \left| \psi^{(0)} (t_{\mathrm{in}}) \right\rangle,
\end{aligned}
\label{eq:evo_full}
\end{equation}
where the unitary operator
\begin{equation}
U_{\mathrm{micro}}(\omega t+\theta,t)= \sum\limits_{n=-\infty}^{+\infty} D^{(n)}(t) \mathrm{e}^{\mathrm{i}n(\omega t+\theta)}
\label{eq:mic}
\end{equation}
is called the micromotion operator. The unitarity of Eq.~(\ref{eq:mic}) follows from Eq.~(\ref{eq:unyt_d}). 
The motivation 
behind such a name reflects the fact that it represents deviations from the effective evolution governed 
by $h_{\mathrm{eff}}(t)$. Importantly, the micromotion depends on the initial phase $\theta$ and is applied only 
at the initial and the final time moments, while the effective evolution is $\theta$-independent but applies 
throughout the time interval. Note that the unitary operator $\mathcal{D}(t)$ as well as $h_{\mathrm{eff}}(t)$ are not easy to find. We will construct their explicit expressions in the high-frequency regime, i.e., by analyzing the high-frequency expansions of the flow equations.

\subsection{A subtlety}

Let us now discuss a special case of a time-dependent perturbation that acts only within a certain time interval
fully contained between the initial and the final times, $t_{\mathrm{in}}$ and $t_{\mathrm{fn}}$, within 
the observed evolution. In other words, the Hamiltonian at the initial and the final time instants is stationary 
$h(\omega t_{\mathrm{in}}+\theta,t_{\mathrm{in}})=h(\omega t_{\mathrm{fn}}+\theta,t_{\mathrm{fn}})=h_0$, however,
at intermediate times --- as the perturbation is gradually turned on and subsequently off --- the Hamiltonian 
is characterized by a periodic time dependence with a time-dependent envelope. 
Since the operator $\mathcal{D}(t)$ seeks to block diagonalize the Kamiltonian~(\ref{eq:kam_sh_form}) and by construction 
$\mathcal{K}(t_{\mathrm{in}})=\mathcal{K}(t_{\mathrm{fn}})$ is already block diagonal, one may be tempted to assume that 
$\mathcal{D}(t_{\mathrm{in}})=\mathcal{D}(t_{\mathrm{fn}})=\mathbf{1}_{\mathscr{L}}$ and, thus the micromotion 
operator at $t_{\mathrm{in}}$ and $t_{\mathrm{fn}}$ is equal to the unit operator. This, however, leads to an obvious 
contradiction: 
if $D(t_{\mathrm{in}}) = D(t_{\mathrm{fn})} = \mathbf{1}_{\mathscr{L}}$, then
according to Eq.~(\ref{eq:evo_full}) the final state does not depend on the initial phase $\theta$, 
while the shape of the perturbation, in general, depends on $\theta$. This contradiction can be resolved by analyzing 
the equation for $\mathcal{D}(t)$. Let us say that the FE Hamiltonian $h_{\mathrm{eff}}(t)$ is known from a different 
source [not from Eq.~(\ref{eq:dkam})], then from Eq.~(\ref{eq:dkam}) one can write
\begin{equation}
\ri\hbar \frac{\mathrm{d}\mathcal{D}}{\mathrm{d}t}=\mathcal{K}\mathcal{D}-\mathcal{D}\mathcal{K}_D.
\label{eq:d_evol}
\end{equation}
Having solved the differential equation (\ref{eq:d_evol}) with the initial condition 
$\mathcal{D}(t_{\mathrm{in}})=\mathbf{1}_{\mathscr{L}}$ we will obtain a particular 
$\mathcal{D}(t_{\mathrm{fn}})\neq \mathbf{1}_{\mathscr{L}}$ which will not be of the block-diagonal form 
--- featuring nonzero $D^{(n)}(t_{\mathrm{fn}})$ for $n \neq 0$ --- and moreover, will be time dependent, 
meaning that $\mathcal{D}(t_{\mathrm{fn}}-\mathrm{d}t)\neq \mathcal{D}(t_{\mathrm{fn}})$ for an infinitesimal $\mathrm{d}t$. 
But $\mathcal{D}(t_{\mathrm{fn}})$ will still be able to block diagonalize the Kamiltonian: After application 
of Eq.~(\ref{eq:dkam}) to the block-diagonal operator $\mathcal{K}(t_{\mathrm{fn}})$ the transformed operator 
$\mathcal{K}_D(t_{\mathrm{fn}})$ will be also block diagonal. Because of such features of the operator 
$\mathcal{D}(t_{\mathrm{fn}})$, the micromotion $U_{\mathrm{micro}}(\omega t_{\mathrm{fn}}+\theta,t_{\mathrm{fn}})$ 
will become $\theta$-dependent, which resolves the contradiction. In fact, this is an open question: When is it 
possible to find such $h_{\mathrm{eff}}(t)$ that both $\mathcal{D}(t_{\mathrm{in}})$ and $\mathcal{D}(t_{\mathrm{fn}})$ 
will have a block-diagonal form (note, that if $\mathcal{D}(t)$ has the block-diagonal form, 
$\mathcal{D}(t)=P_0 \otimes D^{(0)}(t)$, then $U_{\mathrm{micro}}(\omega t+\theta,t)=D^{(0)}(t)$ 
does not depend on $\theta$)? At least one satisfactory case is when the high-frequency expansion 
is applicable~\cite{Novicenko2017}. From the high-frequency expansion it is intuitively clear that 
the envelope of the perturbation changes slowly over one period of periodic signal, and thus, the final 
state~(\ref{eq:evo_full}) does not depend on the initial phase $\theta$.

\subsection{Summary so far}

Let us summarize this rather lengthy section. We distinguish two different approaches to describe a periodically driven 
quantum system: one based on the Floquet stroboscopic Hamiltonian and another based on the Floquet effective Hamiltonian. 
The FS Hamiltonian can be defined for purely periodic systems while the FE Hamiltonian is possible 
to define even when the original Hamiltonian is modulated by a time-dependent envelope. The FS Hamiltonian depends on the 
initial phase of the periodic perturbation and describes the evolution of the quantum system on a time interval which 
is an integer multiple of the period. In contrast, the FE Hamiltonian does not depend on the initial phase and gives 
the evolution of the quantum system on an arbitrary time interval. In principle, both Hamiltonians can be used
to describe the evolution over arbitrary time intervals: The effective evolution should be sandwiched 
by the micromotion operators which, in general, depend on the initial phase, whereas the stroboscopic description
also comes with its own version of phase-dependent micromotion \cite{Holthaus2016tutorial}. The FE Hamiltonian is most useful
when the micromotion operator does not depend on the initial phase, for example in the high-frequency limit, 
when the periodic perturbation with a slowly-modulated amplitude does not act at the beginning and at the end 
of the time interval. Both Hamiltonians (FS and FE) are well defined but difficult to find analytically 
for an arbitrary frequency, thus, one often resorts to 
a high-frequency expansion. 
For the FS Hamiltonian the high-frequency expansion implies that $\omega$ sets the dominant scale in the sense that any matrix element $h_{ij}(\omega t) = \left\langle \psi_i \right| h(\omega t) \left| \psi_j \right\rangle$ when expanded into the Fourier series $h_{ij}(\omega t)=\sum_m h_{ij}^{(m)} \exp(\mathrm{i}m\omega t)$ is characterized by Fourier amplitudes $|h_{ij}^{(m)}|$ that are smaller than the characteristic energy of the periodic drive $\hbar\omega$. For the FE Hamiltonian the high-frequency expansion additionally assumes that the time-dependent Fourier amplitudes $|h_{ij}^{(m)}(t)|$ only slowly depend on time, $\mathrm{d}|h_{ij}^{(m)}(t)|/\mathrm{d}t \ll \omega |h_{ij}^{(m)}(t)|$. 
In such expansion formulas, the FE Hamiltonian reads simpler 
than the FS Hamiltonian due to the independence on the initial phase, 
and hereafter we will focus exclusively on the FE Hamiltonian. In Sec.~\ref{sec:mint} and Sec.~\ref{sec:toda} we derive flow equations that could, in principle, be used to find the FE Hamiltonian for an arbitrary frequency. Yet, the solution of the flow equations is as difficult to obtain as it is to find the unitary operator $\mathcal{D}(t)$ that block-diagonalizes the Kamiltonian. Thus Appendices~\ref{sec:App-B} and~\ref{sec:App-C} and Sec.~\ref{sec:lie} are devoted to the solution of the flow equations in the form of the high-frequency expansion. Finally in Sec.~\ref{sec:fast} we perform the high-frequency expansion by relaxing the requirement of the slow time dependence, i.e., the inequality $\mathrm{d}|h_{ij}^{(m)}(t)|/\mathrm{d}t \ll \omega |h_{ij}^{(m)}(t)|$ no longer imposed.

%
\section{Flow towards diagonalization}
\label{sec:FA}

Originally introduced in the context of many-particle problems~\cite{Wegner1994}, flow equations 
establish a method to gradually bring a Hamiltonian closer to the diagonal form by applying 
a sequence of specifically tailored unitary transformations. In practice, a time-independent Hamiltonian $H$ 
is represented as a finite or infinite Hermitian matrix, $H_{ij} = H_{ji}^*$, and endowed with an auxiliary 
dependence on the flow variable $s$. Application of successive transformations is associated with the forward 
motion along the $s$ axis such that at the beginning of the flow $H(s = 0)$ matches the original Hamiltonian, 
while at the end of the flow the matrix $H(s\rightarrow +\infty)$ assumes the diagonal form.

Two kinds of flows can be distinguished: In a discrete flow, the flow variable follows a sequence 
of integer values $s = 0, 1, 2,\ldots$, and in a continuous (or differential) flow, the value of $s$ 
grows continuously. In the former case, the equation for a single step reads
\begin{equation}
  H(s+1) = U(s)H(s)U^{\dagger}(s) = \mathrm{e}^{\eta(s)}H(s)\mathrm{e}^{-\eta(s)},
\label{eq:flow_discr}
\end{equation}
where the anti-Hermitian operator $\eta(s) = -\eta^{\dagger}(s)$ is known as the flow generator. The generator 
must be constructed in such a way that with each consecutive iteration the Hamiltonian $H(s+1)$ becomes 
closer, according to some measure,
to the diagonal form than its predecessor $H(s)$. Perhaps the simplest way to diagonalize a finite Hermitian 
matrix in an iterative manner is to use the Jacobi rotations \cite{Golub2000diag,Monthus2016}. The algorithm is based 
on the idea that at each step $s$ one may restrict transformations to a two-dimensional vector space spanned 
by two basis vectors $\left|i\right\rangle$ and $\left|j\right\rangle$, and perform a rotation in this subspace so 
that the off-diagonal elements $H_{ij}(s+1)$ and $H_{ji}(s+1)$ vanish, and their weight is absorbed into the diagonal 
elements, $H_{ii}(s+1)$ and $H_{jj}(s+1)$. Even though subsequent transformations will partially restore previously 
eliminated matrix elements, all off-diagonal entries will eventually decay to zero.

In analogy to the discrete flow~(\ref{eq:flow_discr}), application of a continuous flow also leads to a recursive 
relation between the Hamiltonians $H(s+\mathrm{d}s)$ and $H(s)$, but with an infinitesimal (close to the identity) 
unitary transformation
\begin{equation}
\begin{aligned}
H(s+\mathrm{d}s) &= \mathrm{e}^{\eta(s)\mathrm{d}s}H(s)\mathrm{e}^{-\eta(s)\mathrm{d}s} \\
&= H(s)+[\eta(s),H(s)]\mathrm{d}s+\mathcal{O}\left((\mathrm{d}s)^2\right).
\end{aligned}
\label{eq:flow_cont}
\end{equation}
From Eq.~(\ref{eq:flow_cont}) we read off the differential equation representing the evolution of the Hamiltonian 
\begin{equation}
\frac{\mathrm{d}H(s)}{\mathrm{d}s} = [\eta(s),H(s)],
\label{eq:flow_diff_eq}
\end{equation}
where the initial condition $H(0)$ is the original matrix to be diagonalized. The flow generator $\eta(s)$ is typically 
constructed from the elements of $H(s)$, but the recipe is not unique.

During the flow, the Hamiltonian undergoes only unitary transformations, thus the trace of an integer power $p$ of the 
matrix $H(s)$ is preserved for any $s$~\cite{Monthus2016}:
\begin{equation}
I_p = \mathrm{Tr}\left( H^p(s) \right),
\label{eq:inv_flow}
\end{equation}
with $I_p$ independent of the flow variable $s$. The case of $p = 2$ (corresponding to the squared Frobenius norm) 
gives
\begin{equation}
\begin{aligned}
I_2 &= \sum\limits_n \sum\limits_m H_{nm}(s) H_{mn}(s) \\
&= \sum\limits_n H_{nn}^2(s)+\sum\limits_n \sum\limits_{m\neq n} |H_{nm}(s)|^2 \\
&= I_2^{\mathrm{diag}}(s)+ I_2^{\mathrm{off}}(s),
\end{aligned}
\label{eq:Ip2}
\end{equation}
and, because $\mathrm{d}I_2/\mathrm{d}s=0$, the diagonal part and the off-diagonal part evolve in opposite ways:
\begin{equation}
  \frac{\mathrm{d}I_2^{\mathrm{diag}}}{\mathrm{d}s}=-\frac{\mathrm{d}I_2^{\mathrm{off}}}{\mathrm{d}s}.
\label{eq:Id_and_off}
\end{equation}
Full diagonalization means $I_2^{\mathrm{off}}=0$ and is assumed to be achieved at $s\rightarrow + \infty$. Thus, 
the natural requirement for the flow towards diagonalization is to have a non-negative derivative of the diagonal 
part 
\begin{equation}
  \frac{\mathrm{d}I_2^{\mathrm{diag}}}{\mathrm{d}s}\geqslant 0.
\label{eq:Id_pos}
\end{equation}
It turns out that the requirement~(\ref{eq:Id_pos}) can be easily fulfilled. Let us write down an explicit expression 
for the derivative in Eq.~(\ref{eq:Id_pos}) in terms of the matrix elements \cite{Monthus2016} 
\begin{equation}
\begin{aligned}
\frac{\mathrm{d}I_2^{\mathrm{diag}}}{\mathrm{d}s} =& \sum\limits_n H_{nn}(s) \frac{\mathrm{d}H_{nn}(s)}{\mathrm{d}s} + \sum\limits_k \frac{\mathrm{d}H_{kk}(s)}{\mathrm{d}s} H_{kk}(s) \\
=& \sum\limits_{k} \sum\limits_{n\neq k} \left[ H_{nn}(s)-H_{kk}(s) \right] \\
&\times \left[ \eta_{nk}(s)H_{kn}(s)+\eta^*_{nk}(s)H^*_{kn}(s) \right].
\end{aligned}
\label{eq:Id}
\end{equation}
Note, that the contribution from $n=k$ can be omitted because for $n=k$ expressions in both square brackets are equal to zero: 
the first bracket is obviously zero while the second one vanishes due to the anti-Hermitian nature of the generator
\begin{equation}
\eta_{nk}(s)=-\eta^*_{kn}(s).
\label{eq:gen_anti}
\end{equation}
One can chose the generator's matrix elements of the form~\cite{Monthus2016}
\begin{equation}
\eta_{nk}(s)=H_{nk}(s)f\left(H_{nn}(s)-H_{kk}(s) \right),
\label{eq:gen_form}
\end{equation}
where $f(x)$ is a real-valued function that must be odd, $-f(x) = f(-x)$, to ensure that the generator 
will be anti-Hermitian, cf.\ Eq.~(\ref{eq:gen_anti}). The purely imaginary diagonal elements of the generator can, 
without loss of generality, be set to zero, because from Eq.~(\ref{eq:flow_diff_eq}) 
one can see that they do not appear in the evolution of the matrix elements of $H(s)$. Combining 
Eqs.~(\ref{eq:gen_form}) and (\ref{eq:Id}) we see that the defining condition~(\ref{eq:Id_pos}) 
is satisfied if $xf(x) \geqslant 0$. Let us list three notable choices: $f(x) = x$ gives the canonical Wegner 
generator~\cite{Wegner1994}, $f(x) = 1/x$ was analyzed by White~\cite{White2002} and $f(x) = \operatorname{sgn}(x)$ 
used in Ref.~\cite{Hergert2014}. We refer to Ref.~\cite{Monthus2016} for the review of pros and cons of each 
generator. Note that the requirement~(\ref{eq:Id_pos}) does not yet guarantee the full diagonalization, because 
at some point the flow may stall, giving 
$\mathrm{d}I_2^{\mathrm{diag}}/\mathrm{d}s = \mathrm{d}I_2^{\mathrm{off}}/\mathrm{d}s = 0$ while $I_2^{\mathrm{off}}$ 
still remains nonzero. For example, all three generators encounter problems in the case of degeneracy, 
i.e.\ when $H_{nn}(s) = H_{kk}(s)$ while $H_{nk}(s) \neq 0$.

The generator form in Eq.~(\ref{eq:gen_form}) has a certain disadvantage. If the initial Hamiltonian was represented 
by a banded matrix (tridiagonal, five diagonal or similar), this sparse structure would be a nice property 
to preserve during the course of the flow. However, generators of Eq.~(\ref{eq:gen_form}) do not respect 
the banded form, and in order to rectify the issue, Mielke~\cite{Mielke1998} proposed the following generator
\begin{equation}
  \eta_{nk}(s) = H_{nk}(s)\,\mathrm{sgn}(n-k).
\label{eq:gen_toda}
\end{equation}
Note, that here the sign function acts not on the matrix elements but on the row and column indices. 
If the initial Hamiltonian is banded, $H_{nk}(0) = 0$ for $|n-k| > n_0$, this structure survives during the flow. 
The generator~(\ref{eq:gen_toda}) is not of the form of Eq.~(\ref{eq:gen_form}), thus the condition of nondecreasing
weight of the diagonal in Eq.~(\ref{eq:Id_pos}) 
is not applicable. However, it was shown~\cite{Mielke1998} that the flow always converges to a final diagonal matrix, 
even if degeneracies are encountered. Moreover, the diagonal elements are automatically sorted in the ascending order, 
$H_{nn}(+\infty) \geqslant H_{kk}(+\infty)$ for $n>k$. For the special case of real tridiagonal matrices, 
the generator~(\ref{eq:gen_toda}) was previously studied in a different context, namely the integrable Toda 
lattice~\cite{Henon1974}. The name stuck. Shortly, we will define and use a generalized version of 
the Toda generator to block diagonalize the Kamiltonian~(\ref{eq:kam_sh_form}), therefore, it is natural 
to call our version the \emph{block-diagonalizing Toda generator}, or briefly just the Toda generator.

%
\section{Block-diagonalization of Kamiltonian using flow approach}
\label{sec:block}

The Floquet effective Hamiltonian, defined in Sec.~\ref{sec:FH}, can be found by block diagonalizing the Kamiltonian, 
see Eq.~(\ref{eq:dkam}). Thus, in the present section we will address the problem of block diagonalization 
of time-dependent Kamiltonians, thus generalizing a flow-based approach of the previous Sec.~\ref{sec:FA} 
which was limited to diagonalization of static Hamiltonians.

There are several differences between diagonalization of a static Hamiltonian and block diagonalization of the Kamiltonian. 
First, block diagonalization is, in fact, a much simpler task than the full diagonalization, because a diagonal matrix can 
be treated as a special case of a block diagonal one but not the other way around. Our goal is to find the FE Hamiltonian 
$h_{\mathrm{eff}}(t)$, which is not necessary diagonal with respect to the basis of the physical space $\mathscr{H}$. 
This is in contrast to Ref.~\cite{Thomson2021} where the static Kamiltonian is fully diagonalized. Second, the Kamiltonian 
has a special structure which allows us to write it in terms of the shift operators~(\ref{eq:kam_sh_form}). Thus, we will 
study only flows that preserve this structure. Third, we consider time-dependent Kamiltonians, therefore, the flow 
equations~(\ref{eq:flow_discr}) and~(\ref{eq:flow_diff_eq}) must be generalized to time-dependent matrices. And finally, 
during the flow we need to keep track of all infinitesimal unitary transformations because they define the overall 
transformation $\mathcal{D}(t)$ which is used to calculate the micromotion operator~(\ref{eq:mic}).

The discrete flow equation~(\ref{eq:flow_discr}) can be generalized for time-dependent Kamiltonians as follows 
\begin{equation}
\begin{split}
  \mathcal{K}(s+1,t) &= \e^{\ri \mathcal{S}(s,t)} \mathcal{K}(s,t) \, \e^{-\ri\mathcal{S}(s,t)} \\
  &- \ri\hbar \, \e^{\ri\mathcal{S}(s,t)} \, \frac{\partial \e^{-\ri\mathcal{S}(s,t)}}{\partial t}.
\label{eq:flow_discr_kam}
\end{split}
\end{equation}
Here the extended-space operator $\mathrm{i}\mathcal{S}$ plays the role of the flow generator, 
and $\mathcal{S}^{\dagger} = \mathcal{S}$. Correspondingly, the continuous flow equation~(\ref{eq:flow_diff_eq}) 
generalizes to \cite{Tomaras2011}
\begin{equation}
  \frac{\partial \mathcal{K}(s,t)}{\partial s} 
  = \ri\left[ \mathcal{S}(s,t),\mathcal{K}(s,t) \right] - \hbar \, \frac{\p \mathcal{S}(s,t)}{\p t}.
\label{eq:flow_diff_eq_kam}
\end{equation}
We focus exclusively on generators of the form
\begin{equation}
  \ri\mathcal{S}(s,t) = \ri \sum\limits_{m=-\infty}^{+\infty} P_m \otimes S^{(m)}(s,t),
\label{eq:gen_form_kam}
\end{equation}
with
\begin{equation}
  \left[ S^{(m)}(s,t) \right]^{\dag} = S^{(-m)}(s,t),
\label{eq:S}
\end{equation}
since this form guarantees that during the flow the Kamiltonian remains of the form set by Eq.~(\ref{eq:kam_sh_form}).

In the case of a discrete flow, the net effect of all unitary transformations can be written as
\begin{equation}
  \mathcal{D}^{\dagger}(t) = \e^{\ri \mathcal{S}(s^{\prime},t)} \e^{\ri\mathcal{S}(s^{\prime}-1,t)} \ldots 
  \e^{\ri\mathcal{S}(2,t)}\e^{\ri\mathcal{S}(1,t)} \e^{\ri\mathcal{S}(0,t)},
\label{eq:flow_D_disc}
\end{equation}
where $s^{\prime}$ can be either a finite or an infinite number. For the continuous flow, the joint action 
of all unitary transformations is written as a flow-ordered integral
\begin{equation}
  \mathcal{D}^{\dag}(t) = \mathcal{T}_s \exp \left[\ri \int\limits_{0}^{+\infty}  \mathcal{S}(s,t) \, 
  \mathrm{d}s \right].
\label{eq:flow_D_cont}
\end{equation}

\subsection{Generator for discrete flow in the high-frequency regime}
\label{sec:discr}

Let us analyze a specific example of a discrete flow, considered within the framework of a high-frequency expansion. 
To be more precise, we assume that $(\hbar \omega)^{-1}$ is a small expansion variable, and our goal is to make 
the Kamiltonian block diagonal up to some order in $(\hbar \omega)^{-1}$. At each step $s$, the Kamiltonian 
can be written as
\begin{equation}
  \mathcal{K}(s,t) = \hbar\omega N \otimes \mathbf{1}_{\mathscr{H}} 
  + \sum\limits_{m=-\infty}^{+\infty} P_m \otimes H^{(m)}(s,t),
\label{eq:kam_disc_expan}
\end{equation}
where each Fourier component $H^{(m)}(s,t)$ is represented as a power series $H^{(m)}(s,t)=\sum_{i=0}^{+\infty} H^{(m)}_i(s,t)$, 
with $H^{(m)}_i(s,t)$ being of the order of $(\hbar \omega)^{-i}$. At the start of the flow, $s = 0$ and the zeroth-order 
terms $H^{(m)}_0(0,t)$ are set equal to the original Fourier components, $H^{(m)}_0(0,t) = h^{(m)}(t)$, while all higher-order 
terms are zero, $H^{(m)}_{i\geqslant 1}(0,t) = 0$. Moreover, we assume that the time dependence of $h^{(m)}(t)$ 
is slow in comparison to $\omega$, 
thus the derivatives behave as $\mathrm{d}^{j} h^{(m)}(t)/\mathrm{d}t^j \sim \mathcal{O}(1)$ for arbitrary $j$.

Before we proceed, a note on notation and terminology 
is in order. We use the lowercase $h$ to refer to the Fourier harmonics of the driven Hamiltonian 
as well as of the effective Hamiltonian. They are, respectively, the input and the output of the flow procedure. 
During the flow, we operate with running (that is, $s$-dependent) Fourier harmonics denoted by uppercase $H$.
We note that Fourier components $H^{(m)}$ with $m \neq 0$ define off-diagonal blocks of the Kamiltonian, thus, we will 
refer to them as the `off-diagonal' Fourier components. Likewise, $H^{(0)}$ appears only in the diagonal block and thus
is the `diagonal' Fourier component. 

Let us assume that it was possible to find a generator that, on each successive 
application, eliminates the leading  term from the expansion of all off-diagonal Fourier components. To be more precise, 
in the initial ($s = 0$) Kamiltonian $\mathcal{K}(0,t)$ the Fourier components  $H^{(m\neq 0)}(0,t)$ are represented
as series that naturally start with $H^{(m \neq 0)}_0(0,t)$, while at $s = 1$ the transformed Kamiltonian 
$\mathcal{K}(1,t)$ contains Fourier components of $H^{(m\neq 0)}(1,t)$  whose expansion starts with the 
terms $H^{(m \neq 0)}_1(1,t)$, because $H^{(m \neq 0)}_0(1,t)$ has been set to zero. After two steps of the flow, 
both $H^{(m \neq 0)}_0(2,t)$ and $H^{(m \neq 0)}_1(2,t)$ have been eliminated, and so on. 
When such a flow reaches a certain value $s$, one can write
\begin{equation}
\begin{split}
  \mathcal{K}(s,t) &= \hbar \omega N \otimes \mathbf{1}_{\mathscr{H}} \\
  &+ P_0 \otimes \sum\limits_{i=0}^{s-1} H^{(0)}_i(s,t) 
  + \mathcal{O}\left(\frac{1}{(\hbar \omega)^{s}}\right),
\end{split}
\label{eq:kam_disc_expan1}
\end{equation}
with the obvious interpretation that the Kamiltonian $\mathcal{K}(s,t)$ is block diagonal up to the order $(s-1)$. 
Thus the FE Hamiltonian reads $h_{\mathrm{eff}}(t)=\sum_{i=0}^{s-1}H^{(0)}_i(s,t)+\mathcal{O}\left(1/(\hbar \omega)^{s} \right)$. 
Surprisingly, this scenario is realized with a very simple choice of the generator
\begin{equation}
\mathrm{i}\mathcal{S}(s,t)=\sum\limits_{m\neq 0} \frac{P_m}{m} \otimes \frac{H^{(m)}_s (s,t)}{\hbar \omega}.
\label{eq:gen_dicsr}
\end{equation}
For example, at $s = 0$ the initial generator reads
\begin{equation}
  \ri \mathcal{S}(0,t) = \sum\limits_{m\neq 0} \frac{P_m}{m} \otimes \frac{h^{(m)}(t)}{\hbar \omega},
\label{eq:gen_dicsr1}
\end{equation}
and after one step of the flow (\ref{eq:flow_discr_kam}) the updated Kamiltonian becomes
\begin{equation}
\mathcal{K}(1,t)=\hbar \omega N \otimes \mathbf{1}_{\mathscr{H}} 
+ P_0 \otimes h^{(0)}(t)+\mathcal{O}\left(\frac{1}{(\hbar \omega)^{1}}\right),
\label{eq:kam_disc_expan2}
\end{equation}
giving the FE Hamiltonian $h_{\mathrm{eff}}(t)=h^{(0)}(t)+\mathcal{O}\left(1/(\hbar \omega)^{1} \right)$.

We have verified that the discrete flow with the generator of the form of Eq.~(\ref{eq:gen_dicsr}) reproduces 
the high-frequency expansion procedure presented in Ref.~\cite{Novicenko2017}. In Appendix~\ref{sec:App-A},
we prove the validity of the generator~(\ref{eq:gen_dicsr}) and also show that the obtained accuracy 
of Eq.~(\ref{eq:kam_disc_expan1}) is, in fact, even better than anticipated.

Let us draw attention to one drawback of the discussed discrete flow driven by the generator~(\ref{eq:gen_dicsr}). Let us 
say that the initial Hamiltonian $h(\omega t, t)$ has a limited number of contributing Fourier harmonics. In other words, 
there exists such a positive $m_0$ that all higher harmonics vanish, i.e.\ $h^{(m)}(t) = 0$ for $|m| > m_0$. One might expect 
that during the flow we can restrict the analysis to the limited Fourier spectrum $H^{(m)}(s)$ with $|m|\leqslant m_0$. 
Unfortunately that is not true, and sooner or later $H^{(m)}(s)$ with $|m| > m_0$ become non-zero. This complicates 
the automated 
implementation of the flow using symbolic computation packages. The same drawback is present in the continuous flow discussed 
in the subsequent subsection Sec.~\ref{sec:mint}. Therefore, in Sec.~\ref{sec:toda} we will proceed to the introduction 
of the Toda generator that is not plagued with this problem.

\subsection{Generator proposed by A.~Verdeny, A.~Mielke and F.~Mintert}
\label{sec:mint}

Let us now turn to continuous flows to implement the block diagonalization of the Kamiltonian. One can adapt 
the generator~(\ref{eq:gen_form}) to work with block matrices. By interpreting the partial inner product 
$\left\langle n \right| \mathrm{i}\mathcal{S} \left| k \right\rangle$ as an element of a block matrix, 
one can rewrite Eq.~(\ref{eq:gen_form}) in terms of block matrices
\begin{equation}
\begin{aligned}
  &\left\langle n \right| \ri \mathcal{S}(s,t) \left| k \right\rangle = \\
  &\left\langle n \right| \mathcal{K}(s,t) \left| k \right\rangle 
  f\left( \left\langle n \right| \mathcal{K}(s,t) \left| n \right\rangle 
  - \left\langle k \right| \mathcal{K}(s,t) \left| k \right\rangle \right).
\label{eq:gen_form1}
\end{aligned}
\end{equation}
Because $\left\langle n \right| \mathcal{K}(s,t) \left| n \right\rangle 
= n \hbar \omega \mathbf{1}_{\mathscr{H}} + H^{(0)}(s,t)$, the difference 
between two diagonal blocks is always proportional to the unit operator. Previously, the function $f(\cdot)$ was defined 
for real numbers. Here we must slightly generalize the function's action to operators defined in the vector space 
$\mathscr{H}$ and proportional to the unit operator as $f\left(x \mathbf{1}_{\mathscr{H}} \right) = \mathbf{1}_{\mathscr{H}} f(x)$. 
Thus, with the Wegner's case of the function $f(\cdot)$, the generator reads
\begin{equation}
  \ri \mathcal{S}(s,t) = \hbar \omega \sum\limits_{m \neq 0} m P_m \otimes H^{(m)}(s,t).
\label{eq:gen_form2}
\end{equation}
This generator was proposed in Ref.~\cite{Verdeny2013}. Note that because the generator~(\ref{eq:gen_form2}) is of the form 
specified by Eqs.~(\ref{eq:gen_form_kam}) and (\ref{eq:S}), the Kamiltonian remains of the form set by Eq.~(\ref{eq:kam_sh_form})
during the flow. By substituting Eq.~(\ref{eq:gen_form2}) into Eq.~(\ref{eq:flow_diff_eq_kam}) we obtain the following flow 
equations for the Fourier components
\begin{subequations}
\label{eq:fw_znz}
\begin{equation}
  \frac{\mathrm{d}H^{(0)}(s,t)}{\mathrm{d}s} 
  = \frac{2}{\hbar \omega} \sum\limits_{m=1}^{+\infty} m \left[ H^{(m)}(s,t),H^{(-m)}(s,t) \right],
\label{eq:fw_z}
\end{equation}
and for $n \neq 0$
\begin{equation}
\begin{aligned}
&\frac{\mathrm{d}H^{(n)}(s,t)}{\mathrm{d}s} = -n^2 H^{(n)}(s,t)+\frac{\mathrm{i}}{\omega} n \dot{H}^{(n)}(s,t) \\
&\quad +\frac{1}{\hbar \omega} \sum\limits_{m \neq n} (m-n) \left[ H^{(m)}(s,t),H^{(n-m)}(s,t) \right].
\label{eq:fw_nz}
\end{aligned}
\end{equation}
\end{subequations}
Note that here we rescaled the flow variable $s \rightarrow s/(\hbar \omega)^2$ for the sake of convenience. 
The initial conditions for Eq.~(\ref{eq:fw_znz}) are $H^{(m)}(0,t)=h^{(m)}(t)$, and the FE Hamiltonian 
is obtained as the limit $\lim_{s \rightarrow +\infty} H^{(0)}(s,t) = h_{\mathrm{eff}}(t)$. Eq.~(\ref{eq:fw_znz})
can be used as a starting point for approximations. For example, in Appendices~\ref{sec:App-B} 
and~\ref{sec:App-C} we show how one can obtain the FE Hamiltonian and the micromotion operator 
in terms of high-frequency expansions when the modulation of the envelope is slow on the scale set by the driving 
frequency, i.e.\ $\mathrm{d}^j h^{(m)}(t)/ \mathrm{d}t^j \sim\mathcal{O}\left( 1 \right)$. 
In principle, such results are a reproduction of the equations obtained in Ref.~\cite{Novicenko2017}. 
A different variant of a high-frequency expansion is obtained in Sec.~\ref{sec:fast}, where we assume 
a rapid variation of the envelope, i.e.\ 
$\mathrm{d} h^{(m)}(t)/ \mathrm{d}t \sim\mathcal{O}\left( \hbar \omega \right)$. As we will see in Sec.~\ref{sec:fast}, 
such an assumption places some restrictions on the behavior of the Fourier components as a function of time. 
The expansion presented in Sec.~\ref{sec:fast} has not been obtained previously, and thus can potentially be useful 
in situations where the envelope of the perturbation is so fast that only a few oscillations are performed 
during the perturbation pulse.

\subsection{Block-diagonalizing Toda generator}
\label{sec:toda}

As discussed previously, the continuous flow defined by Eq.~(\ref{eq:fw_znz}) as well as the discrete 
flow presented in Sec.~\ref{sec:discr} lead to a proliferation of Fourier harmonics $H^{(m)}(s,t)$. 
In order to overcome this obstacle, we will employ a generator akin of the Toda generator in Eq.~(\ref{eq:gen_toda}). 
In fact, we use the same Eq.~(\ref{eq:gen_form1}), but this time with the function $f(x) = \operatorname{sgn}(x)$, thus
\begin{equation}
  \ri\m{S}(s,t) = \sum\limits_{m \neq 0} \frac{\operatorname{sgn}(m)}{\hbar \omega}  P_m \otimes H^{(m)}(s,t).
\label{eq:gen_form4}
\end{equation}
Here we rescale the flow variable $s \rightarrow s/(\hbar \omega)^2$ as before. 
The flow equations produced by the Toda generator~(\ref{eq:gen_form4}) are similar to Eq.~(\ref{eq:fw_znz})
%
\begin{subequations}
\label{eq:fw_znz1}
\begin{equation}
\frac{\mathrm{d}H^{(0)}(s,t)}{\mathrm{d}s}=\frac{2}{\hbar \omega} \sum\limits_{m=1}^{+\infty} \left[ H^{(m)}(s,t),H^{(-m)}(s,t) \right],
\label{eq:fw_z1}
\end{equation}
and for $n \neq 0$
\begin{equation}
\begin{aligned}
&\frac{\mathrm{d}H^{(n)}(s,t)}{\mathrm{d}s} = - n\cdot \mathrm{sgn}(n) H^{(n)}(s,t)+\frac{\mathrm{i}}{\omega} \mathrm{sgn}(n) \dot{H}^{(n)}(s,t) \\
&+\frac{1}{\hbar \omega} \sum\limits_{m \neq n} \mathrm{sgn}(m-n) \left[ H^{(m)}(s,t),H^{(n-m)}(s,t) \right].
\label{eq:fw_nz1}
\end{aligned}
\end{equation}
\end{subequations}
In Appendix~\ref{sec:App-E} we show that if the initial condition $H^{(n)}(0,t)=0$ for $|n| > n_0$, then $H^{(n)}(s,t)=0$ 
for all $s\in [0,+\infty)$. This feature facilitates an automated generation of a high-frequency expansion from the flow
equations~(\ref{eq:fw_znz1}).

\section{Automated expansion for a finite closed algebra}
\label{sec:lie}

Appendices~\ref{sec:App-B} and \ref{sec:App-C} describe the procedure that allows to derive the FE Hamiltonian 
and the micromotion operator starting from the equations~(\ref{eq:fw_znz1}) in the high-frequency regime, 
including the possibility of a slow modulation of the driving amplitude. 
However, manual implementation of this procedure for expansions of higher order is too tedious. 
Mikami et al.~\cite{Mikami2016PRB} were able to write out such an expansion up to the fourth order 
for the simpler case of unmodulated periodic driving. 
The authors of Refs.~\cite{Zeuch2020,Zeuch2020refuting} were able to proceed to even higher orders 
in within the framework of the Magnus-Taylor expansion and specializing to a two-level system.
Here we demonstrate that an automated expansion to an arbitrary order (limited only by the available computational resources)
can be achieved with the help of a symbolic computational package. However, for this scheme to succeed, we need 
to place several restrictions on the initial Hamiltonian $h(\omega t+\theta,t)$. Namely, we assume that the Hamiltonian 
has a limited number of Fourier harmonics, that is $h^{(n)}(t) = 0$ for $|n| > n_0$, and the Hamiltonian is written 
in terms of a finite $L$-dimensional Lie algebra spanned by the Hermitian generators $G_l$ with $l = 1, 2, \ldots, L$. Thus
\begin{equation}
  h^{(n)}(t)=\sum\limits_{l=1}^L c_l^{(n)}(t) \, G_l,
\label{eq:alg}
\end{equation}
and the functions $c_l^{(n)}(t)$ are complex-valued functions of the time. In the absence of the additional
modulation of the driving signals they reduce to mere complex numbers. The algebra is closed with respect 
to the commutator
\begin{equation}
  \left[G_l,G_m \right] = \sum\limits_{n=1}^L \gamma_{lmn} G_n,
\label{eq:algx}
\end{equation}
here $\gamma_{lmn}$ are the structure constants defined by the algebra. Note, that in general the algebra
can be closed in an approximate sense: certain terms produced by the commutator (\ref{eq:alg}) are ignored
as not belonging to the considered model. The simplest nontrivial example is provided by a two-level system 
or a spin-$\tfrac{1}{2}$ particle in a magnetic field. Here one operates within the 
algebra with $L = 3$ generators $\sigma_{\{x,y,z\}}$ represented by the Pauli matrices, and 
the structure constant $\gamma_{lmn} = 2\ri \, \epsilon_{lmn}$ proportional to the Levi-Civita antisymmetric tensor. 
Thus the commutators in Eq.~(\ref{eq:fw_znz1}) can be expressed in terms of products of coefficients 
$c_l^{(n)}(t)$. By expanding the coefficients in power series of the inverse frequency, 
$c_l^{(n)}(t) = c_{l,0}^{(n)}(t) + c_{l,1}^{(n)}(t) + \cdots$, a symbolic computation package can collect 
terms of the same order on both sides of the flow equations, 
and solve the ensuing differential equations order by order for $c_{l,j}^{(n)}(s,t)$. 
Note that the differential equations produced during this procedure are nothing more than linear inhomogeneous first-order 
differential equations [see for example Eq.~(\ref{eq:fw_hf_1})], thus easily solvable with a symbolic computation package. 
The micromotion operator is obtained in a similar way; in this case one computes the integrals generated by the Magnus 
expansion. Each consecutive term in the expansion is obtained from the exact recursive relations given in 
Ref.~\cite{Blanes2009}.

To give a concrete example of an automated construction of a high-frequency expansion, we consider the case 
of the $\mathsf{su(2)}$ algebra, specifically, the Rabi model subject to a linearly polarized drive (alternative
expansions for the same model were analyzed in Refs.~\cite{Zeuch2020,Zeuch2020refuting})
\begin{equation}
  h_{\mathrm{lab}}(\omega t,t) = \frac{\hbar\Omega}{2}\sigma_z + 2g(t)\cos(\omega t+ \phi)\sigma_x.
\label{eq:rabi}
\end{equation}
Here we assume that $\Omega$ is comparable to $\omega$, more precisely, we assume a small detuning 
$(\hbar \Omega-\hbar\omega) = \Delta \sim \mathcal{O}(1)$. Therefore the original Hamiltonian~(\ref{eq:rabi}) 
is not suitable to proceed with the inverse frequency expansion, and we first transition into a moving frame 
with the help of the unitary transformation $\tilde{U}(t) = \exp[-\ri \omega t \sigma_z /2]$. 
The transformed Hamiltonian reads
\begin{equation}
\begin{aligned}
  h(\omega t,t) &=& \tilde{U}^{\dagger}(t) h_{\mathrm{lab}}(\omega t,t) \tilde{U}(t)
    - \ri \hbar \tilde{U}^{\dagger}(t) \frac{\mathrm{d} \tilde{U}(t)}{ \mathrm{d} t} \\
  &=& \frac{\Delta}{2}\sigma_z + g(t)\left[\cos(\phi)+\cos(2\omega t+ \phi) \right]\sigma_x\\
  & &+g(t)\left[\sin(\phi)-\sin(2\omega t+ \phi) \right]\sigma_y,
\label{eq:rabi1}
\end{aligned}
\end{equation}
and the contributing non-zero Fourier components are
\begin{subequations}
\label{eq:fc}
\begin{align}
  h^{(0)} &= \frac{\Delta}{2}\sigma_z + g(t)\cos(\phi)\sigma_x + g(t)\sin(\phi)\sigma_y, \label{eq:fc_1} \\
  h^{(2)} &= \frac{g(t)}{2}\e^{\ri\phi}\sigma_x + \frac{g(t)}{2}\ri\e^{\ri\phi}\sigma_y
  = \left[ h^{(-2)}\right]^{\dagger}. \label{eq:fc_3}
\end{align} 
\end{subequations}
Explicit expansion formulas for the FE Hamiltonian are quite bulky, therefore, in Appendix~\ref{sec:App-F}
we present a truncated expansion up to the fourth-order with $\phi = 0$. Here, let us limit ourselves to 
the second-order expansions. The effective Hamiltonian reads
\begin{subequations}
\label{eq:numeric12}
\begin{equation}
\label{eq:numeric1}
\begin{split}
  h_{\eff} &= \sigma_x \cdot \left[ g(t) \cos\phi - \frac{g^3(t) \cos\phi}{4 (\hbar\omega)^2} \right] \\
  &+ \sigma_y \cdot \left[ g(t) \sin\phi - \frac{g^3(t) \sin\phi}{4 (\hbar\omega)^2} \right] \\
  &+ \sigma_z \cdot \left[\frac{\Delta}{2} + \frac{g^2(t)}{2 \hbar\omega } - \frac{\Delta  g^2(t)}{4 (\hbar\omega)^2} \right],
\end{split}
\end{equation}
and the micromotion is given by $U_{\mathrm{micro}}(t) = \e^{-\ri S(t)}$ with
\begin{widetext}
\begin{equation}
\label{eq:numeric2}
\begin{split}
  S(t) &= \sigma_x \left[ \frac{ g(t) \sin (2 \omega t + \phi)}{2\hbar\omega}
    - \frac{ \Delta  g(t) \sin (2 \omega t + \phi )}{4 (\hbar\omega)^2} 
    + \frac{ \hbar g'(t) \cos (2 \omega t + \phi)}{4 (\hbar\omega)^2} \right] \\
  &+ \sigma_y \left[ \frac{ g(t) \cos (2 \omega t + \phi)}{2\hbar\omega}
    - \frac{ \Delta  g(t) \cos (2 \omega t + \phi )}{4 (\hbar\omega)^2} 
    - \frac{ \hbar g'(t) \sin (2 \omega t + \phi)}{4 (\hbar\omega)^2} \right] 
  + \sigma_z \frac{g^2(t) \sin (2 \omega t + 2 \phi )}{2(\hbar\omega)^2}. 
\end{split}
\end{equation}
\end{widetext}
\end{subequations}
In contrast to Refs.~\cite{Zeuch2020,Zeuch2020refuting}, in our expansions 
the first- and second-order FE Hamiltonian does not depend on the time-derivative $g^{\prime}(t)$. 
The reason for this is that the expansion is performed for different Hamiltonians: FE Hamiltonian in our case
and FS Hamiltonian in the example of Refs.~\cite{Zeuch2020,Zeuch2020refuting}. 

\begin{figure}
\includegraphics[width=84mm]{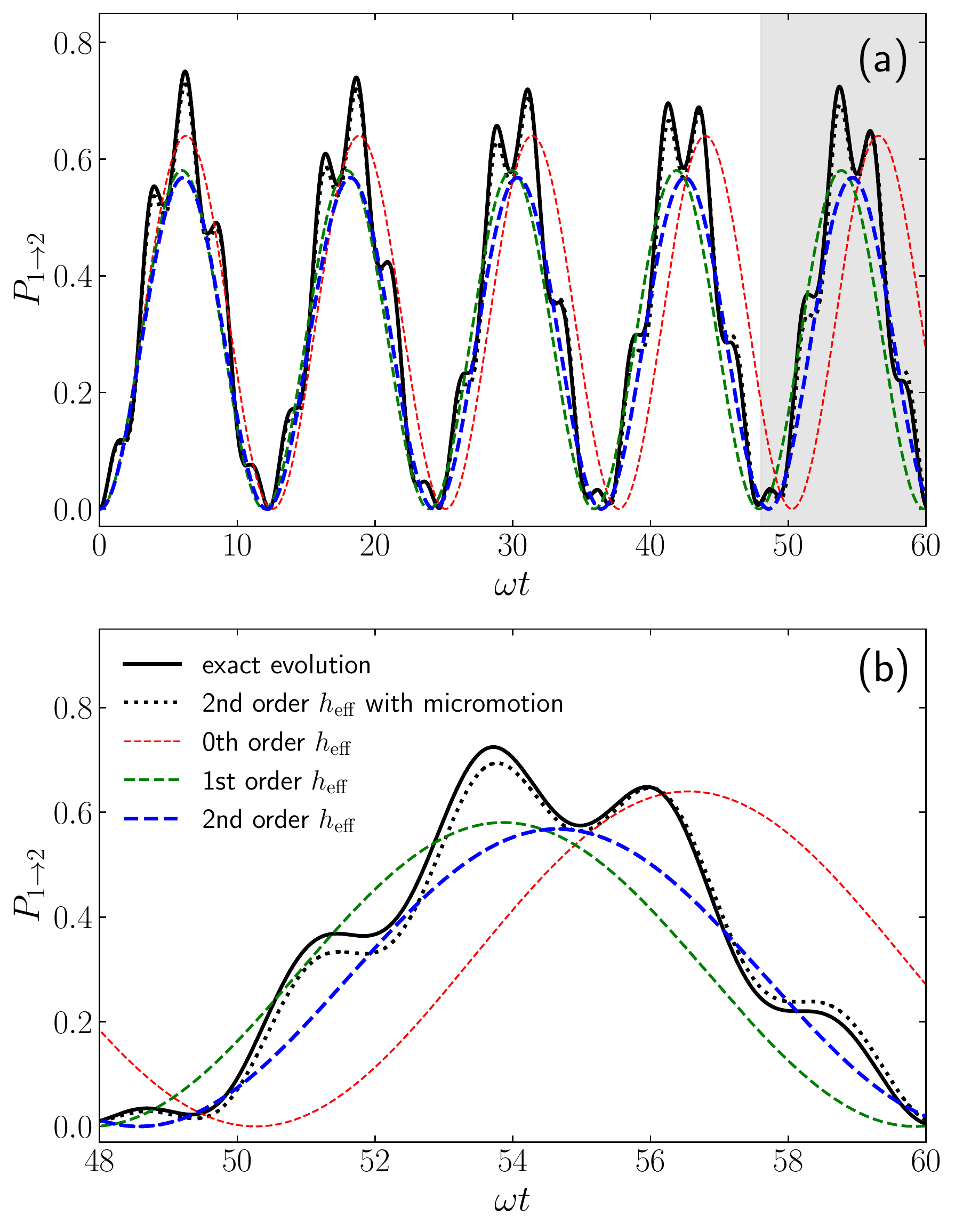}
\caption{\label{fig1} (Color online.) Probability of transition to the excited state in a driven two-level 
system, calculated using various approximations. The numerically exact evolution (solid line) is reliably 
reproduced only when the micromotion is included (dotted line). Approximations that do not take 
micromotion into account (dashed lines) consistently fail to reproduce full details of the evolution.}
\end{figure}

Figure~\ref{fig1} shows the results of a numerical simulation based on the obtained high-frequency expansion
given in Eq.~(\ref{eq:numeric12}). We focus on a two-level system described by the Hamiltonian~(\ref{eq:rabi1}) with 
the parameters: $g(t) = \mathrm{const.}$,
$\Delta / \hbar\omega = 0.3$, $g / \hbar\omega = 0.2$, and $\phi = 0$ and initially prepared 
in the ground state. Using a number of different approximations, we compute the probability $P_{1 \to 2}$ of transition 
to the excited state as a function of the scaled time $\omega t$. In Fig.~\ref{fig1}(a) we cover the time interval 
$\omega t \in [0, 60]$ that roughly corresponds to five Rabi cycles, and Fig.~\ref{fig1}(b) shows a magnification
of the last cycle marked by the shaded area in Fig.~\ref{fig1}(a). The black solid line shows the exact evolution, obtained from numerical time propagation of the time-dependent Schr\"odinger equation using a fourth-order Runge-Kutta scheme. 
The black dotted line corresponds to the approximate evolution calculated using the complete information available 
from Eq.~(\ref{eq:numeric12}), i.e., including both the effective Hamiltonian (\ref{eq:numeric1}) 
and the micromotion (\ref{eq:numeric2}). To elucidate the role of particular terms, we include three dashed lines 
corresponding to approximations that neglect the micromotion and truncate the effective Hamiltonian. The different
levels of approximation are encoded using the color and thickness of the dashed lines. The thinnest red line
depicts the evolution generated by including only the zeroth-order terms in the effective Hamiltonian, and the
green line of intermediate thickness corresponds to the case when the zeroth- and first-order terms are included. 
Finally, the thickest dashed blue line corresponds to the second-order approximation for the effective 
Hamiltonian~(\ref{eq:numeric1}), however, still not taking the micromotion into account. We see that improved accuracy
of the effective Hamiltonian translates to a more reliable calculation of the period of the Rabi cycle, however, 
inclusion of the micromotion is indeed essential for the reliable description of the complete quantum evolution.

%
\section{Case of fast amplitude modulation}
\label{sec:fast}

Let us now go back to the flow equations~(\ref{eq:fw_znz}) and perform the inverse-frequency expansion analogous 
to that presented in Appendices~\ref{sec:App-B} and \ref{sec:App-C}, however, focusing on the case of fast 
amplitude modulation. We now assume that the rate of change of the Fourier components of the driven Hamiltonian 
is comparable to the driving frequency, formally expressed as
$\mathrm{d}^j h^{(m)}(t)/ \mathrm{d}t^j \sim\mathcal{O}\left( \omega^j \right)$. Although it may seem 
that in such a regime the frequency can no longer be considered `high', we stress that the amplitudes 
of the Fourier components are still assumed to be low, $h^{(m)}(t) \sim \mathcal{O}(1)$.

Repeating the steps in Appendix~\ref{sec:App-B}, we expand each Fourier component into an inverse-frequency 
power series $H^{(n)}(s,t) = H^{(n)}_0(s,t) + H^{(n)}_1(s,t) + \cdots$, insert the expansions into 
the flow equations~(\ref{eq:fw_znz}), and collect terms of the same order on both sides. 
In the zeroth order we get a trivial equation for the zeroth Fourier component,
\begin{subequations}
\begin{equation}
  \frac{\mathrm{d}H^{(0)}_0}{\mathrm{d}s}=0,
\label{eq:triv}
\end{equation}
and a modified equation for the $n \neq 0$ Fourier components
\begin{equation}
  \frac{\mathrm{d}H^{(n)}_0(s,t)}{\mathrm{d}s} 
  = -n^2 H^{(n)}_0(s,t) + \frac{\mathrm{i}}{\omega}n \dot{H}^{(n)}_0(s,t).
\label{eq:fast}
\end{equation}
\end{subequations}
From the trivial equation~(\ref{eq:triv}) we find the zeroth-order FE Hamiltonian $h_{\mathrm{eff}(0)}(t) = h^{(0)}(t)$, 
while the first-order FE Hamiltonian $h_{\mathrm{eff}(1)}(t)$ can be obtained by solving Eq.~(\ref{eq:fast}). 
In principle, Eq.~(\ref{eq:fast}) is a partial differential equation, because derivatives with respect to $s$ and $t$ 
act on the same quantity. Such an equation can be solved using an additional Fourier expansion with respect to $t$. 
We are interested in the time interval $t\in [t_{\mathrm{in}},t_{\mathrm{fn}}]$ and assume that the driven Hamiltonian 
is the same on the boundaries of the interval, $h^{(n)}(t_{\mathrm{in}}) = h^{(n)}(t_{\mathrm{fn}})$. Thus the Fourier 
expansion with respect to $t$ reads
\begin{equation}
  H^{(n)}_0(s,t) = \sum\limits_{j=-\infty}^{+\infty} H^{(n,j)}_0(s) \exp[\mathrm{i}j\Omega t],
\label{eq:fast1}
\end{equation}
where $\Omega=2\pi/(t_{\mathrm{fn}}-t_{\mathrm{in}})$ is the characteristic frequency of the envelope. 
In view of $\left[H_0^{(n)}\right]^{\dagger} = H_0^{(-n)}$ we have
\begin{equation}
\left[H_0^{(n,j)}(s)\right]^{\dagger} = H_0^{(-n,-j)}(s).
\label{eq:fast2}
\end{equation}
Each Fourier component of the original Hamiltonian can be expanded into a Fourier series, 
$h^{(n)}(t) = \sum_j h^{(n,j)}\exp[\mathrm{i}j \Omega t]$, thus at the beginning of the flow we have 
$H^{(n,j)}_0(0)=h^{(n,j)}$. Substituting the ansatz~(\ref{eq:fast1}) into the flow equation~(\ref{eq:fast}) 
one can see that different Fourier components (for different $j$) do not couple
\begin{equation}
  \frac{\mathrm{d}H^{(n,j)}_0(s)}{\mathrm{d}s} 
  = -n^2 H^{(n,j)}_0(s)-\frac{\Omega}{\omega}jn H^{(n,j)}_0(s),
\label{eq:fast3}
\end{equation}
and are described by the exponential solutions
\begin{equation}
  H^{(n,j)}_0(s)=h^{(n,j)}\exp\left[-\left(n^2+\frac{\Omega}{\omega}jn \right)s \right].
\label{eq:fast4}
\end{equation}
The high-frequency expansion converges only if $H^{(n,j)}_0(s \rightarrow +\infty) = 0$. Therefore one should 
place a restriction on the index $j$. We assume that the expansion of the envelope~(\ref{eq:fast1}) 
does not have high harmonics
\begin{equation}
  H^{(n)}_0(s,t)=\sum\limits_{j=-J}^{J} H^{(n,j)}_0(s) \exp[\mathrm{i}j\Omega t],
\label{eq:fast1_1}
\end{equation}
and the positive integer $J$ satisfies
\begin{equation}
  J<\frac{\omega}{\Omega}.
\label{eq:fast1_2}
\end{equation}
This means that if, over the interval $[t_{\mathrm{in}},t_{\mathrm{fn}}]$, the high-frequency modulation 
oscillated $J+\varepsilon$ times [here $\varepsilon \in (0,1)$] then the expansion of the envelope 
can not have harmonic number higher than $J$. For example, if the high frequency modulation oscillated $3.5$ 
times then the shape of the envelope expanded into the Fourier series can not possess the fourth harmonic.

Next, using the expansion~(\ref{eq:fast1_1}), one can write down the first-order flow equation for the zeroth Fourier component
\begin{equation}
\begin{aligned}
  \frac{\mathrm{d}H^{(0)}_1(s,t)}{\mathrm{d}s} =& \frac{2}{\hbar \omega} \sum\limits_{m=1}^{+\infty} \sum\limits_{j=-J}^{J} 
  \sum\limits_{j^{\prime}=-J}^{J} m \left[ h^{(m,j)},h^{(-m,j^{\prime})} \right] \\
  & \times \mathrm{e}^{\mathrm{i}(j+j^{\prime})\Omega t} \mathrm{e}^{-\left(2m^2+m\frac{\Omega}{\omega}(j-j^{\prime}) \right)s}.
\label{eq:fast5}
\end{aligned}
\end{equation}
Integrating this equation and taking into account the initial condition $H_1^{(0)}(0,t) = 0$ we obtain the solution. 
Calculation of the limit $s\rightarrow +\infty$ yields the first-order FE Hamiltonian
\begin{equation}
\begin{aligned}
  &h_{\mathrm{eff}(1)}(t) = \\
  &\frac{1}{\hbar \omega} \sum\limits_{m=1}^{+\infty} \sum\limits_{j=-J}^{J} 
  \sum\limits_{j^{\prime}=-J}^{J} \frac{\left[ h^{(m,j)},h^{(-m,j^{\prime})} \right]}{m+\frac{\Omega}{2\omega}(j-j^{\prime})}
  \mathrm{e}^{\mathrm{i}(j+j^{\prime})\Omega t}.
\label{eq:fast6}
\end{aligned}
\end{equation}

We obtained the FE Hamiltonian up to the first order. Let us now calculate the corresponding first-order expansion 
of the micromotion operator. The procedure is similar to that presented in Appendix~\ref{sec:App-C}, because 
we use the same generator form~(\ref{eq:gen_form3}) as in Appendix~\ref{sec:App-C}. The equations up to Eq.~(\ref{eq:ss1}) 
are identical, thus $\mathcal{S}_0 = 0$. Yet the first-order correction, taking into account the solution~(\ref{eq:fast4}), 
reads
\begin{equation}
  \mathcal{S}_1(t) = \frac{1}{\mathrm{i}\hbar \omega} \sum\limits_{m \neq 0} P_m 
  \otimes \sum\limits_{j=-J}^{J} \frac{h^{(m,j)}\mathrm{e}^{\mathrm{i}j\Omega t}}{m+\frac{\Omega}{\omega}j}.
\label{eq:mfc}
\end{equation}
\begin{widetext}
The micromotion operator in the physical space $\mathscr{H}$ reads
\begin{equation}
  U_{\mathrm{micro}}^{\dagger}(\omega t+\theta,t) 
  = \exp\left[ \frac{1}{\hbar \omega} \sum\limits_{m \neq 0} 
  \left\lbrace \frac{\mathrm{e}^{\mathrm{i}m(\omega t+\theta)}}{m}  
  \sum\limits_{j=-J}^{J} \frac{h^{(m,j)}\mathrm{e}^{\mathrm{i}j\Omega t}}{1+\frac{\Omega}{\omega}\frac{j}{m}} \right\rbrace 
  + \mathcal{O}\left(\frac{1}{(\hbar\omega)^2}\right)\right].
\label{eq:mfc_ps}
\end{equation}
Because of Eq.~(\ref{eq:fast1_2}) we have $|\,\Omega j / \omega m| < 1$ and using the geometric progression series 
$1/(1+x)=1-x+x^2-\cdots$ we can rewrite Eq.~(\ref{eq:mfc_ps}) in terms of the full Fourier harmonics, 
$h^{(m)}(t)$ and its derivatives as
\begin{equation}
U_{\mathrm{micro}}^{\dagger}(\omega t+\theta,t) = \exp\left[ \frac{1}{\hbar \omega} \sum\limits_{m \neq 0} \left\lbrace \frac{\mathrm{e}^{\mathrm{i}m(\omega t+\theta)}}{m}  \sum\limits_{l=0}^{+\infty} \left( \frac{\mathrm{i}}{\omega m} \frac{\mathrm{d}}{\mathrm{d}t} \right)^l h^{(m)}(t) \right\rbrace +\mathcal{O}\left(\frac{1}{(\hbar\omega)^2}\right)\right].
\label{eq:mfc_ps1}
\end{equation}
From the last expression one can see that if at the time moments $t=\{t_{\mathrm{in}},t_{\mathrm{fn}}\}$ the amplitude 
of the envelope equals to zero, $h^{(m\neq 0)}(t) = 0$, then the micromotion operator still does not equal unity, 
because of the derivatives of the Fourier components $\mathrm{d}^l h^{(m \neq 0)}(t)/\mathrm{d}t^l \neq 0$
of various orders. 
This result differs from the expansion presented in Appendix~\ref{sec:App-C}, where the first-order micromotion 
operator~(\ref{eq:ss1}) does not contain derivatives, the second-order micromotion operator~(\ref{eq:ss4}) has only 
first derivative and so on.

Similarly to Eq.~(\ref{eq:mfc_ps1}), one can rewrite the first order FE Hamiltonian~(\ref{eq:fast6}) in the terms of 
the full Fourier component $h^{(m)}(t)$ instead of the double-indexed Fourier components $h^{(m,j)}$. 
To do that we again use the geometric series for the fraction
\begin{equation}
  \frac{1}{m+\frac{\Omega}{2\omega}(j-j^{\prime})} 
  = \frac{1}{m} \left[1+\frac{\mathrm{i}^2 \Omega (j-j^{\prime})}{2\omega m}
  + \left(\frac{\mathrm{i}^2 \Omega (j-j^{\prime})}{2\omega m} \right)^2 +\ldots \right].
\label{eq:gs}
\end{equation}
Then, in the terms of $h^{(m)}(t)$ and its derivatives, the first order FE Hamiltonian reads
\begin{equation}
  h_{\mathrm{eff}(1)}(t) = \frac{1}{\hbar \omega} \sum\limits_{m\neq 0}\frac{1}{2m} 
  \sum\limits_{l=0}^{+\infty} \left( \frac{\mathrm{i}}{2\omega m} \right)^l 
  \sum\limits_{r=0}^{l} \frac{l! (-1)^r}{r!(l-r)!}
  \left[ \frac{\mathrm{d}^{(l-r)}h^{(m)}(t)}{\mathrm{d}t^{(l-r)}} ,\frac{\mathrm{d}^r h^{(-m)}(t)}{\mathrm{d}t^r}  \right].
\label{eq:fast7}
\end{equation}
\end{widetext}
The last expression is consistent with Eq.~(\ref{eq:fw_hfsolzz_2}). To be more precise, we can reproduce the term proportional 
to $[\dot{h}^{(m)},h^{(-m)}]$. Let us take only $l=0$ and $l=1$ in Eq.~(\ref{eq:fast7}). Then it reads
\begin{equation}
\begin{aligned}
  &h_{\mathrm{eff}(1)}(t) \approx \sum\limits_{m=1}^{+\infty} \frac{\left[ h^{(m)}(t),h^{(-m)}(t)  \right]}{m\hbar\omega}+\frac{1}{(2\hbar \omega)^2} \\
  & \times \sum\limits_{m\neq 0}\frac{\mathrm{i}\hbar}{m^2 } \left(\left[ \dot{h}^{(m)}(t) ,h^{(-m)}(t) \right]-\left[ h^{(m)}(t) ,\dot{h}^{(-m)}(t) \right] \right) \\
  &=\sum\limits_{m=1}^{+\infty} \frac{\left[ h^{(m)}(t),h^{(-m)}(t)  \right]}{m\hbar\omega}
  +\sum\limits_{m\neq 0}\frac{\mathrm{i}\hbar \left[ \dot{h}^{(m)}(t) ,h^{(-m)}(t) \right]}{2(\hbar \omega)^2 m^2 } .
\label{eq:fast8}
\end{aligned}
\end{equation}
If we assume that the additional time dependence is slow, such that $\dot{h}^{(m)}\sim\mathcal{O}(1)$, 
the second term of the last equation becomes second order, and it exactly equal to the term responsible 
for the non-Abelian geometric phase~\cite{Novicenko2017} in quantum systems where $[\dot{h}^{(m)},h^{(-m)}]\neq 0$. 
Note that Eq.~(\ref{eq:fast7}) contains not only first order but also higher order derivatives, 
thus it can be used to engineer unusual effects.

\section{Examples}
\label{sec:ex}

In this section, we describe several examples of high-frequency expansion for specific operator algebras.

\subsection{Spin-$\tfrac{1}{2}$ in an oscillating and slowly rotating magnetic field}

We begin by revisiting the model discussed in Refs.~\cite{Novicenko2017,Novicenko2019} as the simplest example
of a driven system where nontrivial behavior emerges from the modulation of the drive. The scaled driven
Hamiltonian reads
\begin{equation}
  h (\omega t, t) = [B_x (t) \sigma_x  + B_y (t) \sigma_y] \cos \omega t,
\end{equation}
and describes a spin-$\tfrac{1}{2}$ particle in a rapidly oscillating magnetic field (the frequency
$\omega$ sets the dominant scale) whose amplitude is also slowly changing as a function of time.
Here, the Hamiltonian and the magnetic field are measured in the units of the frequency.
The envelope is parametrized by the two projections $B_x (t)$ and $B_y (t)$. Its magnitude 
$[B_x^2 (t) + B_y^2 (t)]^{1/2}$ is of little interest to our purposes, but the direction must be changing
as a function of time. Thus, $B_x(t)$ and $B_y(t)$ must be distinct functions, i.e.\ we will take that
they are not equal or proportional. Automated generation of the effective Hamiltonian up to the fourth
order in the inverse frequency gives
\begin{subequations}
\label{pra17:123}
\begin{align}
  h_{\eff(0)} &= h_{\eff(1)} = h_{\eff(3)} = 0, \label{pra17:1}\\
  h_{\eff(2)} (t) &= \sigma_z \cdot 2 \left[B_x(t) B_y'(t) -  B_y(t) B_x'(t)\right] \cdot \omega^{-2}\label{pra17:2},\\
  h_{\eff(4)} (t) &= \sigma_z \cdot \left\{ \frac{1}{2} B_x''' B_y - \frac{1}{2} B_x B_y''' \right. \nonumber\\
  &+ B_x' \left[2 B_x^2 B_y + \frac{3 B_y''}{2} + 2 B_y^3 \right] \label{pra17:3}\nonumber\\
  &- \left. B_y' \left[2 B_x B_y^2 + \frac{3 B_x''}{2} + 2 B_x^3 \right] \right\}\cdot\omega^{-4}.
\end{align}
\end{subequations}
Here, in the last subequation we skipped the time argument for brevity. As the zeroth and the first terms
vanish, the leading term is given by Eq.~(\ref{pra17:2}). This term can be expressed as
\begin{equation}
  H = 2 \omega^{-2} (\bm{B} \times \bm{\dot{B}}) \cdot \bm{\sigma}, \;\text{with}\quad
  \bm{B} = \bm{e}_x B_x(t) + \bm{e}_y B_y(t),
\end{equation}
and reproduces the result of Ref.~\cite{Novicenko2017}. The availability of automated expansion allows
us obtain also the subleading term behaving as $\omega^{-4}$ and  featuring higher-order derivatives 
and their combinations. We note that all terms in Eq.~(\ref{pra17:123}) are proportional to derivatives, and vanish 
identically in the absence of temporal modulation.

\subsection{Minimal many-body system}
In this example, we discuss a minimal many-body system of two interacting bosonic particles populating a dimer consisting 
of two lattice sites. We introduce the creation (annihilation) operators $c_j^{\dag}$ ($c_j$)
with the subscript $j \in \{1, 2\}$ corresponding to the site number, and write the driven Hamiltonian as
\begin{equation}
\label{eq:mb_ham}
  h (\omega t, t) = J (\omega t, t) \, \tau_1 + \frac{\Delta (\omega t, t)}{2} \, \tau_3 + \frac{U}{2} \, \tau_4,
\end{equation}
with the operators
\begin{subequations}
\begin{align}
  \tau_1 &= \ca_1 \co_2 + \ca_2 \co_1, \\
  \tau_3 &= \ca_2 \co_2 - \ca_1 \co_1, \\
  \tau_4 &= \ca_1 \ca_1 \co_1 \co_1 + \ca_2 \ca_2 \co_2 \co_2.
\end{align}
The three parts of the Hamiltonian (\ref{eq:mb_ham}) describe, respectively, the hopping transitions with 
a time-dependent transition strength $J$, the difference of onsite energies $\Delta$, and bosonic onsite 
interaction of strength $U$. Let us list also the remaining operators that will be needed to describe 
the  effective dynamics:
\begin{align}
  \tau_2 &= i(\ca_1 \co_2 - \ca_2 \co_1),\\
  \tau_5 
         &= i (n_2 \ca_1 \co_2 + n_1 \ca_2 \co_1) + \mathrm{h.c.}, \\
  \tau_6 
         &= n_2 \ca_1 \co_2 - n_1 \ca_2 \co_1 + \mathrm{h.c.}, \\
  \tau_7 &= n_1 n_2, \\
  \tau_8 &= \ca_1 \ca_1 \co_2 \co_2 + \ca_2 \ca_2 \co_1 \co_1,\\
  \tau_9 &= i (\ca_2 \ca_2 \co_1 \co_1 - \ca_1 \ca_1 \co_2 \co_2 ),
\end{align}
\end{subequations}
here $n_j = \ca_j \co_j$.

The collection of operators $\tau_n$ is closely related, albeit not in one-to-one correspondence, 
to the $\mathsf{su(3)}$ algebra spanned by the Gell-Mann matrices. To see this, we define the basis
of three two-particle states
\begin{subequations}
\begin{align}
  | 1 \rangle &= \tfrac{1}{\sqrt{2}} \,\ca_1 \ca_1 | \emptyset \rangle,\\
  | 2 \rangle &=  \,\ca_1 \ca_2 | \emptyset \rangle,\\
  | 3 \rangle &= \tfrac{1}{\sqrt{2}} \,\ca_2 \ca_2 | \emptyset \rangle,
\end{align}
\end{subequations}
here $| \emptyset \rangle$ denotes the vacuum state. In this basis, the operators $\tau_n$ are
encoded as matrices that are simple combinations of the Gell-Mann matrices $\lambda_m$ and the $3 \times 3$
unit matrix. For example, $\tau_1$ and $\tau_6$ are represented as
\begin{equation}
  \tau_{1|6} = \begin{pmatrix}  0 & \pm\sqrt{2} & 0 \\ \pm\sqrt{2} & 0 & \sqrt{2} \\ 0 & \sqrt{2} & 0 \end{pmatrix}
  = \frac{(\lambda_6 \pm \lambda_1)}{\sqrt{2}},
\end{equation}
with the upper (lower) sign for $\tau_1$ ($\tau_6$). Note that the physical content of the model requires that the unit 
matrix is also included: Interactions are described in terms of matrices $\tau_4$ (onsite interaction) and $\tau_7$ 
(neighbor interactions) that have nonzero traces. The introduced matrix representations allow for an easy calculation
of the commutator algebra and its implementation in a symbolic computation system.

To give an example, we assume that the shaking protocol is such that
\begin{subequations}
\label{eq:prot12}
\begin{align}
  J (\omega t, t) &= j_0 + 2 j_1 (t) \cos \omega t, \label{eq:prot1}\\
  \Delta (\omega t, t) &= \delta_0 = \mathrm{const}., \label{eq:prot2}
\end{align}
\end{subequations}
that is, the onsite energy splitting is a constant set to $\delta_0$ whereas the hopping matrix element 
is harmonically modulated in time around some average value $j_0$. The factor $2$ is included in Eq.~(\ref{eq:prot1})
to account for the fact that $2 \cos \omega t = \e^{\ri \omega t} + \e^{-\ri \omega t}$ so that $j_1 (t)$ has the
meaning of the modulated first Fourier component.

We obtain the effective Hamiltonian in the form of a high-frequency expansion
\begin{equation}
  h_{\eff} = \sum_{l=1}^9 \tau_l \cdot \left( c_{l,0} + c_{l,2} + c_{l,4} + \cdots \right),
\end{equation}
with $c_{l,i} \sim \m{O} (\omega^{-i})$. Note that first- and third-order terms vanish. In the zeroth order,
quite obviously, only time averages are present
\begin{equation}
  c_{1,0} = j_0, \quad c_{3,0} = \delta_0, \quad c_{4,0} = \frac{U}{2},
\end{equation}
and the remaining coefficients $c_{l,0} = 0$ with $l \in \{2,5,6,7,8,9 \}$.

In the second order, we find four nonzero contributions
\begin{subequations}
\label{eq:modify}
\begin{align}
  c_{3,2} &= -\frac{4 \delta_0 j_1^2(t)}{(\hbar\omega)^2}, \label{eq:modify1} \\
  c_{4,2} &= -\frac{2 U j_1^2(t)}{(\hbar\omega)^2}, \label{eq:modify2} \\
  c_{7,2} &= \frac{8 U j_1^2(t)}{(\hbar\omega)^2}, \label{eq:modify3} \\
  c_{8,2} &= -\frac{2 U j_1^2(t)}{(\hbar\omega)^2} \label{eq:modify4}.
\end{align}
\end{subequations}
Interestingly, Eq,~(\ref{eq:modify2}) shows that shaking has led to a renormalization of the onsite interaction 
strength from $U$ to ($U - \Delta U$) with $\Delta U = 4 U j_1^2(t)/(\hbar\omega)^2$. This is easy to interpret
in terms of a process where a particle is able to jump twice during the period of the drive: Two particles
that share a site, in fact spend some fraction of the period sitting on separate sites and thus not interacting.
Likewise, as a consequence of the same process Eq.~(\ref{eq:modify3}) shows that shaking has led to the engineering
of hitherto absent nearest-neighbor interactions: Two particles sitting on neighboring sites experience 
an interaction energy of magnitude $V = 2 \Delta U$. The factor $2$ comes from the fact that in the considered 
dimer two sites share a single link. The appearance of modified interactions with a similar sum rule can be generalized
to large lattices as discussed in Refs.~\cite{Eckardt2015,Anisimovas2015}. A similar expression in Eq.~(\ref{eq:modify4})
shows that the described process is also responsible for the appearance of pair tunnelings expressed by 
the operator $\tau_8$. Finally, from Eq.~(\ref{eq:modify1}) we also learn that the difference of onsite energies
is also weakened. This effect is not related to interactions, thus the modification is proportional to $\delta_0$
rather than $U$ (as seen in previous cases). However, the proportionality to $j_1^2(t)$ is still present, revealing
that a sequence of two hopping transitions during the period of the drive is involved. The remaining second-order
contributions are zero, thus $c_{l,2} = 0$ for $l \in \{1, 2, 5, 6, 9 \}$.

In the fourth order of the high-frequency expansion we find further modifications to the terms already affected 
in the second order --- the coefficients $c_{3,4}$, $c_{4,4}$, $c_{7,4}$, and $c_{8,4}$ are all nonzero.
However, these effects are not the leading ones, they are dwarfed by the presence of already discussed processes,
and therefore, are not so interesting. Let us just mention, that they feature combinations of the time derivatives
such as $j''_1(t)$ and $[j'_1(t)]^2$ as a key prediction of the presented high-frequency expansion for systems
with the time-modulated drive.

In addition, we find two more fourth-order contributions to the effective Hamiltonian,
\begin{subequations}
\begin{align}
  c_{1,4} &= - \frac{12 j_0 j_1^2 (t) (\delta_0^2 + U^2)}{(\hbar\omega)^4}, \\
  c_{6,4} &= - \frac{10 j_0 j_1^2 (t) \delta_0 U}{(\hbar\omega)^4}.
\end{align}
\end{subequations}
The former signals the reduction of the hopping matrix element, while the latter is an indication of 
density-assisted tunneling events captured by $\tau_6$. These processes are, however, best explored
on large lattices, that is beyond, the overly restrictive dimer model.

The consideration of an alternative [to Eq.~(\ref{eq:prot12})] shaking protocol that modifies onsite
energies, i.e.\ $\Delta$, rather than the hopping strengths, 
\begin{subequations}
\label{eq:alt12}
\begin{align}
  J (\omega t, t) &= j_0 = \mathrm{const}., \label{eq:alt1}\\
  \Delta (\omega t, t) &= \delta_0 + 2 \delta_1 (t) \cos \omega t, \label{eq:alt2}
\end{align}
\end{subequations}
is conceptually similar. In the fourth
order it delivers a number of contributions that are leading, i.e.\ not dwarfed by the second-order contributions.
(The first and the third order still vanish.) For example, the renormalized onsite interaction energy is
\begin{equation}
  U_{\eff} = U - \frac{12 j_0^2 U \delta_1^2(t)}{(\hbar\omega)^4} + \m{O} ((\hbar\omega)^{-6}).
\end{equation}
In the second order, the only effect is the well-known \cite{Eckardt2017review} modification of the tunneling strength
\begin{equation}
  j_{\eff} = j_0 \left[ 1 - \left(\frac{2 \delta_1(t)}{\hbar\omega} \right)^2 + \m{O} ((\hbar\omega)^{-4}) \right],
\end{equation}
by the Bessel function of the zeroth-order, identifiable by the leading terms of its power series.

\section{Conclusions}

To summarize, the flow-equation approach is a versatile tool useful in the study 
of driven quantum systems. This method accomplishes a gradual transformation
of the given Hamiltonian from its original form to the desired (diagonal
or block diagonal) final
form. The flow is implemented by introducing an auxiliary flow variable and specifying 
an anti-Hermitian generator that expresses the law of motion along the flow variable.
In our work, we demonstrated the applicability of such a scheme to the block diagonalization 
of Kamiltonian (quasienergy) matrices that describe driven quantum systems
in the extended-space formalism. Importantly, we consider situations where
the time dependence is twofold: the system is driven be a force of a certain
frequency and its amplitude is additionally modulated as a function of time.
Thus, the drive is described as a superposition of a number of Fourier harmonics
whose amplitudes are time dependent. The outcome of the flow procedure is the
Floquet effective Hamiltonian and the complementary micromotion operator that
allow to faithfully describe the system's time evolution. Alongside a previously 
proposed Verdeny-Mielke-Minter generator \cite{Verdeny2013}, we introduced  
a modified (Toda) generator~(\ref{eq:gen_form4}) that leads to similar flow 
equations~(\ref{eq:fw_znz1}), 
however, also guarantees that the spectrum of the contributing Fourier harmonics does not 
broaden beyond what was present in the driven Hamiltonian. We note, that the above
procedure, in principle, applies to both slow and rapid modulation of the amplitude
of the drive. However, concrete expressions for the effective Hamiltonian and the
micromotion operator are obtained from the high-frequency expansion that assumes
that driving frequency sets the dominant scale. The feature of nonproliferation 
of the Fourier harmonics during the flow is particularly useful 
when enlisting the help of computer algebra systems to obtain automated high-frequency 
expansions for the effective Hamiltonian and the micromotion operator. 
Such expansions are cumbersome to perform manually 
beyond a few leading terms. We showed that in the case when the Hamiltonian is spanned by 
a finite Lie algebra, the derivation of the high-frequency expansion can be straightforwardly
implemented based on automated solution of a set of linear inhomogeneous differential 
equations. 
The applicability of such a strategy is supported by studying a selection 
of examples: a spin-$\tfrac{1}{2}$ in oscillating and slowly rotating magnetic field,
a minimal many-body system consisting of two interacting bosons on a driven dimer,
and a two-level system subject to a linearly polarized drive.

\acknowledgments
This research was funded by European Social Fund under Grant No.\ 09.3.3-LMT-K-712-01-0051.
The authors are grateful to Th.~Gajdosik, V.~Regelskis, and G.Juzeliūnas for enlightening discussions.

\appendix
\section*{Appendix}

\section{Validity of the generator for the discrete flow}
\label{sec:App-A}

Let us consider a discrete flow indexed by an integer-valued variable running from zero to infinity, 
and assume that having reached a particular value of $s$ the Kamiltonian has the form
\begin{equation}
\begin{aligned}
  \mathcal{K}(s) =& \hbar \omega N \otimes \mathbf{1}_{\mathscr{H}} 
  + P_0 \otimes \sum\limits_{i=0}^{+\infty} H^{(0)}_i(s) \\
  &+\sum\limits_{m\neq 0} \left( P_m \otimes \sum\limits_{i=s}^{+\infty} H^{(m)}_i(s) \right).
\label{eq:kdf}
\end{aligned}
\end{equation}
Here, in the off-diagonal part of the Kamiltonian the power series of Fourier components $H^{(m)}$ with 
$m \neq 0$ start from the order $s$, that is, the terms that behave as $(\hbar\omega)^{-i}$ with
$i \in [0, s-1]$ have already been eliminated. For the sake of simplicity, in this section 
we will not write the explicit dependence on time but will keep it in mind. Now we want to demonstrate 
that the application of the following step of the flow governed by Eq.~(\ref{eq:flow_discr_kam}) with
the generator~(\ref{eq:gen_dicsr}) will move the Kamiltonian forward along the flow: The flow variable 
will reach the value $s+1$, and the power series of the Fourier components will start from the order $(s+1)$. 
Thus, the general form of Eq.~(\ref{eq:kdf}) will be maintained subject to relabeling $s \to (s +1)$.
Note that the generator~(\ref{eq:gen_dicsr}) is of the order $(\hbar\omega)^{-(s+1)}$, but is combined 
with the first term on the right hand side of Eq.~(\ref{eq:kdf}) which is proportional to $\hbar\omega$.
This allows to eliminate the $s$-th order terms. Of course, during the considered $s \to (s+1)$ step, 
the flow will modify not only term of order $s$ and $(s+1)$, but higher order terms as well. However, 
in order to prove our statement it is sufficient to focus on terms that behave as $(\hbar\omega)^{-s}$
and $(\hbar\omega)^{-(s+1)}$. Let us expand the first term in Eq.~(\ref{eq:flow_discr_kam})
\begin{equation}
\begin{aligned}
  &\e^{\ri\mathcal{S}(s)}\mathcal{K}(s)\e^{-\ri\mathcal{S}(s)} 
    = \mathcal{K}(s)+\left[ \ri\mathcal{S}(s),\mathcal{K}(s) \right] \\
  &+\frac{1}{2!}\left[\ri \mathcal{S}(s),\left[ \ri \mathcal{S}(s),\mathcal{K}(s) \right] \right] + \cdots.
\label{eq:fir_term_disc}
\end{aligned}
\end{equation}
The first commutator can be written as
\begin{widetext}
\begin{equation}
\begin{aligned}
  &\left[\ri \mathcal{S}(s),\mathcal{K}(s) \right] 
    = -\sum\limits_{m\neq 0} P_m \otimes H^{(m)}_s(s)
  + \sum\limits_{m\neq 0} \left( \frac{P_m}{m} \otimes \sum\limits_{i=0}^{+\infty} 
    \frac{\left[ H^{(m)}_s(s),H^{(0)}_i(s) \right]}{\hbar \omega} \right)\\
  &+ P_0 \otimes \sum\limits_{m \neq 0} \sum\limits_{i=s}^{+\infty} 
    \frac{\left[H^{(m)}_s(s), H^{(-m)}_i(s) \right]}{m\hbar \omega} 
  +\sum\limits_{m \neq 0} \sum\limits_{k \neq \{0,m\}}  P_k 
    \otimes  \sum\limits_{i=s}^{+\infty} \frac{\left[H^{(m)}_s(s), H^{(k-m)}_i(s) \right]}{m\hbar \omega} .
\label{eq:com}
\end{aligned}
\end{equation}
\end{widetext}
Now we need to consider separately two cases: $s=0$ and $s\geqslant 1$. For $s\geqslant 1$, one may drop 
the two last terms from Eq.~(\ref{eq:com}):
\begin{equation}
\begin{aligned}
  &\left[\ri \mathcal{S}(s),\mathcal{K}(s) \right] 
    = -\sum\limits_{m\neq 0} P_m \otimes H^{(m)}_s(s) \\
  &+ \sum\limits_{m\neq 0}  P_m \otimes \frac{\left[ H^{(m)}_s(s),H^{(0)}_0(s) \right]}{m\hbar \omega} 
    + \mathcal{O}\left( \frac{1}{(\hbar \omega)^{s+2}} \right).
\label{eq:com1}
\end{aligned}
\end{equation}
The leading term here is the first term on the right hand side; when added to the Kamiltonian, it will eliminate 
the $s$-th order term of the $m \neq 0$ Fourier components that appear in the block off-diagonal part. The double 
commutator $\left[\mathrm{i} \mathcal{S},\left[ \mathrm{i} \mathcal{S},\mathcal{K} \right] \right]$ and subsequent 
nested commutators in Eq.~(\ref{eq:fir_term_disc}) contribute to orders higher than $s+1$, and need not 
be considered. The second term on the right hand side of Eq.~(\ref{eq:flow_discr_kam}) reads
\begin{equation}
\begin{aligned}
  &-\ri\hbar \mathrm{e}^{\ri\mathcal{S}(s)}\frac{\mathrm{d} \mathrm{e}^{-\ri\mathcal{S}(s)}}{\mathrm{d} t} 
    = \ri\hbar \left( \ri\dot{\mathcal{S}}(s)\right)+\frac{\ri\hbar }{2!}
    \left[ \ri\mathcal{S}(s),\ri\dot{\mathcal{S}}(s) \right] + \ldots \\
  &= \ri \hbar \sum\limits_{m\neq 0} P_m \otimes \frac{\dot{H}^{(m)}_s (s)}{m\hbar \omega}
    + \mathcal{O}\left( \frac{1}{(\hbar \omega)^{s+2}} \right).
\label{eq:td}
\end{aligned}
\end{equation}
Summarizing Eqs.~(\ref{eq:fir_term_disc}), (\ref{eq:com1}) and (\ref{eq:td}) one can write down recursion expressions 
for the transformed Kamiltonian $ \mathcal{K} (s+1)$. The Fourier components read: 
$H^{(m\neq 0)}_i(s+1) = 0$ for $i \in [0, s]$ and
\begin{equation}
H^{(m\neq 0)}_{s+1}(s+1)=H^{(m)}_{s+1}(s)+\frac{\left[ H^{(m)}_s(s),H^{(0)}_0(s) \right]}{m\hbar \omega}+\frac{\mathrm{i}\dot{H}^{(m)}_s (s)}{m \omega}.
\label{eq:nrec}
\end{equation}
The relevant contributions to the zeroth Fourier component remain unchanged, $H^{(0)}_i(s+1)=H^{(0)}_i(s)$ for $i\in [0,s+1]$.

Proceeding to the case $s = 0$, one should repeat the corresponding analysis starting from Eq.~(\ref{eq:com}). 
The analog of Eq.~(\ref{eq:com1}) reads
\begin{widetext}
\begin{equation}
\begin{aligned}
&\left[\mathrm{i} \mathcal{S}(0),\mathcal{K}(0) \right] = -\sum\limits_{m\neq 0} P_m \otimes H^{(m)}_0(0) 
+ \sum\limits_{m\neq 0}  P_m \otimes \frac{\left[ H^{(m)}_0(0),H^{(0)}_0(0) \right]}{\hbar \omega} \\
&+ P_0 \otimes \sum\limits_{m \neq 0} \frac{\left[H^{(m)}_0(0), H^{(-m)}_0(0) \right]}{m\hbar \omega} 
+\sum\limits_{m \neq 0} \sum\limits_{k \neq \{0,m\}}  P_k \otimes  \frac{\left[H^{(m)}_0(0), H^{(k-m)}_0(0) \right]}{m\hbar \omega} 
+\mathcal{O}\left( \frac{1}{(\hbar \omega)^2} \right).
\label{eq:com2}
\end{aligned}
\end{equation}
Now also the double commutator from Eq.~(\ref{eq:fir_term_disc}) has to be taken into account:
\begin{equation}
\begin{aligned}
  &\frac{\left[\ri \mathcal{S}(s),\left[ \ri \mathcal{S}(s),\mathcal{K}(s) \right]\right]}{2!}
    = -P_0 \otimes \sum\limits_{m \neq 0} \frac{\left[H^{(m)}_0(0), H^{(-m)}_0(0) \right]}{2m\hbar \omega} \\
  &-\sum\limits_{m \neq 0} \sum\limits_{k \neq \{0,m\}}  P_k 
    \otimes  \frac{\left[H^{(m)}_0(0), H^{(k-m)}_0(0) \right]}{2m\hbar \omega} 
  + \mathcal{O}\left( \frac{1}{(\hbar \omega)^2} \right).
\label{eq:dcom}
\end{aligned}
\end{equation}

Thus the recursive expressions for the $m \neq 0$ Fourier components read: $H^{(m\neq 0)}_0(1)=0$ and
\begin{equation}
\begin{aligned}
H^{(m\neq 0)}_{1}(1)=H^{(m)}_{1}(0)+\frac{\left[ H^{(m)}_0(0),H^{(0)}_0(0) \right]}{m\hbar \omega}  
+ \sum\limits_{k \neq \{0,m\}} \frac{\left[H^{(k)}_0(0), H^{(m-k)}_0(0) \right]}{2k\hbar \omega}+\frac{\mathrm{i}\dot{H}^{(m)}_0 (0)}{m \omega}.
\label{eq:nrec1}
\end{aligned}
\end{equation}
\end{widetext}%
While for the zeroth Fourier component we have $H^{(0)}_0(1)=H^{(0)}_0(0)$ and
\begin{equation}
H^{(0)}_1(1)=H^{(0)}_1(1)+\sum\limits_{m \neq 0} \frac{\left[H^{(m)}_0(0), H^{(-m)}_0(0) \right]}{2m\hbar \omega}.
\label{eq:drec1}
\end{equation}
The last two equations can be simplified taking into account that $H^{(m)}_1(0)=0$ and $H^{(m)}_0(0)=h^{(m)}$,
thus in terms of the Fourier components of the driven Hamiltonian
\begin{equation}
\begin{aligned}
H^{(m\neq 0)}_{1}(1) =& \frac{\left[ h^{(m)},h^{(0)} \right]}{m\hbar \omega} + \sum\limits_{k \neq \{0,m\}} \frac{\left[h^{(k)}, h^{(m-k)} \right]}{2k\hbar \omega}\\ 
&+\frac{\mathrm{i}\dot{h}^{(m)}}{m \omega},
\label{eq:nrec2}
\end{aligned}
\end{equation}
and
\begin{equation}
H^{(0)}_1(1)=\sum\limits_{m \neq 0} \frac{\left[h^{(m)}, h^{(-m)} \right]}{2m\hbar \omega}.
\label{eq:drec2}
\end{equation}

Now let us return to the case $s\geqslant 1$. By looking at Eqs.~(\ref{eq:com1}) and (\ref{eq:td}) one can see 
that at the flow step $s \to (s + 1)$ the diagonal Fourier components $H^{(0)}_i$ are not affected for $i \in [0, s+1]$. 
Then one can ask, at which lowest order the diagonal Fourier components are affected? From the third term 
of Eq.~(\ref{eq:com}) one can see that the diagonal Fourier component is modified only with terms proportional 
to $(\hbar\omega)^{-(2s+1)}$. This means that having obtained the Kamiltonian $K(s)$, we can claim that its diagonal 
blocks have converged up to the order $2s$, because further steps of the flow will not modify these terms. The upshot
of the argument is that from the Kamiltonian $K(s)$ we can extract the FE Hamiltonian as
\begin{equation}
  h_{\mathrm{eff}}= \sum\limits_{i=0}^{2s} H^{(0)}_i(s)+\mathcal{O}\left( \frac{1}{(\hbar \omega)^{2s+1}} \right),
\label{eq:fe_extr}
\end{equation}
that is, the accuracy is much better than anticipated from Eq.~(\ref{eq:kam_disc_expan1}).

\section{Solution of the flow equations~(\ref{eq:fw_znz}) in terms of a high-frequency expansion}
\label{sec:App-B}

In this section, we solve Eq.~(\ref{eq:fw_znz}) in terms of power series in the inverse frequency.
We thus assume that the amplitude of the Fourier components $h^{(m)}(t)$, as well as their time derivatives 
$\mathrm{d}^j h^{(m)}(t)/ \mathrm{d}t^j$, are of the order of $\mathcal{O}(1)$. For the sake of simplicity, 
in this section we will not write the explicit dependence on time but will keep it in mind. Each running 
Fourier component $H^{(m)}(s,t)$ is expanded as
\begin{equation}
 H^{(m)}(s) = H^{(m)}_0(s)+H^{(m)}_1(s)+H^{(m)}_2(s)+\ldots,
\label{eq:fexp}
\end{equation}
where the expansion terms $H^{(m)}_j(s) \sim \mathcal{O}(1/(\hbar \omega)^j)$. The initial conditions are
\begin{equation}
H^{(m)}_0(0)=h^{(m)},
\label{eq:int_cond}
\end{equation}
while for any $j\geqslant 1$ we have
\begin{equation}
H^{(m)}_j(0)=0.
\label{eq:int_cond1}
\end{equation}
Substituting Eq.~(\ref{eq:fexp}) into the left and right hand side of Eq.~(\ref{eq:fw_znz}) we obtain 
the zeroth-order flow equations
\begin{equation}
  \frac{\mathrm{d}H^{(0)}_0(s)}{\mathrm{d}s}=0, 
  \quad\text{and}\quad 
  \frac{\mathrm{d}H^{(n)}_0(s)}{\mathrm{d}s}=-n^2 H^{(n)}_0(s).
\label{eq:fw_hf_0}
\end{equation}
The solutions to these equations are
\begin{equation}
H^{(0)}_0(s)=h^{(0)} \quad \mathrm{and} \quad H^{(n)}_0(s)=h^{(n)}\mathrm{e}^{-n^2 s}.
\label{eq:fw_hfsol_0}
\end{equation}
Here we can see that in the limit $s\rightarrow +\infty$ the off-diagonal Fourier components vanish while 
the diagonal Fourier component gives the zeroth-order approximation of the FE Hamiltonian, 
$h_{\mathrm{eff}(0)}=h^{(0)}$.

In the first order, Eq.~(\ref{eq:fw_z}) gives
\begin{equation}
\frac{\mathrm{d}H^{(0)}_1(s)}{\mathrm{d}s}=\frac{2}{\hbar \omega} \sum\limits_{m=1}^{+\infty} m \left[ h^{(m)},h^{(-m)} \right] \mathrm{e}^{-2m^2 s}.
\label{eq:fw_hfz_1}
\end{equation}
By taking into account Eq.~(\ref{eq:int_cond1}), the solution to Eq.~(\ref{eq:fw_hfz_1}) reads
\begin{equation}
H^{(0)}_1(s)=\sum\limits_{m=1}^{+\infty} \frac{\left[ h^{(m)},h^{(-m)} \right]}{m\hbar \omega} \left(1- \mathrm{e}^{-2m^2 s} \right).
\label{eq:fw_hfsolz_1}
\end{equation}
Taking the limit $s \to \infty$ one obtains the first-order approximation of the FE Hamiltonian
\begin{equation}
h_{\mathrm{eff}(1)}= \lim\limits_{s \rightarrow +\infty} H^{(0)}_1(s)=\sum\limits_{m=1}^{+\infty} \frac{\left[ h^{(m)},h^{(-m)} \right]}{m\hbar \omega}.
\label{eq:fw_hfsolzz_1}
\end{equation}
In the first order Eq.~(\ref{eq:fw_nz}) gives the following linear inhomogeneous differential equation
\begin{equation}
\begin{aligned}
&\frac{\mathrm{d}H^{(n)}_1(s)}{\mathrm{d}s}=-n^2 H^{(n)}_1(s)+\frac{\mathrm{i}n}{\omega}  \dot{h}^{(n)} \mathrm{e}^{-n^2 s}\\
&+\sum\limits_{m \neq n} \frac{(m-n)}{\hbar \omega} \left[ h^{(m)},h^{(n-m)}\right]\mathrm{e}^{-\left[ m^2+(n-m)^2 \right]s}.
\label{eq:fw_hf_1}
\end{aligned}
\end{equation}
The solution to Eq.~(\ref{eq:fw_hf_1}) corresponding to the initial conditions~(\ref{eq:int_cond1}) reads
\begin{equation}
\begin{aligned}
&H^{(n)}_1(s)= n \left( \frac{\mathrm{i}}{\omega}  \dot{h}^{(n)} - \frac{\left[ h^{(0)},h^{(n)}\right]}{\hbar \omega}  \right) s\mathrm{e}^{-n^2 s}\\
&+\sum\limits_{m \neq \{0,n\}} \frac{\left[ h^{(m)},h^{(n-m)}\right]}{2m\hbar \omega} \left( \mathrm{e}^{-n^2 s} -\mathrm{e}^{-\left[m^2+ (n-m)^2 \right] s} \right) .
\label{eq:fw_hfsol_1}
\end{aligned}
\end{equation}
One can see that in the limit $s \rightarrow +\infty$ the term $H^{(n)}_1(s)$ vanishes.

\begin{widetext}
To proceed further, we collect the second-order terms of Eq.~(\ref{eq:fw_z}) 
\begin{equation}
\begin{aligned}
  &\frac{\mathrm{d}H^{(0)}_2(s)}{\mathrm{d}s} = 
  \sum\limits_{n=1}^{+\infty} \frac{2n}{\hbar \omega} \left\lbrace \left[H^{(n)}_1(s),H^{(-n)}_0(s)\right] + \left[H^{(n)}_0(s),H^{(-n)}_1(s)\right] \right\rbrace 
=\sum\limits_{n \neq 0} \frac{\left[ h^{(n)} , \left[ h^{(0)},h^{(-n)} \right] \right]}{(\hbar\omega)^2}  2n^2 s \mathrm{e}^{-2n^2 s} \\
&+\sum\limits_{n\neq 0} \sum\limits_{m\neq \{0,n\}} \frac{\left[ h^{(-n)} , \left[ h^{(n-m)},h^{(m)} \right] \right]}{(\hbar\omega)^2} \frac{n}{m} \left( \mathrm{e}^{-2n^2 s} \right. 
 \left.-\mathrm{e}^{-\left[m^2+n^2+(n-m)^2\right] s} \right)+2\mathrm{i}\hbar \sum\limits_{n\neq 0} \frac{\left[ \dot{h}^{(n)},h^{(-n)} \right]}{(\hbar \omega)^2} n^2 s \mathrm{e}^{-2n^2 s}.
\label{eq:fw_hfz_2}
\end{aligned}
\end{equation}
The solution corresponding to the initial conditions~(\ref{eq:int_cond1}) reads
\begin{equation}
\begin{aligned}
&H^{(0)}_2(s)=\sum\limits_{n \neq 0} \frac{\left[ h^{(n)} , \left[ h^{(0)},h^{(-n)} \right] \right]}{(\hbar\omega)^2} \frac{1-\mathrm{e}^{-2n^2 s}\left(2n^2 s+1 \right)}{2n^2} \\
& +\sum\limits_{n\neq 0} \sum\limits_{m\neq \{0,n\}} \frac{\left[ h^{(-n)} , \left[ h^{(n-m)},h^{(m)} \right] \right]}{(\hbar\omega)^2} \left\lbrace \frac{1-\mathrm{e}^{-2n^2 s}}{2nm} \right. 
 \left. -\frac{n\left(1- \mathrm{e}^{-\left[m^2+n^2+(n-m)^2\right] s}\right)}{m\left(m^2+n^2+(n-m)^2\right)} \right\rbrace \\
&+\mathrm{i}\hbar\sum\limits_{n\neq 0} \frac{\left[ \dot{h}^{(n)},h^{(-n)} \right]}{(\hbar \omega)^2} \frac{1-\mathrm{e}^{-2n^2 s}\left(1+2n^2 s \right)}{2n^2}.
\label{eq:fw_hfsolz_2}
\end{aligned}
\end{equation}
Taking the limit, one obtains the second-order approximation of the FE Hamiltonian
\begin{equation}
\begin{aligned}
&h_{\mathrm{eff}(2)}= \lim\limits_{s \rightarrow +\infty} H^{(0)}_2(s)= \\
&\frac{1}{(\hbar\omega)^2}\sum\limits_{n\neq 0} \left\lbrace \frac{\left[ h^{(n)} , \left[ h^{(0)},h^{(-n)} \right] \right] +\mathrm{i}\hbar \left[ \dot{h}^{(n)},h^{(-n)} \right]}{2n^2} \right. 
\left. +\sum\limits_{m\neq \{0,n\}}\frac{\left[ h^{(-n)} , \left[ h^{(n-m)},h^{(m)} \right] \right](m-n)}{n\left(m^2+n^2+(n-m)^2\right)}\right\rbrace.
\label{eq:fw_hfsolzz_2}
\end{aligned}
\end{equation}
\end{widetext}
Comparing the obtained results with those given in Ref.~\cite{Novicenko2017}, one does not immediately 
appreciate their equivalence. Indeed, the third term of Eq.~(\ref{eq:fw_hfsolzz_2}) reads
\begin{equation}
A= \sum\limits_{n \neq 0} \sum\limits_{m\neq \{0,n\}} \left[ h^{(-n)} , \left[ h^{(n-m)},h^{(m)} \right] \right] f(n,m),
\label{eq:A}
\end{equation}
with $f(n,m)=(m-n)/\left( n\left[m^2+n^2+(n-m)^2\right]\right)$. However, while the corresponding term 
from Ref.~\cite{Novicenko2017} reads
\begin{equation}
B= \sum\limits_{n \neq 0} \sum\limits_{m\neq \{0,n\}} \left[ h^{(-n)} , \left[ h^{(n-m)},h^{(m)} \right] \right] g(n,m),
\label{eq:B}
\end{equation}
with $g(n,m)=1/(3nm)$. In Appendix~\ref{sec:App-D} we show that $A = B$, thus up to the second order the FE Hamiltonian 
obtained from the flow equations~(\ref{eq:fw_znz}) coincides with the results in Ref.~\cite{Novicenko2017}
as well as in Refs.~\cite{Goldman2014,Eckardt2015}.

%
\section{Micromotion operator obtained from the flow equations~(\ref{eq:fw_znz}) in terms of a high-frequency expansion}
\label{sec:App-C}

In this section, we will find the micromotion operator~(\ref{eq:mic}) in the exponential form $U_{\mathrm{micro}}=\exp(-\mathrm{i}S)$, 
where the Hermitian operator $S = S_0 + S_1 + S_2 + \ldots$ is obtained as a series expansion with respect to the inverse frequency. 
With respect to Eq.~(\ref{eq:flow_D_cont}), the extended-space unitary transformation $\mathcal{D}^{\dagger}(t)$ 
can be obtained with the help of the Magnus expansion in the exponential form 
$\mathcal{D}^{\dagger}(t)=\exp\left[ \mathrm{i}(\mathcal{S}_0+\mathcal{S}_1+\mathcal{S}_2+\ldots) \right]$ where the exponent 
is expanded in power series with respect to the inverse frequency. The relation between the physical-space operator $S_j$ 
and the extended-space operator $\mathcal{S}_j=\sum_m P_m \otimes S^{(m)}_j$ is established by simply replacing the shift 
operator $P_m$ by the exponent $\mathrm{e}^{\mathrm{i}m(\omega t+\theta)}$: 
$S_j(\omega t+\theta, t)=\sum_m S_j^{(m)}(t)\mathrm{e}^{\mathrm{i}m(\omega t+\theta)}$. In terms of the generator 
$\mathrm{i}\mathcal{S}(s)$, the logarithm of the operator $\mathcal{D}^{\dagger}$ reads (for the simplicity, in this section 
we will not write the explicit dependence on time but will keep it in mind):
\begin{equation}
\begin{aligned}
&\mathrm{i}\left(\mathcal{S}_0+\mathcal{S}_1+\mathcal{S}_2+\ldots \right)= \\
& \int\limits_0^{+\infty} \mathrm{i}\mathcal{S}(s_1) \mathrm{d}s_1+\frac{\mathrm{1}}{2}\int\limits_0^{+\infty} \int\limits_0^{s_1} \left[\mathrm{i} \mathcal{S}(s_1),\mathrm{i}\mathcal{S}(s_2) \right]\mathrm{d}s_2 \mathrm{d}s_1+\ldots,
\label{eq:mag}
\end{aligned}
\end{equation}
where the generator reads
\begin{equation}
\mathrm{i}\mathcal{S}(s) = \sum\limits_{m \neq 0} \frac{m}{\hbar \omega} P_m \otimes H^{(m)}(s).
\label{eq:gen_form3}
\end{equation}
Note that Eq.~(\ref{eq:gen_form2}) differs from the obtained expression~(\ref{eq:gen_form3}) because similarly to Eq.~(\ref{eq:fw_znz}) 
here we use the rescaled flow variable, $s \rightarrow s/(\hbar\omega)^2$. Since the expansion of the Fourier components 
$H^{(m)}(s) = H^{(m)}_0(s)+H^{(m)}_1(s)+H^{(m)}_2(s)+\ldots$ was find analytically in Appendix~\ref{sec:App-B} one can 
derive the analytical expressions for $\mathcal{S}_0$, $\mathcal{S}_1$ and $\mathcal{S}_2$. 
Let us substitute Eq.~(\ref{eq:gen_form3}) into the right hand side of Eq.~(\ref{eq:mag})
\begin{equation}
\begin{aligned}
& \sum\limits_{m \neq 0} \left( \frac{m}{\hbar \omega} P_m \otimes \int\limits_0^{+\infty} H^{(m)}_0(s_1) \mathrm{d}s_1 \right) \\
&+ \sum\limits_{m \neq 0} \left( \frac{m}{\hbar \omega} P_m \otimes \int\limits_0^{+\infty} H^{(m)}_1(s_1) \mathrm{d}s_1 \right)\\
&+ \sum\limits_{m \neq 0} \sum\limits_{n \neq 0} \left( \frac{mn}{2(\hbar\omega)^2} P_{m+n} \right.\\
&\left.\otimes \int\limits_0^{+\infty} \int\limits_0^{s_1} \left[H^{(m)}_0(s_1),H^{(n)}_0(s_2) \right]\mathrm{d}s_2 \mathrm{d}s_1 \right)+\ldots.
\label{eq:mag1}
\end{aligned}
\end{equation}
This shows that $\mathcal{S}_0=0$. Using Eq.~(\ref{eq:fw_hfsol_0}) we get
\begin{equation}
\mathcal{S}_1 = \frac{1}{\mathrm{i}\hbar \omega} \sum\limits_{m \neq 0} \frac{P_m}{m} \otimes h^{(m)}.
\label{eq:ss1}
\end{equation}
In order to obtain $\mathcal{S}_2$, let us first calculate the double commutator
\begin{equation}
\begin{aligned}
&\int\limits_0^{+\infty} \int\limits_0^{s_1} \left[H^{(m)}_0(s_1),H^{(n)}_0(s_2) \right]\mathrm{d}s_2 \mathrm{d}s_1= \\
&\left[h^{(m)},h^{(n)}\right] \int\limits_0^{+\infty} \mathrm{e}^{-m^2 s_1} \int\limits_0^{s_1} \mathrm{e}^{-n^2s_2}\mathrm{d}s_2 \mathrm{d}s_1 =\\
&\left[h^{(m)},h^{(n)}\right] \left(\frac{1}{m^2 n^2}-\frac{1}{n^2\left(m^2+n^2\right)} \right).
\label{eq:ss2}
\end{aligned}
\end{equation}
Next, by using Eq.~(\ref{eq:fw_hfsol_1}), let us evaluate the integral
\begin{equation}
\begin{aligned}
& \int\limits_0^{+\infty} H^{(m)}_1(s_1) \mathrm{d}s_1 = \frac{\mathrm{i}}{\omega m^3}  \dot{h}^{(m)} - \frac{\left[ h^{(0)},h^{(m)}\right]}{m^3\hbar \omega}\\
&+\sum\limits_{n \neq \{0,m\}} \frac{\left[ h^{(n)},h^{(m-n)}\right]}{2n\hbar \omega} \left( \frac{1}{m^2}-\frac{1}{n^2+(m-n)^2} \right).
\label{eq:ss3}
\end{aligned}
\end{equation}
Now one can collect the second and the third terms of Eq.~(\ref{eq:mag1})
\begin{equation}
\begin{aligned}
& \mathcal{S}_2=\frac{1}{2\mathrm{i}(\hbar\omega)^2} \sum\limits_{m \neq 0} P_m \otimes \left[ 2\frac{\mathrm{i}\hbar \dot{h}^{(m)}}{m^2} + 2\frac{\left[ h^{(m)},h^{(0)}\right]}{m^2} \right. \\
&\left. +\sum\limits_{n \neq \{0,m\}}\left[ h^{(n)},h^{(m-n)}\right] \left( \frac{1}{nm}-\frac{1}{(m-n)^2+n^2} \right) \right] \\
&+\frac{\mathrm{i}}{4(\hbar\omega)^2} P_0 \sum\limits_{m\neq 0} \frac{\left[ h^{(m)},h^{(-m)}\right]}{m^2}.
\label{eq:ss4}
\end{aligned}
\end{equation}
The obtained result can be simplified. The term proportional to $P_0$ is zero because any positive-$m$ term
will be compensated by a corresponding negative-$m$ term. Another simplification results from the fact that
\begin{equation}
A=\sum\limits_{m \neq 0}\sum\limits_{n \neq \{0,m\}} P_m \otimes \frac{\left[ h^{(n)},h^{(m-n)}\right]}{(m-n)^2+n^2}
\label{eq:aa}
\end{equation}
is equal to zero. Indeed, by applying the transformation of the second variable $n=m-n^{\prime}$ we get
\begin{equation}
A=-\sum\limits_{m \neq 0}\sum\limits_{n^{\prime} \neq \{0,m\}} P_m \otimes \frac{\left[ h^{(n^{\prime})},h^{(m-n^{\prime})}\right]}{(m-n^{\prime})^2+n^{\prime 2}},
\label{eq:aa1}
\end{equation}
which shows that $A=-A$, thus $A=0$. Therefore Eq.~(\ref{eq:ss4}) is simplified to
\begin{equation}
\begin{aligned}
& \mathcal{S}_2=\frac{1}{2\mathrm{i}(\hbar\omega)^2} \sum\limits_{m \neq 0} P_m \otimes \left[ 2\frac{\mathrm{i}\hbar}{m^2}\dot{h}^{(m)} + \frac{\left[ h^{(m)},h^{(0)}\right]}{m^2} \right. \\
&\left. +\sum\limits_{n \neq 0} \frac{\left[ h^{(n)},h^{(m-n)}\right]}{nm}\right].
\label{eq:ss5}
\end{aligned}
\end{equation}
Comparing~(\ref{eq:ss1}) and (\ref{eq:ss5}) with the micromotion operator obtained in Ref.~\cite{Novicenko2017} 
we conclude that they coincide.

%
\section{Proof that two quantities (\ref{eq:A}) and (\ref{eq:B}) are equal}
\label{sec:App-D}

Let us verify that the quantities (\ref{eq:A}) and (\ref{eq:B}) are equal. Expression (\ref{eq:A}) reads
\begin{equation}
\label{ea:1}
  A = \sum_{n \neq 0} \sum_{m \neq \{0,n \}}
  \left[ h^{(-n)}, \left[ h^{(n-m)}, h^{(m)} \right] \right] f(n,m),
\end{equation}
whereas expression (\ref{eq:B}) is of the same form but in place of $f(n,m)$ features a different function $g (n,m)$.
Such an ambiguity arises because the summation over the specified values $n$ and $m$ contributes the same commutators
from different terms. For example, a simple inspection shows that, already for $n = 1$ terms with $ m = -1$ and $m = 2$ 
both contribute to the same commutator $\left[h^{(-1)}, \left[h^{(2)}, h^{(-1)} \right] \right]$. In fact, the same 
commutator is produced two more times from the remaining terms of the double sum.

To rectify the situation, let us now show that Eq.~(\ref{ea:1}) can be written in an unambiguous way in terms of commutators
of the form $\left[h^{(-j)}, \left[h^{(j+\ell)}, h^{(-\ell)} \right] \right]$ and 
$\left[h^{(j)}, \left[h^{(-j-\ell)}, h^{(j)} \right] \right]$ with positive $j$ and $\ell$. In other words, all 
commutators are thus cast into a `standard' form where a positive-indexed Fourier component is flanked by two negatived-indexed
ones, or vice versa. Note that the Fourier indices of all three components participating in a triple commutator sum to zero,
and zero indices are excluded. Thus one always has two negative indices and a positive one or the other way around. As we
will show shortly, the desired transformation is achieved by: (i) exchanging the order of the two Fourier components in the
inner commutator, and (ii) applying the Jacobi identity.

Let us start by observing that the summation
\begin{equation*}
  \sum_{n \neq 0} \sum_{m \neq \{0,n \}}
\end{equation*}
covers six sectors:
\begin{alignat*}{2}
 &(1)\quad m < 0 < n, \qquad &&(4)\quad m < n < 0,\\
 &(2)\quad 0 < m < n, \qquad &&(5)\quad n < m < 0,\\
 &(3)\quad 0 < n < m, \qquad &&(6)\quad n < 0 < m.
\end{alignat*}
Listed in the left (right) column are the three cases that correspond to the three possible ways 
to locate $m$ relative to zero and positive (negative) $n$.

Case (1) is immediately in the form that we want. We relabel $n = j$ and $m = -\ell$ and write
\begin{equation}
\begin{split}
  &\sum_{n > 0} \sum_{m < 0} \left[ h^{(-n)}, \left[ h^{(n-m)}, h^{(m)} \right] \right] f(n,m) \\
  &\quad= \sum_{j > 0} \sum_{\ell > 0} \left[ h^{(-j)}, \left[ h^{(j+\ell)}, h^{(-\ell)} \right] \right] f(j,-\ell).
\end{split}
\end{equation}
Case (3) is one that requires the inversion of the order of operators in the inner commutator. We relabel
$n = j$ and $m = j + \ell$ and obtain
\begin{equation}
\begin{split}
  &\sum_{n > 0} \sum_{m > n} \left[ h^{(-n)}, \left[ h^{(n-m)}, h^{(m)} \right] \right] f(n,m) \\
  &\quad= - \sum_{j > 0} \sum_{\ell > 0} \left[ h^{(-j)}, \left[ h^{(j+\ell)}, h^{(-\ell)} \right] \right] f(j,j+\ell).
\end{split}
\end{equation}
Finally, case (5) is one that requires the application of the Jacobi identity and subsequent inversion of the inner
commutator in one of the two resulting terms. Carrying out the algebra and introducing positive indices $j$ and $\ell$
as above, we find
\begin{equation}
\begin{split}
  &\sum_{m < 0} \sum_{n < m} \left[ h^{(-n)}, \left[ h^{(n-m)}, h^{(m)} \right] \right] f(n,m) \\
  &\quad= \sum_{j > 0} \sum_{\ell > 0} \left[ h^{(-j)}, \left[ h^{(j+\ell)}, h^{(-\ell)} \right] \right] f(-j-\ell,-\ell) \\
  &\quad- \sum_{j > 0} \sum_{\ell > 0} \left[ h^{(-j)}, \left[ h^{(j+\ell)}, h^{(-\ell)} \right] \right] f(-j-\ell,-j).
\end{split}
\end{equation}
The considered cases (1), (3), and (5) cover the terms that lead to a positive indexed Fourier component 
flanked by two negative indexed ones. All in all, they sum up to the final result
\begin{equation}
\begin{split}
  &\sum_{m < 0} \sum_{n < m} \left[ h^{(-n)}, \left[ h^{(n-m)}, h^{(m)} \right] \right] f(n,m) \\
  &\quad= \sum_{j > 0} \sum_{\ell > 0} \left[ h^{(-j)}, \left[ h^{(j+\ell)}, h^{(-\ell)} \right] \right] \tilde{f}(j,\ell),
\end{split}
\end{equation}
with
\begin{equation}
\begin{split}
 \tilde{f} (j, \ell) &= f (j, -\ell) - f (j, j + \ell) \\
 &+ f (-j-\ell, -\ell) - f (-j-\ell, -j).
\end{split}
\end{equation}

Using
\begin{equation}
  f (n,m) = \frac{m-n}{n \left[ n^2 + m^2 + (m-n)^2 \right]},
\end{equation}
we find
\begin{equation}
  \tilde{f} (j, \ell) = \frac{-1}{j (j + \ell)},
\end{equation}
and using
\begin{equation}
  g (n,m) = \frac{1}{3nm},
\end{equation}
we find 
\begin{equation}
  \tilde{g} (j, \ell) = \frac{-1}{j (j + \ell)}.
\end{equation}
The consideration of cases (2), (4), and (6) is identical in spirit and restructures the complementary
case where a single negative-indexed Fourier component is flanked by to positive-indexed ones. 

%
\section{The Toda flow equations for the case of a limited number of harmonics}
\label{sec:App-E}

In this section we prove the statement made in the main text: If the initial condition for Eq.~(\ref{eq:fw_nz1})
has a limited number of non-vanishing Fourier harmonics at $s = 0$, i.e.\ $H^{(n)}(0,t) = 0$ for all $|n| > n_0$, 
then these high harmonics (with $|n| > n_0$) may be neglected altogether since they will not be populated 
in the course of the flow. This implies that the flow needs to be analyzed only for the limited spectrum 
with harmonics $|n| \leqslant n_0$. For the notational simplicity, we will not write the explicit dependence 
on time but will keep it in mind.

Let us first rewrite Eq.~(\ref{eq:fw_nz1}) as
\begin{equation}
\begin{aligned}
&\frac{\mathrm{d}H^{(n)}(s)}{\mathrm{d}s} = - n\cdot \mathrm{sgn}(n) H^{(n)}(s)+\frac{\mathrm{i}}{\omega} \mathrm{sgn}(n) \dot{H}^{(n)}(s) \\
&- \frac{\mathrm{sgn}(n)}{\hbar \omega} \left[ H^{(0)}(s),H^{(n)}(s) \right] \\
&+ \sum\limits_{m \neq \{0,n\}} \frac{\mathrm{sgn}(m-n)}{\hbar \omega} \left[ H^{(m)}(s),H^{(n-m)}(s) \right],
\label{eq:fw_nz2}
\end{aligned}
\end{equation}
and focus on the last term of Eq.~(\ref{eq:fw_nz2}). Let us distinguish two cases: when $n$ is odd and when $n$ is even. 
In both cases, instead of $m$ we introduce new index $l = 2m - n$. For an odd $n$ we have $l = \{\pm 1, \pm 3,\ldots \}$ 
whereas for even $n$ the permissible values of $l$ read $l=\{0, \pm 2, \pm 4,\ldots \}$. From the condition 
$m\neq \{0,n\}$ we obtain $l \neq \pm n$. Thus for the case of an odd value of $n$ the term in question reads
\begin{equation}
  \sum\limits_{l = \{\pm 1, \pm 3,\ldots ,\neq \pm n, \ldots \}} 
  \mathrm{sgn} \left(\frac{l-n}{2}\right) 
  \left[ H^{\left(\frac{n+l}{2}\right)}(s),H^{\left(\frac{n-l}{2}\right)}(s) \right] \\
\label{eq:fw_nz3}
\end{equation}
The denominator $2$ in the argument of the sign function $\mathrm{sgn}(\cdot)$ can be omitted. We split Eq.~(\ref{eq:fw_nz3}) into 
two terms separating positive and negative values of $l$, and introduce the replacement $l^{\prime} = -l$ for the negative $l$'s 
with the result
\begin{equation}
\begin{aligned}
&\sum\limits_{l^{\prime} = \{1, 3,\ldots ,\neq |n|, \ldots \}} \mathrm{sgn}(-l^{\prime}-n) \left[ H^{\left(\frac{n-l^{\prime}}{2}\right)}(s),H^{\left(\frac{n+l^{\prime}}{2}\right)}(s) \right]= \\
&\sum\limits_{l^{\prime} = \{1, 3,\ldots ,\neq |n|, \ldots \}} \mathrm{sgn}(l^{\prime}+n) \left[ H^{\left(\frac{n+l^{\prime}}{2}\right)}(s),H^{\left(\frac{n-l^{\prime}}{2}\right)}(s) \right].
\label{eq:fw_nz4}
\end{aligned}
\end{equation}
We now sum the obtained expression with its counterpart for positive values of $l$ and obtain
\begin{equation}
\begin{aligned}
&\sum\limits_{l = \{1, 3,\ldots ,\neq |n|, \ldots \}} \left\lbrace \mathrm{sgn}(n+l) -\mathrm{sgn}(n-l)\right\rbrace \\
& \times  \left[ H^{\left(\frac{n+l}{2}\right)}(s),H^{\left(\frac{n-l}{2}\right)}(s) \right].
\label{eq:fw_nz5}
\end{aligned}
\end{equation}
Let us now consider the combination of sign functions in the curly braces. For $l < |n|$ the expression in the braces yield zero: 
If $n$ is positive, then $(n+l)>0$ and $(n-l)>0$, and if $n$ is negative then $(n+l) < 0$ and $(n-l) < 0$. On the other hand, 
for $l > |n|$ the curved brackets equal to $2$. Thus, the expression in Eq.~(\ref{eq:fw_nz5}) can be simplified to
\begin{equation}
2\sum\limits_{l = \{|n|+2, |n|+4, \ldots \}} \left[ H^{\left(\frac{n+l}{2}\right)}(s),H^{\left(\frac{n-l}{2}\right)}(s) \right].
\label{eq:fw_nz6}
\end{equation}
By performing a similar procedure that lead from Eq.~(\ref{eq:fw_nz3}) to Eq.~(\ref{eq:fw_nz6}) for the case of $n = \mathrm{even}$, one arrives 
to the same expression~(\ref{eq:fw_nz6}). Therefore, Eq.~(\ref{eq:fw_nz2}) transforms to
\begin{equation}
\begin{aligned}
&\frac{\mathrm{d}H^{(n)}(s)}{\mathrm{d}s} = - n\cdot \mathrm{sgn}(n) H^{(n)}(s)+\frac{\mathrm{i}}{\omega} \mathrm{sgn}(n) \dot{H}^{(n)}(s) \\
&- \frac{\mathrm{sgn}(n)}{\hbar \omega} \left[ H^{(0)}(s),H^{(n)}(s) \right] \\
&+ \frac{2}{\hbar \omega} \sum\limits_{l = \{|n|+2, |n|+4, \ldots \}}  \left[ H^{\left(\frac{n+l}{2}\right)}(s),H^{\left(\frac{n-l}{2}\right)}(s) \right].
\label{eq:fw_nz7}
\end{aligned}
\end{equation}
From Eq.~(\ref{eq:fw_nz7}) one can see: if $n >|n_0|$, even with smallest possible $l = |n|+2$ one of the indices of the Fourier harmonics 
$(n+l)/2 = (n+|n|)/2+1$ and $(n-l)/2=(n-|n|)/2-1$ will have an absolute value higher than $n_0$, and thus the commutator will vanish. 
All the other terms on the right hand side of Eq.~(\ref{eq:fw_nz7}) also vanish for $n > |n_0|$.

For harmonics with small indices $n\leq |n_0|$, the index $l$ in Eq.~(\ref{eq:fw_nz7}) can be restricted to $l=\{|n|+2, |n|+4, \ldots, |n|+2(n_0-|n|) \}$. 
Finally, Eq.~(\ref{eq:fw_nz7}) reads [here we rename the index $l \to (l-|n|)/2$]:
\begin{widetext}
\begin{equation}
\frac{\mathrm{d}H^{(n)}(s)}{\mathrm{d}s} = 
\begin{cases}
- n H^{(n)}(s)+\frac{\mathrm{i}}{\omega}\dot{H}^{(n)}(s)+ \frac{\left[ H^{(n)}(s),H^{(0)}(s) \right]}{\hbar \omega} + \frac{2}{\hbar \omega} \sum\limits_{l = 1}^{n_0-n}  \left[ H^{(n+l)}(s),H^{(-l)}(s) \right]  & \mathrm{for}\; 0<n\leq n_0 \\
 n H^{(n)}(s)-\frac{\mathrm{i}}{\omega}\dot{H}^{(n)}(s)+ \frac{\left[ H^{(0)}(s),H^{(n)}(s) \right]}{\hbar \omega} + \frac{2}{\hbar \omega} \sum\limits_{l = 1}^{n_0+n}  \left[ H^{(l)}(s),H^{(n-l)}(s) \right] & \mathrm{for}\;  -n_0 \leq n<0 \\
 0 & \mathrm{for}\; |n|> n_0
\end{cases}
\label{eq:fw_nz8}
\end{equation}
The flow equation for the zeroth Fourier harmonic reads
\begin{equation}
\frac{\mathrm{d}H^{(0)}(s)}{\mathrm{d}s}=\frac{2}{\hbar \omega} \sum\limits_{m=1}^{n_0} \left[ H^{(m)}(s),H^{(-m)}(s) \right].
\label{eq:fw_z9}
\end{equation}
%

%
\section{Automated expansion of the FE Hamiltonian and micromotion operator for the Rabi linearly polarized drive}
\label{sec:App-F}

We perform the automated solution of the flow equations~(\ref{eq:fw_znz1}) with the initial conditions~(\ref{eq:fc}).
For $\phi = 0$, the effective Hamiltonian reads ($\hbar = 1$)
\begin{equation}
\begin{split}
  h_{\eff} &= \sigma_x \cdot \left[ g(t) - \frac{g(t)^3}{4 \omega ^2} +\frac{3\Delta  g(t)^3}{16 \omega ^3}       
    + \frac{-6 g(t) g'(t)^2+7 g(t)^2 g''(t)-7 \Delta ^2 g(t)^3-8 g(t)^5}{64 \omega ^4} \right] \\
  &+ \sigma_y \left[ -\frac{g(t)^2 g'(t)}{16 \omega ^3}  +\frac{\Delta g(t)^2 g'(t)}{16 \omega ^4} \right] \\
  &+ \sigma_z \left[\frac{\Delta}{2} + \frac{g(t)^2 }{2 \omega } + -\frac{\Delta  g(t)^2}{4 \omega ^2}
        \frac{g'(t)^2-g(t) g''(t) + 2 \Delta ^2 g(t)^2}{16 \omega ^3} 
        +\frac{-3 \Delta  g'(t)^2+3 \Delta  g(t) g''(t)-2 \Delta^3 g(t)^2+\Delta  g(t)^4}{32 \omega ^4}
  \right].
\end{split}
\end{equation}
Higher-order contributions were obtained (also in the general case $\phi \neq 0$) but are not suitable 
for reproduction in a journal page. The \textsc{mathematica} script used to generate expansions up to the
fifth order is available at \cite{script}.




\end{widetext}

\bibliography{todarefs}

\begin{thebibliography}{72}%
\makeatletter
\providecommand \@ifxundefined [1]{%
 \@ifx{#1\undefined}
}%
\providecommand \@ifnum [1]{%
 \ifnum #1\expandafter \@firstoftwo
 \else \expandafter \@secondoftwo
 \fi
}%
\providecommand \@ifx [1]{%
 \ifx #1\expandafter \@firstoftwo
 \else \expandafter \@secondoftwo
 \fi
}%
\providecommand \natexlab [1]{#1}%
\providecommand \enquote  [1]{``#1''}%
\providecommand \bibnamefont  [1]{#1}%
\providecommand \bibfnamefont [1]{#1}%
\providecommand \citenamefont [1]{#1}%
\providecommand \href@noop [0]{\@secondoftwo}%
\providecommand \href [0]{\begingroup \@sanitize@url \@href}%
\providecommand \@href[1]{\@@startlink{#1}\@@href}%
\providecommand \@@href[1]{\endgroup#1\@@endlink}%
\providecommand \@sanitize@url [0]{\catcode `\\12\catcode `\$12\catcode
  `\&12\catcode `\#12\catcode `\^12\catcode `\_12\catcode `\%12\relax}%
\providecommand \@@startlink[1]{}%
\providecommand \@@endlink[0]{}%
\providecommand \url  [0]{\begingroup\@sanitize@url \@url }%
\providecommand \@url [1]{\endgroup\@href {#1}{\urlprefix }}%
\providecommand \urlprefix  [0]{URL }%
\providecommand \Eprint [0]{\href }%
\providecommand \doibase [0]{https://doi.org/}%
\providecommand \selectlanguage [0]{\@gobble}%
\providecommand \bibinfo  [0]{\@secondoftwo}%
\providecommand \bibfield  [0]{\@secondoftwo}%
\providecommand \translation [1]{[#1]}%
\providecommand \BibitemOpen [0]{}%
\providecommand \bibitemStop [0]{}%
\providecommand \bibitemNoStop [0]{.\EOS\space}%
\providecommand \EOS [0]{\spacefactor3000\relax}%
\providecommand \BibitemShut  [1]{\csname bibitem#1\endcsname}%
\let\auto@bib@innerbib\@empty
\bibitem [{\citenamefont {H{\" a}nggi}(1998)}]{Hanggi1998}%
  \BibitemOpen
  \bibfield  {author} {\bibinfo {author} {\bibfnamefont {P.}~\bibnamefont {H{\"
  a}nggi}},\ }\bibfield  {title} {\bibinfo {title} {Driven quantum systems},\
  }in\ \href@noop {} {\emph {\bibinfo {booktitle} {Quantum transport and
  dissipation}}},\ \bibinfo {editor} {edited by\ \bibinfo {editor}
  {\bibfnamefont {T.}~\bibnamefont {Dittrich}}, \bibinfo {editor}
  {\bibfnamefont {P.}~\bibnamefont {H{\"a}nggi}}, \bibinfo {editor}
  {\bibfnamefont {G.-L.}\ \bibnamefont {Ingold}}, \bibinfo {editor}
  {\bibfnamefont {B.}~\bibnamefont {Kramer}}, \bibinfo {editor} {\bibfnamefont
  {G.}~\bibnamefont {Sch{\"o}n}},\ and\ \bibinfo {editor} {\bibfnamefont
  {W.}~\bibnamefont {Zwerger}}}\ (\bibinfo  {publisher} {Wiley-VCH},\ \bibinfo
  {address} {Weinheim},\ \bibinfo {year} {1998})\ Chap.~\bibinfo {chapter} {5},
  pp.\ \bibinfo {pages} {249--286}\BibitemShut {NoStop}%
\bibitem [{\citenamefont {Holthaus}(2016)}]{Holthaus2016tutorial}%
  \BibitemOpen
  \bibfield  {author} {\bibinfo {author} {\bibfnamefont {M.}~\bibnamefont
  {Holthaus}},\ }\bibfield  {title} {\bibinfo {title} {Floquet engineering with
  quasienergy bands of periodically driven optical lattices},\ }\href
  {https://doi.org/10.1088/0953-4075/49/1/013001} {\bibfield  {journal}
  {\bibinfo  {journal} {J. Phys. B: At. Mol. Opt. Phys.}\ }\textbf {\bibinfo
  {volume} {49}},\ \bibinfo {pages} {013001} (\bibinfo {year} {2016})},\
  \Eprint {https://arxiv.org/abs/1510.09042} {arXiv:1510.09042 [quant-ph]}
  \BibitemShut {NoStop}%
\bibitem [{\citenamefont {Eckardt}(2017)}]{Eckardt2017review}%
  \BibitemOpen
  \bibfield  {author} {\bibinfo {author} {\bibfnamefont {A.}~\bibnamefont
  {Eckardt}},\ }\bibfield  {title} {\bibinfo {title} {Colloquium: {A}tomic
  quantum gases in periodically driven optical lattices},\ }\href
  {https://doi.org/10.1103/RevModPhys.89.011004} {\bibfield  {journal}
  {\bibinfo  {journal} {Rev. Mod. Phys.}\ }\textbf {\bibinfo {volume} {89}},\
  \bibinfo {pages} {011004} (\bibinfo {year} {2017})},\ \Eprint
  {https://arxiv.org/abs/1606.08041} {arXiv:1606.08041 [cond-mat.quant-gas]}
  \BibitemShut {NoStop}%
\bibitem [{\citenamefont {Howland}(1974)}]{Howland1974}%
  \BibitemOpen
  \bibfield  {author} {\bibinfo {author} {\bibfnamefont {J.~S.}\ \bibnamefont
  {Howland}},\ }\bibfield  {title} {\bibinfo {title} {Stationary scattering
  theory for time-dependent {H}amiltonians},\ }\href
  {https://doi.org/10.1007/BF01351346} {\bibfield  {journal} {\bibinfo
  {journal} {Math. Ann.}\ }\textbf {\bibinfo {volume} {207}},\ \bibinfo {pages}
  {315} (\bibinfo {year} {1974})}\BibitemShut {NoStop}%
\bibitem [{\citenamefont {Peskin}\ and\ \citenamefont
  {Moiseyev}(1993)}]{Peskin1993}%
  \BibitemOpen
  \bibfield  {author} {\bibinfo {author} {\bibfnamefont {U.}~\bibnamefont
  {Peskin}}\ and\ \bibinfo {author} {\bibfnamefont {N.}~\bibnamefont
  {Moiseyev}},\ }\bibfield  {title} {\bibinfo {title} {The solution of the
  time‐dependent {S}chr{\" o}dinger equation by the (t,t’) method:
  {T}heory, computational algorithm and applications},\ }\href
  {https://doi.org/10.1063/1.466058} {\bibfield  {journal} {\bibinfo  {journal}
  {J. Chem. Phys.}\ }\textbf {\bibinfo {volume} {99}},\ \bibinfo {pages} {4590}
  (\bibinfo {year} {1993})}\BibitemShut {NoStop}%
\bibitem [{\citenamefont {Giscard}\ and\ \citenamefont
  {Bonhomme}(2020)}]{Giscard2020}%
  \BibitemOpen
  \bibfield  {author} {\bibinfo {author} {\bibfnamefont {P.-L.}\ \bibnamefont
  {Giscard}}\ and\ \bibinfo {author} {\bibfnamefont {C.}~\bibnamefont
  {Bonhomme}},\ }\bibfield  {title} {\bibinfo {title} {Dynamics of quantum
  systems driven by time-varying {H}amiltonians: {S}olution for the
  {B}loch-{S}iegert {H}amiltonian and applications to {NMR}},\ }\href
  {https://doi.org/10.1103/PhysRevResearch.2.023081} {\bibfield  {journal}
  {\bibinfo  {journal} {Phys. Rev. Research}\ }\textbf {\bibinfo {volume}
  {2}},\ \bibinfo {pages} {023081} (\bibinfo {year} {2020})},\ \Eprint
  {https://arxiv.org/abs/1905.04024} {arXiv:1905.04024 [quant-ph]} \BibitemShut
  {NoStop}%
\bibitem [{\citenamefont {Shirley}(1965)}]{Shirley1965}%
  \BibitemOpen
  \bibfield  {author} {\bibinfo {author} {\bibfnamefont {J.~H.}\ \bibnamefont
  {Shirley}},\ }\bibfield  {title} {\bibinfo {title} {Solution of the
  {S}chr{\"o}dinger equation with a {H}amiltonian periodic in time},\ }\href
  {https://doi.org/10.1103/PhysRev.138.B979} {\bibfield  {journal} {\bibinfo
  {journal} {Phys. Rev.}\ }\textbf {\bibinfo {volume} {138}},\ \bibinfo {pages}
  {B979} (\bibinfo {year} {1965})}\BibitemShut {NoStop}%
\bibitem [{\citenamefont {Sambe}(1973)}]{Sambe1973}%
  \BibitemOpen
  \bibfield  {author} {\bibinfo {author} {\bibfnamefont {H.}~\bibnamefont
  {Sambe}},\ }\bibfield  {title} {\bibinfo {title} {Steady states and
  quasienergies of a quantum-mechanical system in an oscillating field},\
  }\href {https://doi.org/10.1103/PhysRevA.7.2203} {\bibfield  {journal}
  {\bibinfo  {journal} {Phys. Rev. A}\ }\textbf {\bibinfo {volume} {7}},\
  \bibinfo {pages} {2203} (\bibinfo {year} {1973})}\BibitemShut {NoStop}%
\bibitem [{\citenamefont {Casas}\ \emph {et~al.}(2001)\citenamefont {Casas},
  \citenamefont {Oteo},\ and\ \citenamefont {Ros}}]{Casas2001}%
  \BibitemOpen
  \bibfield  {author} {\bibinfo {author} {\bibfnamefont {F.}~\bibnamefont
  {Casas}}, \bibinfo {author} {\bibfnamefont {J.~A.}\ \bibnamefont {Oteo}},\
  and\ \bibinfo {author} {\bibfnamefont {J.}~\bibnamefont {Ros}},\ }\bibfield
  {title} {\bibinfo {title} {{F}loquet theory: exponential perturbative
  treatment},\ }\href {https://doi.org/10.1088/0305-4470/34/16/305} {\bibfield
  {journal} {\bibinfo  {journal} {J. Phys. A: Math. Gen.}\ }\textbf {\bibinfo
  {volume} {34}},\ \bibinfo {pages} {3379} (\bibinfo {year}
  {2001})}\BibitemShut {NoStop}%
\bibitem [{\citenamefont {Oka}\ and\ \citenamefont {Aoki}(2009)}]{Oka2009}%
  \BibitemOpen
  \bibfield  {author} {\bibinfo {author} {\bibfnamefont {T.}~\bibnamefont
  {Oka}}\ and\ \bibinfo {author} {\bibfnamefont {H.}~\bibnamefont {Aoki}},\
  }\bibfield  {title} {\bibinfo {title} {Photovoltaic {H}all effect in
  graphene},\ }\href {https://doi.org/10.1103/PhysRevB.79.081406} {\bibfield
  {journal} {\bibinfo  {journal} {Phys. Rev. B}\ }\textbf {\bibinfo {volume}
  {79}},\ \bibinfo {pages} {081406} (\bibinfo {year} {2009})},\ \Eprint
  {https://arxiv.org/abs/0807.4767} {arXiv:0807.4767} \BibitemShut {NoStop}%
\bibitem [{\citenamefont {Lindner}\ \emph {et~al.}(2011)\citenamefont
  {Lindner}, \citenamefont {Refael},\ and\ \citenamefont
  {Galitski}}]{Lindner2011}%
  \BibitemOpen
  \bibfield  {author} {\bibinfo {author} {\bibfnamefont {N.~H.}\ \bibnamefont
  {Lindner}}, \bibinfo {author} {\bibfnamefont {G.}~\bibnamefont {Refael}},\
  and\ \bibinfo {author} {\bibfnamefont {V.}~\bibnamefont {Galitski}},\
  }\bibfield  {title} {\bibinfo {title} {Floquet topological insulator in
  semiconductor quantum wells},\ }\href {https://doi.org/10.1038/nphys1926}
  {\bibfield  {journal} {\bibinfo  {journal} {Nat. Phys.}\ }\textbf {\bibinfo
  {volume} {7}},\ \bibinfo {pages} {490} (\bibinfo {year} {2011})},\ \Eprint
  {https://arxiv.org/abs/1008.1792} {arXiv:1008.1792} \BibitemShut {NoStop}%
\bibitem [{\citenamefont {Gu}\ \emph {et~al.}(2011)\citenamefont {Gu},
  \citenamefont {Fertig}, \citenamefont {Arovas},\ and\ \citenamefont
  {Auerbach}}]{Gu2011}%
  \BibitemOpen
  \bibfield  {author} {\bibinfo {author} {\bibfnamefont {Z.}~\bibnamefont
  {Gu}}, \bibinfo {author} {\bibfnamefont {H.~A.}\ \bibnamefont {Fertig}},
  \bibinfo {author} {\bibfnamefont {D.~P.}\ \bibnamefont {Arovas}},\ and\
  \bibinfo {author} {\bibfnamefont {A.}~\bibnamefont {Auerbach}},\ }\bibfield
  {title} {\bibinfo {title} {Floquet spectrum and transport through an
  irradiated graphene ribbon},\ }\href
  {https://doi.org/10.1103/PhysRevLett.107.216601} {\bibfield  {journal}
  {\bibinfo  {journal} {Phys. Rev. Lett.}\ }\textbf {\bibinfo {volume} {107}},\
  \bibinfo {pages} {216601} (\bibinfo {year} {2011})},\ \Eprint
  {https://arxiv.org/abs/1106.0302} {arXiv:1106.0302} \BibitemShut {NoStop}%
\bibitem [{\citenamefont {Kitagawa}\ \emph {et~al.}(2011)\citenamefont
  {Kitagawa}, \citenamefont {Oka}, \citenamefont {Brataas}, \citenamefont
  {Fu},\ and\ \citenamefont {Demler}}]{Kitagawa2011}%
  \BibitemOpen
  \bibfield  {author} {\bibinfo {author} {\bibfnamefont {T.}~\bibnamefont
  {Kitagawa}}, \bibinfo {author} {\bibfnamefont {T.}~\bibnamefont {Oka}},
  \bibinfo {author} {\bibfnamefont {A.}~\bibnamefont {Brataas}}, \bibinfo
  {author} {\bibfnamefont {L.}~\bibnamefont {Fu}},\ and\ \bibinfo {author}
  {\bibfnamefont {E.}~\bibnamefont {Demler}},\ }\bibfield  {title} {\bibinfo
  {title} {Transport properties of nonequilibrium systems under the application
  of light: {P}hotoinduced quantum {H}all insulators without {L}andau levels},\
  }\href {https://doi.org/10.1103/PhysRevB.84.235108} {\bibfield  {journal}
  {\bibinfo  {journal} {Phys. Rev. B}\ }\textbf {\bibinfo {volume} {84}},\
  \bibinfo {pages} {235108} (\bibinfo {year} {2011})},\ \Eprint
  {https://arxiv.org/abs/1104.4636} {arXiv:1104.4636} \BibitemShut {NoStop}%
\bibitem [{\citenamefont {Bukov}\ \emph {et~al.}(2015)\citenamefont {Bukov},
  \citenamefont {D'Alessio},\ and\ \citenamefont {Polkovnikov}}]{Bukov2015}%
  \BibitemOpen
  \bibfield  {author} {\bibinfo {author} {\bibfnamefont {M.}~\bibnamefont
  {Bukov}}, \bibinfo {author} {\bibfnamefont {L.}~\bibnamefont {D'Alessio}},\
  and\ \bibinfo {author} {\bibfnamefont {A.}~\bibnamefont {Polkovnikov}},\
  }\bibfield  {title} {\bibinfo {title} {Universal high-frequency behavior of
  periodically driven systems: from dynamical stabilization to {F}loquet
  engineering},\ }\href {https://doi.org/10.1080/00018732.2015.1055918}
  {\bibfield  {journal} {\bibinfo  {journal} {Adv. Phys.}\ }\textbf {\bibinfo
  {volume} {64}},\ \bibinfo {pages} {139} (\bibinfo {year} {2015})},\ \Eprint
  {https://arxiv.org/abs/1407.4803} {arXiv:1407.4803 [cond-mat.quantum-gas]}
  \BibitemShut {NoStop}%
\bibitem [{\citenamefont {Oka}\ and\ \citenamefont
  {Kitamura}(2019)}]{Oka2019Review}%
  \BibitemOpen
  \bibfield  {author} {\bibinfo {author} {\bibfnamefont {T.}~\bibnamefont
  {Oka}}\ and\ \bibinfo {author} {\bibfnamefont {S.}~\bibnamefont {Kitamura}},\
  }\bibfield  {title} {\bibinfo {title} {Floquet engineering of quantum
  materials},\ }\href
  {https://doi.org/10.1146/annurev-conmatphys-031218-013423} {\bibfield
  {journal} {\bibinfo  {journal} {Annu. Rev. Condens. Matter Phys.}\ }\textbf
  {\bibinfo {volume} {10}},\ \bibinfo {pages} {387} (\bibinfo {year} {2019})},\
  \Eprint {https://arxiv.org/abs/1804.03212} {arXiv:1804.03212 [cond-mat]}
  \BibitemShut {NoStop}%
\bibitem [{\citenamefont {Rudner}\ and\ \citenamefont
  {Lindner}(2020)}]{Rudner2020}%
  \BibitemOpen
  \bibfield  {author} {\bibinfo {author} {\bibfnamefont {M.~S.}\ \bibnamefont
  {Rudner}}\ and\ \bibinfo {author} {\bibfnamefont {N.~H.}\ \bibnamefont
  {Lindner}},\ }\bibfield  {title} {\bibinfo {title} {Band structure
  engineering and non-equilibrium dynamics in floquet topological insulators},\
  }\href {https://doi.org/10.1038/s42254-020-0170-z} {\bibfield  {journal}
  {\bibinfo  {journal} {Nat. Rev. Phys.}\ }\textbf {\bibinfo {volume} {2}},\
  \bibinfo {pages} {229} (\bibinfo {year} {2020})}\BibitemShut {NoStop}%
\bibitem [{\citenamefont {Bajpai}\ \emph {et~al.}(2020)\citenamefont {Bajpai},
  \citenamefont {Ku},\ and\ \citenamefont {Nikoli\'c}}]{Bajpai2020}%
  \BibitemOpen
  \bibfield  {author} {\bibinfo {author} {\bibfnamefont {U.}~\bibnamefont
  {Bajpai}}, \bibinfo {author} {\bibfnamefont {M.~J.~H.}\ \bibnamefont {Ku}},\
  and\ \bibinfo {author} {\bibfnamefont {B.~K.}\ \bibnamefont {Nikoli\'c}},\
  }\bibfield  {title} {\bibinfo {title} {Robustness of quantized transport
  through edge states of finite length: {I}maging current density in {F}loquet
  topological versus quantum spin and anomalous {H}all insulators},\ }\href
  {https://doi.org/10.1103/PhysRevResearch.2.033438} {\bibfield  {journal}
  {\bibinfo  {journal} {Phys. Rev. Research}\ }\textbf {\bibinfo {volume}
  {2}},\ \bibinfo {pages} {033438} (\bibinfo {year} {2020})},\ \Eprint
  {https://arxiv.org/abs/2006.16999} {arXiv:2006.16999} \BibitemShut {NoStop}%
\bibitem [{\citenamefont {Rodriguez-Vega}\ \emph {et~al.}(2021)\citenamefont
  {Rodriguez-Vega}, \citenamefont {Vogl},\ and\ \citenamefont
  {Fiete}}]{Rodriguez2021}%
  \BibitemOpen
  \bibfield  {author} {\bibinfo {author} {\bibfnamefont {M.}~\bibnamefont
  {Rodriguez-Vega}}, \bibinfo {author} {\bibfnamefont {M.}~\bibnamefont
  {Vogl}},\ and\ \bibinfo {author} {\bibfnamefont {G.~A.}\ \bibnamefont
  {Fiete}},\ }\bibfield  {title} {\bibinfo {title} {Low-frequency and
  {M}oir{\'e}-{F}loquet engineering: {A} review},\ }\href
  {https://doi.org/10.1016/j.aop.2021.168434} {\bibfield  {journal} {\bibinfo
  {journal} {Annals of Physics}\ }\textbf {\bibinfo {volume} {435}},\ \bibinfo
  {pages} {168434} (\bibinfo {year} {2021})},\ \Eprint
  {https://arxiv.org/abs/2011.11079} {arXiv:2011.11079 [cond-mat.mes-hall]}
  \BibitemShut {NoStop}%
\bibitem [{\citenamefont {Aidelsburger}\ \emph {et~al.}(2011)\citenamefont
  {Aidelsburger}, \citenamefont {Atala}, \citenamefont {Nascimb\`{e}ne},
  \citenamefont {Trotzky}, \citenamefont {Chen},\ and\ \citenamefont
  {Bloch}}]{Aidelsburger2011}%
  \BibitemOpen
  \bibfield  {author} {\bibinfo {author} {\bibfnamefont {M.}~\bibnamefont
  {Aidelsburger}}, \bibinfo {author} {\bibfnamefont {M.}~\bibnamefont {Atala}},
  \bibinfo {author} {\bibfnamefont {S.}~\bibnamefont {Nascimb\`{e}ne}},
  \bibinfo {author} {\bibfnamefont {S.}~\bibnamefont {Trotzky}}, \bibinfo
  {author} {\bibfnamefont {Y.-A.}\ \bibnamefont {Chen}},\ and\ \bibinfo
  {author} {\bibfnamefont {I.}~\bibnamefont {Bloch}},\ }\bibfield  {title}
  {\bibinfo {title} {Experimental realization of strong effective magnetic
  fields in an optical lattice},\ }\href
  {https://doi.org/10.1103/PhysRevLett.107.255301} {\bibfield  {journal}
  {\bibinfo  {journal} {Phys. Rev. Lett.}\ }\textbf {\bibinfo {volume} {107}},\
  \bibinfo {pages} {255301} (\bibinfo {year} {2011})},\ \Eprint
  {https://arxiv.org/abs/1110.5314} {arXiv:1110.5314 [cond-mat.quant-gas]}
  \BibitemShut {NoStop}%
\bibitem [{\citenamefont {Aidelsburger}\ \emph {et~al.}(2013)\citenamefont
  {Aidelsburger}, \citenamefont {Atala}, \citenamefont {Lohse}, \citenamefont
  {Barreiro}, \citenamefont {Paredes},\ and\ \citenamefont
  {Bloch}}]{Aidelsburger2013}%
  \BibitemOpen
  \bibfield  {author} {\bibinfo {author} {\bibfnamefont {M.}~\bibnamefont
  {Aidelsburger}}, \bibinfo {author} {\bibfnamefont {M.}~\bibnamefont {Atala}},
  \bibinfo {author} {\bibfnamefont {M.}~\bibnamefont {Lohse}}, \bibinfo
  {author} {\bibfnamefont {J.~T.}\ \bibnamefont {Barreiro}}, \bibinfo {author}
  {\bibfnamefont {B.}~\bibnamefont {Paredes}},\ and\ \bibinfo {author}
  {\bibfnamefont {I.}~\bibnamefont {Bloch}},\ }\bibfield  {title} {\bibinfo
  {title} {Realization of the {H}ofstadter {H}amiltonian with ultracold atoms
  in optical lattices},\ }\href
  {https://doi.org/10.1103/PhysRevLett.111.185301} {\bibfield  {journal}
  {\bibinfo  {journal} {Phys. Rev. Lett.}\ }\textbf {\bibinfo {volume} {111}},\
  \bibinfo {pages} {185301} (\bibinfo {year} {2013})},\ \Eprint
  {https://arxiv.org/abs/1308.0321} {arXiv:1308.0321 [cond-mat.quantum-gas]}
  \BibitemShut {NoStop}%
\bibitem [{\citenamefont {Struck}\ \emph {et~al.}(2013)\citenamefont {Struck},
  \citenamefont {Weinberg}, \citenamefont {\"Olschl\"ager}, \citenamefont
  {Windpassinger}, \citenamefont {Simonet}, \citenamefont {Sengstock},
  \citenamefont {H\"oppner}, \citenamefont {Hauke}, \citenamefont {Eckardt},
  \citenamefont {Lewenstein},\ and\ \citenamefont {Mathey}}]{Struck2013}%
  \BibitemOpen
  \bibfield  {author} {\bibinfo {author} {\bibfnamefont {J.}~\bibnamefont
  {Struck}}, \bibinfo {author} {\bibfnamefont {M.}~\bibnamefont {Weinberg}},
  \bibinfo {author} {\bibfnamefont {C.}~\bibnamefont {\"Olschl\"ager}},
  \bibinfo {author} {\bibfnamefont {P.}~\bibnamefont {Windpassinger}}, \bibinfo
  {author} {\bibfnamefont {J.}~\bibnamefont {Simonet}}, \bibinfo {author}
  {\bibfnamefont {K.}~\bibnamefont {Sengstock}}, \bibinfo {author}
  {\bibfnamefont {R.}~\bibnamefont {H\"oppner}}, \bibinfo {author}
  {\bibfnamefont {P.}~\bibnamefont {Hauke}}, \bibinfo {author} {\bibfnamefont
  {A.}~\bibnamefont {Eckardt}}, \bibinfo {author} {\bibfnamefont
  {M.}~\bibnamefont {Lewenstein}},\ and\ \bibinfo {author} {\bibfnamefont
  {L.}~\bibnamefont {Mathey}},\ }\bibfield  {title} {\bibinfo {title}
  {Engineering {Ising-XY} spin models in a triangular lattice via tunable
  artificial gauge fields},\ }\href {https://doi.org/10.1038/nphys2750}
  {\bibfield  {journal} {\bibinfo  {journal} {Nat. Phys.}\ }\textbf {\bibinfo
  {volume} {9}},\ \bibinfo {pages} {738} (\bibinfo {year} {2013})},\ \Eprint
  {https://arxiv.org/abs/1304.5520} {arXiv:1304.5520 [cond-mat.quant-gas]}
  \BibitemShut {NoStop}%
\bibitem [{\citenamefont {Aidelsburger}\ \emph {et~al.}(2015)\citenamefont
  {Aidelsburger}, \citenamefont {Lohse}, \citenamefont {Schweizer},
  \citenamefont {Atala}, \citenamefont {Barreiro}, \citenamefont
  {Nascimb\`{e}ne}, \citenamefont {Cooper}, \citenamefont {Bloch},\ and\
  \citenamefont {Goldman}}]{Aidelsburger2015}%
  \BibitemOpen
  \bibfield  {author} {\bibinfo {author} {\bibfnamefont {M.}~\bibnamefont
  {Aidelsburger}}, \bibinfo {author} {\bibfnamefont {M.}~\bibnamefont {Lohse}},
  \bibinfo {author} {\bibfnamefont {C.}~\bibnamefont {Schweizer}}, \bibinfo
  {author} {\bibfnamefont {M.}~\bibnamefont {Atala}}, \bibinfo {author}
  {\bibfnamefont {J.~T.}\ \bibnamefont {Barreiro}}, \bibinfo {author}
  {\bibfnamefont {S.}~\bibnamefont {Nascimb\`{e}ne}}, \bibinfo {author}
  {\bibfnamefont {N.~R.}\ \bibnamefont {Cooper}}, \bibinfo {author}
  {\bibfnamefont {I.}~\bibnamefont {Bloch}},\ and\ \bibinfo {author}
  {\bibfnamefont {N.}~\bibnamefont {Goldman}},\ }\bibfield  {title} {\bibinfo
  {title} {Measuring the {C}hern number of {H}ofstadter bands with ultracold
  bosonic atoms},\ }\href {https://doi.org/10.1038/nphys3171} {\bibfield
  {journal} {\bibinfo  {journal} {Nat. Phys.}\ }\textbf {\bibinfo {volume}
  {11}},\ \bibinfo {pages} {162} (\bibinfo {year} {2015})},\ \Eprint
  {https://arxiv.org/abs/1407.4205} {arXiv:1407.4205 [cond-mat.quantum-gas]}
  \BibitemShut {NoStop}%
\bibitem [{\citenamefont {Fl{\" a}schner}\ \emph {et~al.}(2016)\citenamefont
  {Fl{\" a}schner}, \citenamefont {Rem}, \citenamefont {Tarnowski},
  \citenamefont {Vogel}, \citenamefont {L{\" u}hmann}, \citenamefont
  {Sengstock},\ and\ \citenamefont {Weitenberg}}]{Flaeschner2016}%
  \BibitemOpen
  \bibfield  {author} {\bibinfo {author} {\bibfnamefont {N.}~\bibnamefont
  {Fl{\" a}schner}}, \bibinfo {author} {\bibfnamefont {B.~S.}\ \bibnamefont
  {Rem}}, \bibinfo {author} {\bibfnamefont {M.}~\bibnamefont {Tarnowski}},
  \bibinfo {author} {\bibfnamefont {D.}~\bibnamefont {Vogel}}, \bibinfo
  {author} {\bibfnamefont {D.-S.}\ \bibnamefont {L{\" u}hmann}}, \bibinfo
  {author} {\bibfnamefont {K.}~\bibnamefont {Sengstock}},\ and\ \bibinfo
  {author} {\bibfnamefont {C.}~\bibnamefont {Weitenberg}},\ }\bibfield  {title}
  {\bibinfo {title} {Experimental reconstruction of the {B}erry curvature in a
  {F}loquet {B}loch band},\ }\href {https://doi.org/10.1126/science.aad4568}
  {\bibfield  {journal} {\bibinfo  {journal} {Science}\ }\textbf {\bibinfo
  {volume} {352}},\ \bibinfo {pages} {1091} (\bibinfo {year} {2016})},\ \Eprint
  {https://arxiv.org/abs/1509.05763} {arXiv:1509.05763 [cond-mat.quant-gas]}
  \BibitemShut {NoStop}%
\bibitem [{\citenamefont {Sch{\" a}fer}\ \emph {et~al.}(2020)\citenamefont
  {Sch{\" a}fer}, \citenamefont {Fukuhara}, \citenamefont {Sugawa},
  \citenamefont {Takasu},\ and\ \citenamefont {Takahashi}}]{Schafer2020}%
  \BibitemOpen
  \bibfield  {author} {\bibinfo {author} {\bibfnamefont {F.}~\bibnamefont
  {Sch{\" a}fer}}, \bibinfo {author} {\bibfnamefont {T.}~\bibnamefont
  {Fukuhara}}, \bibinfo {author} {\bibfnamefont {S.}~\bibnamefont {Sugawa}},
  \bibinfo {author} {\bibfnamefont {Y.}~\bibnamefont {Takasu}},\ and\ \bibinfo
  {author} {\bibfnamefont {Y.}~\bibnamefont {Takahashi}},\ }\bibfield  {title}
  {\bibinfo {title} {Tools for quantum simulation with ultracold atoms in
  optical lattices},\ }\href {https://doi.org/10.1038/s42254-020-0195-3}
  {\bibfield  {journal} {\bibinfo  {journal} {Nat. Rev. Phys.}\ }\textbf
  {\bibinfo {volume} {2}},\ \bibinfo {pages} {411} (\bibinfo {year} {2020})},\
  \Eprint {https://arxiv.org/abs/2006.06120} {arXiv:2006.06120
  [cond-mat.quant-gas]} \BibitemShut {NoStop}%
\bibitem [{\citenamefont {{McIver}}\ \emph {et~al.}(2020)\citenamefont
  {{McIver}}, \citenamefont {{Schulte}}, \citenamefont {{Stein}}, \citenamefont
  {{Matsuyama}}, \citenamefont {{Jotzu}}, \citenamefont {{Meier}},\ and\
  \citenamefont {{Cavalleri}}}]{McIver2020}%
  \BibitemOpen
  \bibfield  {author} {\bibinfo {author} {\bibfnamefont {J.~W.}\ \bibnamefont
  {{McIver}}}, \bibinfo {author} {\bibfnamefont {B.}~\bibnamefont {{Schulte}}},
  \bibinfo {author} {\bibfnamefont {F.-U.}\ \bibnamefont {{Stein}}}, \bibinfo
  {author} {\bibfnamefont {T.}~\bibnamefont {{Matsuyama}}}, \bibinfo {author}
  {\bibfnamefont {G.}~\bibnamefont {{Jotzu}}}, \bibinfo {author} {\bibfnamefont
  {G.}~\bibnamefont {{Meier}}},\ and\ \bibinfo {author} {\bibfnamefont
  {A.}~\bibnamefont {{Cavalleri}}},\ }\bibfield  {title} {\bibinfo {title}
  {Light-induced anomalous {H}all effect in graphene},\ }\href
  {https://doi.org/10.1038/s41567-019-0698-y} {\bibfield  {journal} {\bibinfo
  {journal} {Nat. Phys.}\ }\textbf {\bibinfo {volume} {16}},\ \bibinfo {pages}
  {38} (\bibinfo {year} {2020})},\ \Eprint {https://arxiv.org/abs/1811.03522}
  {arXiv:1811.03522} \BibitemShut {NoStop}%
\bibitem [{\citenamefont {Kitagawa}\ \emph {et~al.}(2010)\citenamefont
  {Kitagawa}, \citenamefont {Berg}, \citenamefont {Rudner},\ and\ \citenamefont
  {Demler}}]{Kitagawa2010}%
  \BibitemOpen
  \bibfield  {author} {\bibinfo {author} {\bibfnamefont {T.}~\bibnamefont
  {Kitagawa}}, \bibinfo {author} {\bibfnamefont {E.}~\bibnamefont {Berg}},
  \bibinfo {author} {\bibfnamefont {M.}~\bibnamefont {Rudner}},\ and\ \bibinfo
  {author} {\bibfnamefont {E.}~\bibnamefont {Demler}},\ }\bibfield  {title}
  {\bibinfo {title} {Topological characterization of periodically driven
  quantum systems},\ }\href {https://doi.org/10.1103/PhysRevB.82.235114}
  {\bibfield  {journal} {\bibinfo  {journal} {Phys. Rev. B}\ }\textbf {\bibinfo
  {volume} {82}},\ \bibinfo {pages} {235114} (\bibinfo {year} {2010})},\
  \Eprint {https://arxiv.org/abs/1010.6126} {arXiv:1010.6126} \BibitemShut
  {NoStop}%
\bibitem [{\citenamefont {Cayssol}\ \emph {et~al.}(2013)\citenamefont
  {Cayssol}, \citenamefont {D\'ora}, \citenamefont {Simon},\ and\ \citenamefont
  {Moessner}}]{Cayssol2013}%
  \BibitemOpen
  \bibfield  {author} {\bibinfo {author} {\bibfnamefont {J.}~\bibnamefont
  {Cayssol}}, \bibinfo {author} {\bibfnamefont {B.}~\bibnamefont {D\'ora}},
  \bibinfo {author} {\bibfnamefont {F.}~\bibnamefont {Simon}},\ and\ \bibinfo
  {author} {\bibfnamefont {R.}~\bibnamefont {Moessner}},\ }\bibfield  {title}
  {\bibinfo {title} {Floquet topological insulators},\ }\href
  {https://doi.org/https://doi.org/10.1002/pssr.201206451} {\bibfield
  {journal} {\bibinfo  {journal} {Phys. Status Solidi -- Rapid Res. Lett.}\
  }\textbf {\bibinfo {volume} {7}},\ \bibinfo {pages} {101} (\bibinfo {year}
  {2013})},\ \Eprint {https://arxiv.org/abs/1211.5623} {arXiv:1211.5623}
  \BibitemShut {NoStop}%
\bibitem [{\citenamefont {Foa~Torres}\ \emph {et~al.}(2014)\citenamefont
  {Foa~Torres}, \citenamefont {Perez-Piskunow}, \citenamefont {Balseiro},\ and\
  \citenamefont {Usaj}}]{FoaTorres2014}%
  \BibitemOpen
  \bibfield  {author} {\bibinfo {author} {\bibfnamefont {L.~E.~F.}\
  \bibnamefont {Foa~Torres}}, \bibinfo {author} {\bibfnamefont {P.~M.}\
  \bibnamefont {Perez-Piskunow}}, \bibinfo {author} {\bibfnamefont {C.~A.}\
  \bibnamefont {Balseiro}},\ and\ \bibinfo {author} {\bibfnamefont
  {G.}~\bibnamefont {Usaj}},\ }\bibfield  {title} {\bibinfo {title}
  {Multiterminal conductance of a {F}loquet topological insulator},\ }\href
  {https://doi.org/10.1103/PhysRevLett.113.266801} {\bibfield  {journal}
  {\bibinfo  {journal} {Phys. Rev. Lett.}\ }\textbf {\bibinfo {volume} {113}},\
  \bibinfo {pages} {266801} (\bibinfo {year} {2014})},\ \Eprint
  {https://arxiv.org/abs/1409.2482} {arXiv:1409.2482} \BibitemShut {NoStop}%
\bibitem [{\citenamefont {Grushin}\ \emph {et~al.}(2014)\citenamefont
  {Grushin}, \citenamefont {G\'omez-Le\'on},\ and\ \citenamefont
  {Neupert}}]{Grushin2014}%
  \BibitemOpen
  \bibfield  {author} {\bibinfo {author} {\bibfnamefont {A.~G.}\ \bibnamefont
  {Grushin}}, \bibinfo {author} {\bibfnamefont {A.}~\bibnamefont
  {G\'omez-Le\'on}},\ and\ \bibinfo {author} {\bibfnamefont {T.}~\bibnamefont
  {Neupert}},\ }\bibfield  {title} {\bibinfo {title} {Floquet fractional
  {C}hern insulators},\ }\href {https://doi.org/10.1103/PhysRevLett.112.156801}
  {\bibfield  {journal} {\bibinfo  {journal} {Phys. Rev. Lett.}\ }\textbf
  {\bibinfo {volume} {112}},\ \bibinfo {pages} {156801} (\bibinfo {year}
  {2014})},\ \Eprint {https://arxiv.org/abs/1309.3571} {arXiv:1309.3571}
  \BibitemShut {NoStop}%
\bibitem [{\citenamefont {Farrell}\ and\ \citenamefont
  {Pereg-Barnea}(2016)}]{Farrell2016}%
  \BibitemOpen
  \bibfield  {author} {\bibinfo {author} {\bibfnamefont {A.}~\bibnamefont
  {Farrell}}\ and\ \bibinfo {author} {\bibfnamefont {T.}~\bibnamefont
  {Pereg-Barnea}},\ }\bibfield  {title} {\bibinfo {title} {Edge-state transport
  in {F}loquet topological insulators},\ }\href
  {https://doi.org/10.1103/PhysRevB.93.045121} {\bibfield  {journal} {\bibinfo
  {journal} {Phys. Rev. B}\ }\textbf {\bibinfo {volume} {93}},\ \bibinfo
  {pages} {045121} (\bibinfo {year} {2016})},\ \Eprint
  {https://arxiv.org/abs/1505.05584} {arXiv:1505.05584} \BibitemShut {NoStop}%
\bibitem [{\citenamefont {Huam\'an}\ and\ \citenamefont
  {Usaj}(2019)}]{Huaman2019}%
  \BibitemOpen
  \bibfield  {author} {\bibinfo {author} {\bibfnamefont {A.}~\bibnamefont
  {Huam\'an}}\ and\ \bibinfo {author} {\bibfnamefont {G.}~\bibnamefont
  {Usaj}},\ }\bibfield  {title} {\bibinfo {title} {Floquet spectrum and
  two-terminal conductance of a transition-metal dichalcogenide ribbon under a
  circularly polarized laser field},\ }\href
  {https://doi.org/10.1103/PhysRevB.99.075423} {\bibfield  {journal} {\bibinfo
  {journal} {Phys. Rev. B}\ }\textbf {\bibinfo {volume} {99}},\ \bibinfo
  {pages} {075423} (\bibinfo {year} {2019})},\ \Eprint
  {https://arxiv.org/abs/1811.12303} {arXiv:1811.12303} \BibitemShut {NoStop}%
\bibitem [{\citenamefont {Rahav}\ \emph {et~al.}(2003)\citenamefont {Rahav},
  \citenamefont {Gilary},\ and\ \citenamefont {Fishman}}]{Rahav2003}%
  \BibitemOpen
  \bibfield  {author} {\bibinfo {author} {\bibfnamefont {S.}~\bibnamefont
  {Rahav}}, \bibinfo {author} {\bibfnamefont {I.}~\bibnamefont {Gilary}},\ and\
  \bibinfo {author} {\bibfnamefont {S.}~\bibnamefont {Fishman}},\ }\bibfield
  {title} {\bibinfo {title} {Effective {H}amiltonians for periodically driven
  systems},\ }\href {https://doi.org/10.1103/PhysRevA.68.013820} {\bibfield
  {journal} {\bibinfo  {journal} {Phys. Rev. A}\ }\textbf {\bibinfo {volume}
  {68}},\ \bibinfo {pages} {013820} (\bibinfo {year} {2003})},\ \Eprint
  {https://arxiv.org/abs/nlin/0301033} {arXiv:nlin/0301033} \BibitemShut
  {NoStop}%
\bibitem [{\citenamefont {Goldman}\ and\ \citenamefont
  {Dalibard}(2014)}]{Goldman2014}%
  \BibitemOpen
  \bibfield  {author} {\bibinfo {author} {\bibfnamefont {N.}~\bibnamefont
  {Goldman}}\ and\ \bibinfo {author} {\bibfnamefont {J.}~\bibnamefont
  {Dalibard}},\ }\bibfield  {title} {\bibinfo {title} {Periodically driven
  quantum systems: {E}ffective {H}amiltonians and engineered gauge fields},\
  }\href {https://doi.org/10.1103/PhysRevX.4.031027} {\bibfield  {journal}
  {\bibinfo  {journal} {Phys. Rev. X}\ }\textbf {\bibinfo {volume} {4}},\
  \bibinfo {pages} {031027} (\bibinfo {year} {2014})},\ \Eprint
  {https://arxiv.org/abs/1404.4373} {arXiv:1404.4373 [cond-mat]} \BibitemShut
  {NoStop}%
\bibitem [{\citenamefont {Eckardt}\ and\ \citenamefont
  {Anisimovas}(2015)}]{Eckardt2015}%
  \BibitemOpen
  \bibfield  {author} {\bibinfo {author} {\bibfnamefont {A.}~\bibnamefont
  {Eckardt}}\ and\ \bibinfo {author} {\bibfnamefont {E.}~\bibnamefont
  {Anisimovas}},\ }\bibfield  {title} {\bibinfo {title} {High-frequency
  approximation for periodically driven quantum systems from a {F}loquet-space
  perspective},\ }\href {https://doi.org/10.1088/1367-2630/17/9/093039}
  {\bibfield  {journal} {\bibinfo  {journal} {New J. Phys.}\ }\textbf {\bibinfo
  {volume} {17}},\ \bibinfo {pages} {093039} (\bibinfo {year} {2015})},\
  \Eprint {https://arxiv.org/abs/1502.06477} {arXiv:1502.06477
  [cond-mat.quant-gas]} \BibitemShut {NoStop}%
\bibitem [{\citenamefont {Mananga}\ and\ \citenamefont
  {Charpentier}(2011)}]{Mananga2011}%
  \BibitemOpen
  \bibfield  {author} {\bibinfo {author} {\bibfnamefont {E.~S.}\ \bibnamefont
  {Mananga}}\ and\ \bibinfo {author} {\bibfnamefont {T.}~\bibnamefont
  {Charpentier}},\ }\bibfield  {title} {\bibinfo {title} {Introduction of the
  {F}loquet-{M}agnus expansion in solid-state nuclear magnetic resonance
  spectroscopy},\ }\href {https://doi.org/10.1063/1.3610943} {\bibfield
  {journal} {\bibinfo  {journal} {J. Chem. Phys.}\ }\textbf {\bibinfo {volume}
  {135}},\ \bibinfo {pages} {044109} (\bibinfo {year} {2011})}\BibitemShut
  {NoStop}%
\bibitem [{\citenamefont {Mananga}\ and\ \citenamefont
  {Charpentier}(2016)}]{Mananga2016}%
  \BibitemOpen
  \bibfield  {author} {\bibinfo {author} {\bibfnamefont {E.~S.}\ \bibnamefont
  {Mananga}}\ and\ \bibinfo {author} {\bibfnamefont {T.}~\bibnamefont
  {Charpentier}},\ }\bibfield  {title} {\bibinfo {title} {On the
  {F}loquet-{M}agnus expansion: {A}pplications in solid-state nuclear magnetic
  resonance and physics},\ }\href
  {https://doi.org/https://doi.org/10.1016/j.physrep.2015.10.005} {\bibfield
  {journal} {\bibinfo  {journal} {Phys. Rep.}\ }\textbf {\bibinfo {volume}
  {609}},\ \bibinfo {pages} {1} (\bibinfo {year} {2016})}\BibitemShut {NoStop}%
\bibitem [{\citenamefont {Verdeny}\ \emph {et~al.}(2013)\citenamefont
  {Verdeny}, \citenamefont {Mielke},\ and\ \citenamefont
  {Mintert}}]{Verdeny2013}%
  \BibitemOpen
  \bibfield  {author} {\bibinfo {author} {\bibfnamefont {A.}~\bibnamefont
  {Verdeny}}, \bibinfo {author} {\bibfnamefont {A.}~\bibnamefont {Mielke}},\
  and\ \bibinfo {author} {\bibfnamefont {F.}~\bibnamefont {Mintert}},\
  }\bibfield  {title} {\bibinfo {title} {Accurate effective {H}amiltonians via
  unitary flow in {F}loquet space},\ }\href
  {https://doi.org/10.1103/PhysRevLett.111.175301} {\bibfield  {journal}
  {\bibinfo  {journal} {Phys. Rev. Lett.}\ }\textbf {\bibinfo {volume} {111}},\
  \bibinfo {pages} {175301} (\bibinfo {year} {2013})},\ \Eprint
  {https://arxiv.org/abs/1304.3584} {arXiv:1304.3584 [quant-ph]} \BibitemShut
  {NoStop}%
\bibitem [{\citenamefont {Mikami}\ \emph {et~al.}(2016)\citenamefont {Mikami},
  \citenamefont {Kitamura}, \citenamefont {Yasuda}, \citenamefont {Tsuji},
  \citenamefont {Oka},\ and\ \citenamefont {Aoki}}]{Mikami2016PRB}%
  \BibitemOpen
  \bibfield  {author} {\bibinfo {author} {\bibfnamefont {T.}~\bibnamefont
  {Mikami}}, \bibinfo {author} {\bibfnamefont {S.}~\bibnamefont {Kitamura}},
  \bibinfo {author} {\bibfnamefont {K.}~\bibnamefont {Yasuda}}, \bibinfo
  {author} {\bibfnamefont {N.}~\bibnamefont {Tsuji}}, \bibinfo {author}
  {\bibfnamefont {T.}~\bibnamefont {Oka}},\ and\ \bibinfo {author}
  {\bibfnamefont {H.}~\bibnamefont {Aoki}},\ }\bibfield  {title} {\bibinfo
  {title} {{B}rillouin-{W}igner theory for high-frequency expansion in
  periodically driven systems: {A}pplication to {F}loquet topological
  insulators},\ }\href {https://doi.org/10.1103/PhysRevB.93.144307} {\bibfield
  {journal} {\bibinfo  {journal} {Phys. Rev. B}\ }\textbf {\bibinfo {volume}
  {93}},\ \bibinfo {pages} {144307} (\bibinfo {year} {2016})},\ \Eprint
  {https://arxiv.org/abs/1511.00755} {arXiv:1511.00755 [cond-mat]} \BibitemShut
  {NoStop}%
\bibitem [{\citenamefont {Vogl}\ \emph {et~al.}(2019)\citenamefont {Vogl},
  \citenamefont {Laurell}, \citenamefont {Barr},\ and\ \citenamefont
  {Fiete}}]{Vogl2019}%
  \BibitemOpen
  \bibfield  {author} {\bibinfo {author} {\bibfnamefont {M.}~\bibnamefont
  {Vogl}}, \bibinfo {author} {\bibfnamefont {P.}~\bibnamefont {Laurell}},
  \bibinfo {author} {\bibfnamefont {A.~D.}\ \bibnamefont {Barr}},\ and\
  \bibinfo {author} {\bibfnamefont {G.~A.}\ \bibnamefont {Fiete}},\ }\bibfield
  {title} {\bibinfo {title} {Flow equation approach to periodically driven
  quantum systems},\ }\href {https://doi.org/10.1103/PhysRevX.9.021037}
  {\bibfield  {journal} {\bibinfo  {journal} {Phys. Rev. X}\ }\textbf {\bibinfo
  {volume} {9}},\ \bibinfo {pages} {021037} (\bibinfo {year} {2019})},\ \Eprint
  {https://arxiv.org/abs/1808.01697} {arXiv:1808.01697 [cond-mat]} \BibitemShut
  {NoStop}%
\bibitem [{\citenamefont {Novi\v{c}enko}\ \emph {et~al.}(2017)\citenamefont
  {Novi\v{c}enko}, \citenamefont {Anisimovas},\ and\ \citenamefont
  {Juzeli\={u}nas}}]{Novicenko2017}%
  \BibitemOpen
  \bibfield  {author} {\bibinfo {author} {\bibfnamefont {V.}~\bibnamefont
  {Novi\v{c}enko}}, \bibinfo {author} {\bibfnamefont {E.}~\bibnamefont
  {Anisimovas}},\ and\ \bibinfo {author} {\bibfnamefont {G.}~\bibnamefont
  {Juzeli\={u}nas}},\ }\bibfield  {title} {\bibinfo {title} {Floquet analysis
  of a quantum system with modulated periodic driving},\ }\href
  {https://doi.org/10.1103/PhysRevA.95.023615} {\bibfield  {journal} {\bibinfo
  {journal} {Phys. Rev. A}\ }\textbf {\bibinfo {volume} {95}},\ \bibinfo
  {pages} {023615} (\bibinfo {year} {2017})},\ \Eprint
  {https://arxiv.org/abs/1608.08420} {arXiv:1608.08420 [cond-mat]} \BibitemShut
  {NoStop}%
\bibitem [{\citenamefont {Haddadfarshi}\ \emph {et~al.}(2015)\citenamefont
  {Haddadfarshi}, \citenamefont {Cui},\ and\ \citenamefont
  {Mintert}}]{Haddafarshi2015}%
  \BibitemOpen
  \bibfield  {author} {\bibinfo {author} {\bibfnamefont {F.}~\bibnamefont
  {Haddadfarshi}}, \bibinfo {author} {\bibfnamefont {J.}~\bibnamefont {Cui}},\
  and\ \bibinfo {author} {\bibfnamefont {F.}~\bibnamefont {Mintert}},\
  }\bibfield  {title} {\bibinfo {title} {Completely positive approximate
  solutions of driven open quantum systems},\ }\href
  {https://doi.org/10.1103/PhysRevLett.114.130402} {\bibfield  {journal}
  {\bibinfo  {journal} {Phys. Rev. Lett.}\ }\textbf {\bibinfo {volume} {114}},\
  \bibinfo {pages} {130402} (\bibinfo {year} {2015})},\ \Eprint
  {https://arxiv.org/abs/1409.6184} {arXiv:1409.6184 [quant-ph]} \BibitemShut
  {NoStop}%
\bibitem [{\citenamefont {Dai}\ \emph {et~al.}(2016)\citenamefont {Dai},
  \citenamefont {Shi},\ and\ \citenamefont {Yi}}]{Dai2016}%
  \BibitemOpen
  \bibfield  {author} {\bibinfo {author} {\bibfnamefont {C.~M.}\ \bibnamefont
  {Dai}}, \bibinfo {author} {\bibfnamefont {Z.~C.}\ \bibnamefont {Shi}},\ and\
  \bibinfo {author} {\bibfnamefont {X.~X.}\ \bibnamefont {Yi}},\ }\bibfield
  {title} {\bibinfo {title} {Floquet theorem with open systems and its
  applications},\ }\href {https://doi.org/10.1103/PhysRevA.93.032121}
  {\bibfield  {journal} {\bibinfo  {journal} {Phys. Rev. A}\ }\textbf {\bibinfo
  {volume} {93}},\ \bibinfo {pages} {032121} (\bibinfo {year} {2016})},\
  \Eprint {https://arxiv.org/abs/1512.05562} {1512.05562 [quant-ph]}
  \BibitemShut {NoStop}%
\bibitem [{\citenamefont {Restrepo}\ \emph {et~al.}(2016)\citenamefont
  {Restrepo}, \citenamefont {Cerrillo}, \citenamefont {Bastidas}, \citenamefont
  {Angelakis},\ and\ \citenamefont {Brandes}}]{Restrepo2016}%
  \BibitemOpen
  \bibfield  {author} {\bibinfo {author} {\bibfnamefont {S.}~\bibnamefont
  {Restrepo}}, \bibinfo {author} {\bibfnamefont {J.}~\bibnamefont {Cerrillo}},
  \bibinfo {author} {\bibfnamefont {V.~M.}\ \bibnamefont {Bastidas}}, \bibinfo
  {author} {\bibfnamefont {D.~G.}\ \bibnamefont {Angelakis}},\ and\ \bibinfo
  {author} {\bibfnamefont {T.}~\bibnamefont {Brandes}},\ }\bibfield  {title}
  {\bibinfo {title} {Driven open quantum systems and {F}loquet stroboscopic
  dynamics},\ }\href {https://doi.org/10.1103/PhysRevLett.117.250401}
  {\bibfield  {journal} {\bibinfo  {journal} {Phys. Rev. Lett.}\ }\textbf
  {\bibinfo {volume} {117}},\ \bibinfo {pages} {250401} (\bibinfo {year}
  {2016})},\ \Eprint {https://arxiv.org/abs/1606.08392} {arXiv:1606.08392
  [quant-phys]} \BibitemShut {NoStop}%
\bibitem [{\citenamefont {Schnell}\ \emph {et~al.}(2021)\citenamefont
  {Schnell}, \citenamefont {Denisov},\ and\ \citenamefont
  {Eckardt}}]{Schnell2021}%
  \BibitemOpen
  \bibfield  {author} {\bibinfo {author} {\bibfnamefont {A.}~\bibnamefont
  {Schnell}}, \bibinfo {author} {\bibfnamefont {S.}~\bibnamefont {Denisov}},\
  and\ \bibinfo {author} {\bibfnamefont {A.}~\bibnamefont {Eckardt}},\
  }\bibfield  {title} {\bibinfo {title} {High-frequency expansions for
  time-periodic {L}indblad generators},\ }\href
  {https://doi.org/10.1103/PhysRevB.104.165414} {\bibfield  {journal} {\bibinfo
   {journal} {Phys. Rev. B}\ }\textbf {\bibinfo {volume} {104}},\ \bibinfo
  {pages} {165414} (\bibinfo {year} {2021})},\ \Eprint
  {https://arxiv.org/abs/2107.10054} {arXiv:2107.10054 [quant-ph]} \BibitemShut
  {NoStop}%
\bibitem [{\citenamefont {Ikeda}\ \emph {et~al.}(2021)\citenamefont {Ikeda},
  \citenamefont {Chinzei},\ and\ \citenamefont {Sato}}]{Ikeda2021lindblad}%
  \BibitemOpen
  \bibfield  {author} {\bibinfo {author} {\bibfnamefont {T.~N.}\ \bibnamefont
  {Ikeda}}, \bibinfo {author} {\bibfnamefont {K.}~\bibnamefont {Chinzei}},\
  and\ \bibinfo {author} {\bibfnamefont {M.}~\bibnamefont {Sato}},\ }\href@noop
  {} {\bibinfo {title} {Nonequilibrium steady states in the
  {F}loquet-{L}indblad systems: van {V}leck's high-frequency expansion
  approach}} (\bibinfo {year} {2021}),\ \Eprint
  {https://arxiv.org/abs/2107.07911} {arXiv:2107.07911 [cond-mat.mes-hall]}
  \BibitemShut {NoStop}%
\bibitem [{\citenamefont {{Dai}}\ \emph {et~al.}(2017)\citenamefont {{Dai}},
  \citenamefont {{Li}}, \citenamefont {{Wang}},\ and\ \citenamefont
  {{Yi}}}]{Dai2017}%
  \BibitemOpen
  \bibfield  {author} {\bibinfo {author} {\bibfnamefont {C.~M.}\ \bibnamefont
  {{Dai}}}, \bibinfo {author} {\bibfnamefont {H.}~\bibnamefont {{Li}}},
  \bibinfo {author} {\bibfnamefont {W.}~\bibnamefont {{Wang}}},\ and\ \bibinfo
  {author} {\bibfnamefont {X.~X.}\ \bibnamefont {{Yi}}},\ }\href@noop {}
  {\bibinfo {title} {Generalized {F}loquet theory for open quantum systems}}
  (\bibinfo {year} {2017}),\ \Eprint {https://arxiv.org/abs/1707.05030}
  {arXiv:1707.05030 [quant-ph]} \BibitemShut {NoStop}%
\bibitem [{\citenamefont {Weinberg}\ \emph {et~al.}(2017)\citenamefont
  {Weinberg}, \citenamefont {Bukov}, \citenamefont {D'Alessio}, \citenamefont
  {Polkovnikov}, \citenamefont {Vajna},\ and\ \citenamefont
  {Kolodrubetz}}]{Weinberg2017}%
  \BibitemOpen
  \bibfield  {author} {\bibinfo {author} {\bibfnamefont {P.}~\bibnamefont
  {Weinberg}}, \bibinfo {author} {\bibfnamefont {M.}~\bibnamefont {Bukov}},
  \bibinfo {author} {\bibfnamefont {L.}~\bibnamefont {D'Alessio}}, \bibinfo
  {author} {\bibfnamefont {A.}~\bibnamefont {Polkovnikov}}, \bibinfo {author}
  {\bibfnamefont {S.}~\bibnamefont {Vajna}},\ and\ \bibinfo {author}
  {\bibfnamefont {M.}~\bibnamefont {Kolodrubetz}},\ }\bibfield  {title}
  {\bibinfo {title} {Adiabatic perturbation theory and geometry of
  periodically-driven systems},\ }\href
  {https://doi.org/10.1016/j.physrep.2017.05.003} {\bibfield  {journal}
  {\bibinfo  {journal} {Phys. Rep.}\ }\textbf {\bibinfo {volume} {688}},\
  \bibinfo {pages} {1} (\bibinfo {year} {2017})},\ \Eprint
  {https://arxiv.org/abs/1606.02229} {arXiv:1606.02229 [cond-mat]} \BibitemShut
  {NoStop}%
\bibitem [{\citenamefont {Zeuch}\ \emph {et~al.}(2020)\citenamefont {Zeuch},
  \citenamefont {Hassler}, \citenamefont {Slim},\ and\ \citenamefont
  {DiVincenzo}}]{Zeuch2020}%
  \BibitemOpen
  \bibfield  {author} {\bibinfo {author} {\bibfnamefont {D.}~\bibnamefont
  {Zeuch}}, \bibinfo {author} {\bibfnamefont {F.}~\bibnamefont {Hassler}},
  \bibinfo {author} {\bibfnamefont {J.~J.}\ \bibnamefont {Slim}},\ and\
  \bibinfo {author} {\bibfnamefont {D.~P.}\ \bibnamefont {DiVincenzo}},\
  }\bibfield  {title} {\bibinfo {title} {Exact rotating wave approximation},\
  }\href {https://doi.org/https://doi.org/10.1016/j.aop.2020.168327} {\bibfield
   {journal} {\bibinfo  {journal} {Ann. Phys.}\ }\textbf {\bibinfo {volume}
  {423}},\ \bibinfo {pages} {168327} (\bibinfo {year} {2020})},\ \Eprint
  {https://arxiv.org/abs/1807.02858} {arXiv:1807.02858 [quant-ph]} \BibitemShut
  {NoStop}%
\bibitem [{\citenamefont {Zeuch}\ and\ \citenamefont
  {DiVincenzo}(2020)}]{Zeuch2020refuting}%
  \BibitemOpen
  \bibfield  {author} {\bibinfo {author} {\bibfnamefont {D.}~\bibnamefont
  {Zeuch}}\ and\ \bibinfo {author} {\bibfnamefont {D.~P.}\ \bibnamefont
  {DiVincenzo}},\ }\href@noop {} {\bibinfo {title} {Refuting a proposed axiom
  for defining the exact rotating wave approximation}} (\bibinfo {year}
  {2020}),\ \Eprint {https://arxiv.org/abs/2010.02751} {arXiv:2010.02751
  [quant-ph]} \BibitemShut {NoStop}%
\bibitem [{\citenamefont {Desbuquois}\ \emph {et~al.}(2017)\citenamefont
  {Desbuquois}, \citenamefont {Messer}, \citenamefont {G\"org}, \citenamefont
  {Sandholzer}, \citenamefont {Jotzu},\ and\ \citenamefont
  {Esslinger}}]{Desbq2017}%
  \BibitemOpen
  \bibfield  {author} {\bibinfo {author} {\bibfnamefont {R.}~\bibnamefont
  {Desbuquois}}, \bibinfo {author} {\bibfnamefont {M.}~\bibnamefont {Messer}},
  \bibinfo {author} {\bibfnamefont {F.}~\bibnamefont {G\"org}}, \bibinfo
  {author} {\bibfnamefont {K.}~\bibnamefont {Sandholzer}}, \bibinfo {author}
  {\bibfnamefont {G.}~\bibnamefont {Jotzu}},\ and\ \bibinfo {author}
  {\bibfnamefont {T.}~\bibnamefont {Esslinger}},\ }\bibfield  {title} {\bibinfo
  {title} {Controlling the {F}loquet state population and observing micromotion
  in a periodically driven two-body quantum system},\ }\href
  {https://doi.org/10.1103/PhysRevA.96.053602} {\bibfield  {journal} {\bibinfo
  {journal} {Phys. Rev. A}\ }\textbf {\bibinfo {volume} {96}},\ \bibinfo
  {pages} {053602} (\bibinfo {year} {2017})},\ \Eprint
  {https://arxiv.org/abs/1703.07767} {arXiv:1703.07767 [cond-mat.quant-gas]}
  \BibitemShut {NoStop}%
\bibitem [{\citenamefont {Wilczek}\ and\ \citenamefont
  {Zee}(1984)}]{Wilczek1984}%
  \BibitemOpen
  \bibfield  {author} {\bibinfo {author} {\bibfnamefont {F.}~\bibnamefont
  {Wilczek}}\ and\ \bibinfo {author} {\bibfnamefont {A.}~\bibnamefont {Zee}},\
  }\bibfield  {title} {\bibinfo {title} {Appearance of gauge structure in
  simple dynamical systems},\ }\href
  {https://doi.org/10.1103/PhysRevLett.52.2111} {\bibfield  {journal} {\bibinfo
   {journal} {Phys. Rev. Lett.}\ }\textbf {\bibinfo {volume} {52}},\ \bibinfo
  {pages} {2111} (\bibinfo {year} {1984})}\BibitemShut {NoStop}%
\bibitem [{\citenamefont {Kehrein}(2006)}]{Kehrein2006book}%
  \BibitemOpen
  \bibfield  {author} {\bibinfo {author} {\bibfnamefont {S.}~\bibnamefont
  {Kehrein}},\ }\href {https://doi.org/10.1007/3-540-34068-8} {\emph {\bibinfo
  {title} {The Flow Equation Approach to Many-Particle Systems}}},\ \bibinfo
  {series} {Springer Tracts in Modern Physics}, Vol.\ \bibinfo {volume} {217}\
  (\bibinfo  {publisher} {Springer Berlin Heidelberg},\ \bibinfo {year}
  {2006})\BibitemShut {NoStop}%
\bibitem [{\citenamefont {Tomaras}\ and\ \citenamefont
  {Kehrein}(2011)}]{Tomaras2011}%
  \BibitemOpen
  \bibfield  {author} {\bibinfo {author} {\bibfnamefont {C.}~\bibnamefont
  {Tomaras}}\ and\ \bibinfo {author} {\bibfnamefont {S.}~\bibnamefont
  {Kehrein}},\ }\bibfield  {title} {\bibinfo {title} {Scaling approach for the
  time-dependent {K}ondo model},\ }\href
  {https://doi.org/10.1209/0295-5075/93/47011} {\bibfield  {journal} {\bibinfo
  {journal} {Europhys. Lett.}\ }\textbf {\bibinfo {volume} {93}},\ \bibinfo
  {pages} {47011} (\bibinfo {year} {2011})},\ \Eprint
  {https://arxiv.org/abs/1011.1281} {arXiv:1011.1281 [cond-mat.str-el]}
  \BibitemShut {NoStop}%
\bibitem [{\citenamefont {Monthus}(2016)}]{Monthus2016}%
  \BibitemOpen
  \bibfield  {author} {\bibinfo {author} {\bibfnamefont {C.}~\bibnamefont
  {Monthus}},\ }\bibfield  {title} {\bibinfo {title} {Flow towards
  diagonalization for many-body-localization models: adaptation of the {T}oda
  matrix differential flow to random quantum spin chains},\ }\href
  {https://doi.org/10.1088/1751-8113/49/30/305002} {\bibfield  {journal}
  {\bibinfo  {journal} {J. Phys. A: Math. Theor.}\ }\textbf {\bibinfo {volume}
  {49}},\ \bibinfo {pages} {305002} (\bibinfo {year} {2016})},\ \Eprint
  {https://arxiv.org/abs/1602.03064} {arXiv:1602.03064 [cond-mat]} \BibitemShut
  {NoStop}%
\bibitem [{\citenamefont {Thomson}\ and\ \citenamefont
  {Schir{\'o}}(2018)}]{Thomson2018}%
  \BibitemOpen
  \bibfield  {author} {\bibinfo {author} {\bibfnamefont {S.~J.}\ \bibnamefont
  {Thomson}}\ and\ \bibinfo {author} {\bibfnamefont {M.}~\bibnamefont
  {Schir{\'o}}},\ }\bibfield  {title} {\bibinfo {title} {Time evolution of
  many-body localized systems with the flow equation approach},\ }\href
  {https://doi.org/10.1103/physrevb.97.060201} {\bibfield  {journal} {\bibinfo
  {journal} {Phys. Rev. B}\ }\textbf {\bibinfo {volume} {97}},\ \bibinfo
  {pages} {060201(R)} (\bibinfo {year} {2018})},\ \Eprint
  {https://arxiv.org/abs/1707.06981} {arXiv:1707.06981 [cond-mat.diss-nn]}
  \BibitemShut {NoStop}%
\bibitem [{\citenamefont {Thomson}\ and\ \citenamefont
  {Schir{\'o}}(2020)}]{Thomson2020}%
  \BibitemOpen
  \bibfield  {author} {\bibinfo {author} {\bibfnamefont {S.~J.}\ \bibnamefont
  {Thomson}}\ and\ \bibinfo {author} {\bibfnamefont {M.}~\bibnamefont
  {Schir{\'o}}},\ }\bibfield  {title} {\bibinfo {title} {Dynamics of disordered
  quantum systems using flow equations},\ }\bibfield  {journal} {\bibinfo
  {journal} {Eur. Phys. J. B}\ }\textbf {\bibinfo {volume} {93}},\ \href
  {https://doi.org/10.1140/epjb/e2019-100476-3} {10.1140/epjb/e2019-100476-3}
  (\bibinfo {year} {2020}),\ \Eprint {https://arxiv.org/abs/1909.13856}
  {arXiv:1909.13856 [cond-mat.dis-nn]} \BibitemShut {NoStop}%
\bibitem [{\citenamefont {Kelly}\ \emph {et~al.}(2020)\citenamefont {Kelly},
  \citenamefont {Nandkishore},\ and\ \citenamefont {Marino}}]{Kelly2020}%
  \BibitemOpen
  \bibfield  {author} {\bibinfo {author} {\bibfnamefont {S.~P.}\ \bibnamefont
  {Kelly}}, \bibinfo {author} {\bibfnamefont {R.}~\bibnamefont {Nandkishore}},\
  and\ \bibinfo {author} {\bibfnamefont {J.}~\bibnamefont {Marino}},\
  }\bibfield  {title} {\bibinfo {title} {Exploring many-body localization in
  quantum systems coupled to an environment via {W}egner-{W}ilson flows},\
  }\href {https://doi.org/10.1016/j.nuclphysb.2019.114886} {\bibfield
  {journal} {\bibinfo  {journal} {Nucl. Phys. B}\ }\textbf {\bibinfo {volume}
  {951}},\ \bibinfo {pages} {114886} (\bibinfo {year} {2020})},\ \Eprint
  {https://arxiv.org/abs/1902.11295} {arXiv:1902.11295 [cond-mat.dis-nn]}
  \BibitemShut {NoStop}%
\bibitem [{\citenamefont {Wegner}(1994)}]{Wegner1994}%
  \BibitemOpen
  \bibfield  {author} {\bibinfo {author} {\bibfnamefont {F.}~\bibnamefont
  {Wegner}},\ }\bibfield  {title} {\bibinfo {title} {Flow-equations for
  {H}amiltonians},\ }\href {https://doi.org/10.1002/andp.19945060203}
  {\bibfield  {journal} {\bibinfo  {journal} {Annalen der Physik}\ }\textbf
  {\bibinfo {volume} {506}},\ \bibinfo {pages} {77} (\bibinfo {year}
  {1994})}\BibitemShut {NoStop}%
\bibitem [{\citenamefont {G\l{}azek}\ and\ \citenamefont
  {Wilson}(1993)}]{Glazek1993}%
  \BibitemOpen
  \bibfield  {author} {\bibinfo {author} {\bibfnamefont {S.~D.}\ \bibnamefont
  {G\l{}azek}}\ and\ \bibinfo {author} {\bibfnamefont {K.~G.}\ \bibnamefont
  {Wilson}},\ }\bibfield  {title} {\bibinfo {title} {Renormalization of
  {H}amiltonians},\ }\href {https://doi.org/10.1103/PhysRevD.48.5863}
  {\bibfield  {journal} {\bibinfo  {journal} {Phys. Rev. D}\ }\textbf {\bibinfo
  {volume} {48}},\ \bibinfo {pages} {5863} (\bibinfo {year}
  {1993})}\BibitemShut {NoStop}%
\bibitem [{\citenamefont {Glazek}\ and\ \citenamefont
  {Wilson}(1994)}]{Glazek1994}%
  \BibitemOpen
  \bibfield  {author} {\bibinfo {author} {\bibfnamefont {S.~D.}\ \bibnamefont
  {Glazek}}\ and\ \bibinfo {author} {\bibfnamefont {K.~G.}\ \bibnamefont
  {Wilson}},\ }\bibfield  {title} {\bibinfo {title} {Perturbative
  renormalization group for {H}amiltonians},\ }\href
  {https://doi.org/10.1103/PhysRevD.49.4214} {\bibfield  {journal} {\bibinfo
  {journal} {Phys. Rev. D}\ }\textbf {\bibinfo {volume} {49}},\ \bibinfo
  {pages} {4214} (\bibinfo {year} {1994})}\BibitemShut {NoStop}%
\bibitem [{\citenamefont {Magnus}(1954)}]{Magnus1954}%
  \BibitemOpen
  \bibfield  {author} {\bibinfo {author} {\bibfnamefont {W.}~\bibnamefont
  {Magnus}},\ }\bibfield  {title} {\bibinfo {title} {On the exponential
  solution of differential equations for a linear operator},\ }\href
  {https://doi.org/https://doi.org/10.1002/cpa.3160070404} {\bibfield
  {journal} {\bibinfo  {journal} {Comm. Pure Appl. Math.}\ }\textbf {\bibinfo
  {volume} {7}},\ \bibinfo {pages} {649} (\bibinfo {year} {1954})}\BibitemShut
  {NoStop}%
\bibitem [{\citenamefont {Blanes}\ \emph {et~al.}(2009)\citenamefont {Blanes},
  \citenamefont {Casas}, \citenamefont {Oteo},\ and\ \citenamefont
  {Ros}}]{Blanes2009}%
  \BibitemOpen
  \bibfield  {author} {\bibinfo {author} {\bibfnamefont {S.}~\bibnamefont
  {Blanes}}, \bibinfo {author} {\bibfnamefont {F.}~\bibnamefont {Casas}},
  \bibinfo {author} {\bibfnamefont {J.}~\bibnamefont {Oteo}},\ and\ \bibinfo
  {author} {\bibfnamefont {J.}~\bibnamefont {Ros}},\ }\bibfield  {title}
  {\bibinfo {title} {The {M}agnus expansion and some of its applications},\
  }\href {https://doi.org/10.1016/j.physrep.2008.11.001} {\bibfield  {journal}
  {\bibinfo  {journal} {Phys. Rep.}\ }\textbf {\bibinfo {volume} {470}},\
  \bibinfo {pages} {151} (\bibinfo {year} {2009})},\ \Eprint
  {https://arxiv.org/abs/0810.5488} {arXiv:0810.5488 [math-ph]} \BibitemShut
  {NoStop}%
\bibitem [{\citenamefont {Guerin}\ and\ \citenamefont
  {Jauslin}(2003)}]{Guerin2003}%
  \BibitemOpen
  \bibfield  {author} {\bibinfo {author} {\bibfnamefont {S.}~\bibnamefont
  {Guerin}}\ and\ \bibinfo {author} {\bibfnamefont {H.~R.}\ \bibnamefont
  {Jauslin}},\ }\bibfield  {title} {\bibinfo {title} {Control of quantum
  dynamics by laser pulses: adiabatic {F}loquet theory},\ }\href
  {https://doi.org/10.1002/0471428027.ch3} {\bibfield  {journal} {\bibinfo
  {journal} {Adv. Chem. Phys.}\ }\textbf {\bibinfo {volume} {125}},\ \bibinfo
  {pages} {147} (\bibinfo {year} {2003})}\BibitemShut {NoStop}%
\bibitem [{\citenamefont {Golub}\ and\ \citenamefont {{van der
  Vorst}}(2000)}]{Golub2000diag}%
  \BibitemOpen
  \bibfield  {author} {\bibinfo {author} {\bibfnamefont {G.~H.}\ \bibnamefont
  {Golub}}\ and\ \bibinfo {author} {\bibfnamefont {H.~A.}\ \bibnamefont {{van
  der Vorst}}},\ }\bibfield  {title} {\bibinfo {title} {Eigenvalue computation
  in the 20th century},\ }\href {https://doi.org/10.1016/S0377-0427(00)00413-1}
  {\bibfield  {journal} {\bibinfo  {journal} {J. Comput. Appl. Math.}\ }\textbf
  {\bibinfo {volume} {123}},\ \bibinfo {pages} {35} (\bibinfo {year} {2000})},\
  \bibinfo {note} {numerical Analysis 2000. Vol. III: Linear
  Algebra}\BibitemShut {NoStop}%
\bibitem [{\citenamefont {White}(2002)}]{White2002}%
  \BibitemOpen
  \bibfield  {author} {\bibinfo {author} {\bibfnamefont {S.~R.}\ \bibnamefont
  {White}},\ }\bibfield  {title} {\bibinfo {title} {Numerical canonical
  transformation approach to quantum many-body problems},\ }\href
  {https://doi.org/10.1063/1.1508370} {\bibfield  {journal} {\bibinfo
  {journal} {J. Chem. Phys.}\ }\textbf {\bibinfo {volume} {117}},\ \bibinfo
  {pages} {7472} (\bibinfo {year} {2002})},\ \Eprint
  {https://arxiv.org/abs/cond-mat/0201346} {arXiv:cond-mat/0201346}
  \BibitemShut {NoStop}%
\bibitem [{\citenamefont {Hergert}\ \emph {et~al.}(2014)\citenamefont
  {Hergert}, \citenamefont {Bogner}, \citenamefont {Morris}, \citenamefont
  {Binder}, \citenamefont {Calci}, \citenamefont {Langhammer},\ and\
  \citenamefont {Roth}}]{Hergert2014}%
  \BibitemOpen
  \bibfield  {author} {\bibinfo {author} {\bibfnamefont {H.}~\bibnamefont
  {Hergert}}, \bibinfo {author} {\bibfnamefont {S.~K.}\ \bibnamefont {Bogner}},
  \bibinfo {author} {\bibfnamefont {T.~D.}\ \bibnamefont {Morris}}, \bibinfo
  {author} {\bibfnamefont {S.}~\bibnamefont {Binder}}, \bibinfo {author}
  {\bibfnamefont {A.}~\bibnamefont {Calci}}, \bibinfo {author} {\bibfnamefont
  {J.}~\bibnamefont {Langhammer}},\ and\ \bibinfo {author} {\bibfnamefont
  {R.}~\bibnamefont {Roth}},\ }\bibfield  {title} {\bibinfo {title} {Ab initio
  multireference in-medium similarity renormalization group calculations of
  even calcium and nickel isotopes},\ }\href
  {https://doi.org/10.1103/PhysRevC.90.041302} {\bibfield  {journal} {\bibinfo
  {journal} {Phys. Rev. C}\ }\textbf {\bibinfo {volume} {90}},\ \bibinfo
  {pages} {041302(R)} (\bibinfo {year} {2014})},\ \Eprint
  {https://arxiv.org/abs/1408.0655} {arXiv:1408.0655 [hep-th]} \BibitemShut
  {NoStop}%
\bibitem [{\citenamefont {Mielke}(1998)}]{Mielke1998}%
  \BibitemOpen
  \bibfield  {author} {\bibinfo {author} {\bibfnamefont {A.}~\bibnamefont
  {Mielke}},\ }\bibfield  {title} {\bibinfo {title} {Flow equations for
  band-matrices},\ }\href {https://doi.org/10.1007/s100510050485} {\bibfield
  {journal} {\bibinfo  {journal} {Eur. Phys. J. B}\ }\textbf {\bibinfo {volume}
  {5}},\ \bibinfo {pages} {605} (\bibinfo {year} {1998})},\ \Eprint
  {https://arxiv.org/abs/quant-ph/9803040} {arXiv:quant-ph/9803040}
  \BibitemShut {NoStop}%
\bibitem [{\citenamefont {H{\'{e}}non}(1974)}]{Henon1974}%
  \BibitemOpen
  \bibfield  {author} {\bibinfo {author} {\bibfnamefont {M.}~\bibnamefont
  {H{\'{e}}non}},\ }\bibfield  {title} {\bibinfo {title} {Integrals of the
  {T}oda lattice},\ }\href {https://doi.org/10.1103/physrevb.9.1921} {\bibfield
   {journal} {\bibinfo  {journal} {Phys. Rev. B}\ }\textbf {\bibinfo {volume}
  {9}},\ \bibinfo {pages} {1921} (\bibinfo {year} {1974})}\BibitemShut
  {NoStop}%
\bibitem [{\citenamefont {Thomson}\ \emph {et~al.}(2021)\citenamefont
  {Thomson}, \citenamefont {Magano},\ and\ \citenamefont
  {Schir\'{o}}}]{Thomson2021}%
  \BibitemOpen
  \bibfield  {author} {\bibinfo {author} {\bibfnamefont {S.}~\bibnamefont
  {Thomson}}, \bibinfo {author} {\bibfnamefont {D.}~\bibnamefont {Magano}},\
  and\ \bibinfo {author} {\bibfnamefont {M.}~\bibnamefont {Schir\'{o}}},\
  }\bibfield  {title} {\bibinfo {title} {Flow equations for disordered
  {F}loquet systems},\ }\href {https://doi.org/10.21468/SciPostPhys.11.2.028}
  {\bibfield  {journal} {\bibinfo  {journal} {SciPost Phys.}\ }\textbf
  {\bibinfo {volume} {11}},\ \bibinfo {pages} {028} (\bibinfo {year} {2021})},\
  \Eprint {https://arxiv.org/abs/2009.03186} {arXiv:2009.03186 [cond-mat]}
  \BibitemShut {NoStop}%
\bibitem [{\citenamefont {Novi\v{c}enko}\ and\ \citenamefont
  {Juzeli{\=u}nas}(2019)}]{Novicenko2019}%
  \BibitemOpen
  \bibfield  {author} {\bibinfo {author} {\bibfnamefont {V.}~\bibnamefont
  {Novi\v{c}enko}}\ and\ \bibinfo {author} {\bibfnamefont {G.}~\bibnamefont
  {Juzeli{\=u}nas}},\ }\bibfield  {title} {\bibinfo {title} {Non-{A}belian
  geometric phases in periodically driven systems},\ }\href
  {https://doi.org/10.1103/PhysRevA.100.012127} {\bibfield  {journal} {\bibinfo
   {journal} {Phys. Rev. A}\ }\textbf {\bibinfo {volume} {100}},\ \bibinfo
  {pages} {012127} (\bibinfo {year} {2019})},\ \Eprint
  {https://arxiv.org/abs/1811.06045} {arXiv:1811.06045 [quant-ph]} \BibitemShut
  {NoStop}%
\bibitem [{\citenamefont {Anisimovas}\ \emph {et~al.}(2015)\citenamefont
  {Anisimovas}, \citenamefont {{\v Z}labys}, \citenamefont {Anderson},
  \citenamefont {Juzeli{\=u}nas},\ and\ \citenamefont
  {Eckardt}}]{Anisimovas2015}%
  \BibitemOpen
  \bibfield  {author} {\bibinfo {author} {\bibfnamefont {E.}~\bibnamefont
  {Anisimovas}}, \bibinfo {author} {\bibfnamefont {G.}~\bibnamefont {{\v
  Z}labys}}, \bibinfo {author} {\bibfnamefont {B.~M.}\ \bibnamefont
  {Anderson}}, \bibinfo {author} {\bibfnamefont {G.}~\bibnamefont
  {Juzeli{\=u}nas}},\ and\ \bibinfo {author} {\bibfnamefont {A.}~\bibnamefont
  {Eckardt}},\ }\bibfield  {title} {\bibinfo {title} {Role of real-space
  micromotion for bosonic and fermionic {F}loquet fractional {C}hern
  insulators},\ }\href {https://doi.org/10.1103/PhysRevB.91.245135} {\bibfield
  {journal} {\bibinfo  {journal} {Phys. Rev. B}\ }\textbf {\bibinfo {volume}
  {91}},\ \bibinfo {pages} {245135} (\bibinfo {year} {2015})},\ \Eprint
  {https://arxiv.org/abs/1504.03583} {arXiv:1504.03583 [cond-mat.quant-gas]}
  \BibitemShut {NoStop}%
\bibitem [{scr()}]{script}%
  \BibitemOpen
  \href@noop {} {}\bibinfo {howpublished}
  {\url{http://www.itpa.lt/~novicenko/files/mathematica_scripts/toda_script.nb}}\BibitemShut
  {NoStop}%
\end{thebibliography}%

\end{document}